\documentclass[a4paper,fleqn,usenatbib]{mnras}

\usepackage{newtxtext,newtxmath}

\usepackage[T1]{fontenc}
\usepackage{ae,aecompl}

\usepackage{graphicx}	
\usepackage{amsmath}	
\usepackage{amssymb}	
\usepackage[]{algorithm2e}
\usepackage[caption=false]{subfig}
\graphicspath{{figures/}}

\title[Pulsar timing in EMRB]{Pulsar timing in extreme mass ratio binaries: a general relativistic approach}

\author[T. Kimpson et al.]{
Tom Kimpson,$^{1}$\thanks{E-mail: tom.kimpson.16@ucl.ac.uk}
Kinwah Wu ,$^{1}$
and Silvia Zane $^{1}$
\\
$^{1}$ Mullard Space Science Laboratory, University College London. Holmbury St. Mary, Dorking, Surrey, RH5 6NT, UK
}

\date{Accepted XXX. Received YYY; in original form ZZZ}

\pubyear{2018}

\begin{document}
\label{firstpage}
\pagerange{\pageref{firstpage}--\pageref{lastpage}}
\maketitle

\begin{abstract}
The detection of a pulsar (PSR) in a tight, relativistic orbit around a supermassive or intermediate mass black hole - such as those in the Galactic centre or in the centre of Globular clusters - would allow for precision tests of general relativity (GR) in the strong-field, non-linear regime. We present a framework for calculating the theoretical time-frequency signal from a PSR in such an Extreme Mass Ratio Binary (EMRB). This framework is entirely relativistic with no weak-field approximations and so able to account for all higher-order strong-field gravitational effects, relativistic spin dynamics, the convolution with astrophysical effects and the combined impact on the PSR timing signal. Specifically we calculate both the spacetime path of the pulsar radio signal and the complex orbital and spin dynamics of a spinning pulsar around a Kerr black hole, accounting for spacetime curvature and frame dragging, relativistic and gravitational time delay, gravitational light bending, temporal and spatial dispersion induced by the presence of plasma along the line of sight and relativistic aberration. This then allows for a consistent time-frequency solution to be generated. Such a framework is key for assessing the use of PSR as probes of strong field GR, helping to inform the detection of an EMRB system hosting a PSR and, most essentially, for providing an accurate theoretical basis to then compare with observations to test fundamental physics.
\end{abstract}

\begin{keywords}
gravitation -- pulsars -- black hole physics
\end{keywords}



\section{Introduction}
The Galactic centre is thought to host a central black hole (BH) of mass $4.3 \times 10^6 M_{\odot}$ \citep{Gillessen2009}, and be surrounded by up to $10^3 - 10^4$ pulsars (PSRs) within the central parsec \citep{Wharton2012, Rajwade2017}. Moreover -  if the ``$M$-$\sigma$ relation'' \citep{Ferrarese2000} holds universally - the centre of globular clusters should each contain a nuclear black hole with an `intermediate' mass (IMBH, $M \sim 10^3 - 10^5\;\!{\rm M}_\odot$, see e.g. \citealt{Feng2011} for discussion of IMBH as ultraluminous X-ray sources and \citealt{Perera2017MNRAS} for the identification of a IMBH candidate in the globular cluster NGC~6624, based on PSR timing observations). The pulsar rate in Globular clusters is thought to be enhanced by a factor of $10^2 - 10^3$ per unit mass compared to the Galactic disk \citep{Freire2013}. Such systems where a pulsar attains a compact orbit around a massive ($\gtrsim 10^3 M_{\odot}$) BH are known as Extreme Mass Ratio Binaries (EMRBs). \newline

\noindent The detection and timing of PSR-EMRB systems presents several challenges. Currently no radio pulsars have been detected within 1 parsec of Sgr A* \citep{Macquart2010, Rajwade2017}. Neither have we detected pulsars in globular clusters at sufficiently compact orbits. A key challenge is the construction of a timing model which can be applied to the highly relativistic regimes which these systems inhabit. An accurate and precise model of the time-frequency behaviour of a PSR-EMRB is essential for both their detection - especially given the large data flow and real-time processing of the next generation radio telescopes like SKA \citep{Norris2011} - and also in order to utilise these systems as a scientific apparatus, and to best exploit the vastity of physical information which may be encoded in observational data.

	The main problem is that, in order to be realistic (and feasible for a concrete application to timing data), an accurate theoretical model of the radio signal of a PSR in an EMRB must consider a series of effects, which can be broadly classified into two categories:
	\begin{enumerate}
		\item \noindent Effects that influence the behaviour of the light through a modification of  the geodesic followed by a radio pulsar light ray while traveling through a curved spacetime.  In this category we can list, e.g., gravitational bending, gravitational and relativistic time dilation, relativistic Doppler frequency shift \citep{Fuerst2004,Saxton2016}, velocity induced intensity boost and also some astrophysical effects \citep[e.g. the temporal and spatial dispersion induced by the interaction with  line-of-sight material: interstellar medium, accretion disks, stellar winds ][]{Fuerst2007,Younsi2012, Kimpson2018}.
		\item \noindent Effects related to the relativistic orbital dynamics of a spinning pulsar around a spinning black hole, e.g. spin coupling, curvature coupling and time dilations (gravitomagnetic, relativistic, gravitational).
	\end{enumerate}
	The two categories are naturally linked:  the orbital dynamics of the pulsar provides varying initial conditions for the light ray trajectory to the observer, and both effects work in tandem to influence the key observables, i.e. the photon time of arrival (ToA) and photon frequency. 
	
	This work provides a first step in this direction. By using a ray tracing method, we develop a 
	fully general relativistic approach to calculate the signal from a PSR in an EMRB, accounting for both spin and orbital motion. The method naturally includes phase-dependent relativistic effects \citep[which are important for the relatively longer orbital periods of EMRB systems, see discussion in][]{Zhang2017} and spin precession. We work in an approximated framework inasmuch we restrict our study to the extreme mass ratio of EMRB systems and so do not consider PSRs in stellar-mass black hole binaries with finite mass ratios \citep[e.g.][]{Blanchet2014,Liu2014}. Also, in this work we do not address the task of how to  perform mock data analysis nor consider external Newtonian perturbations due to the gravitational foreground from other bodies(see \S~6 for a discussion of both of these extensions). \newline
	
	\noindent Beside being important ``per se'', the discovery of a PSR in an EMRB may provide us with a powerful laboratory to investigate fundamental aspects of general relativity (GR) in regimes of large gravitational potential, $\epsilon = {GM}/{rc^2} $, and spacetime curvature, $\xi_c = GM/r^3 c^2$ \citep[see e.g. ][]{Kramer2004, Wang2008, Wang2009, Liu2012,Remmen2013, Nampalliwar2013, Singh2014,Kramer2016, Saxton2016, Li2018}. This may be achievable  thanks to both,  the extreme gyroscopic stability of pulsars \citep[e.g.][]{Verbiest2009} and the high-precision measurements that are attainable with radio pulsar timing observations \citep[e.g.][]{Liu2011, Desvignes2016,Lazarus2016, Liu2018}. Several aspects of GR have been tested experimentally, with electromagnetic observations of the solar system \citep{Will2014}, binary pulsar systems \citep{Lorimer2008} and also with the recent breakthrough observation via gravitational radiation of binary BH-BH and BH-Neutron Star coalescing systems \citep{AbbotBHb, AbbottNS}. However there remain open questions in GR, particularly concerning the non-uniqueness of the Einstein-Hilbert action \citep{Psaltis2008}, the existence of singular matter densities and curvatures \citep{Pachner1971}, in addition to problems regarding spinning objects in gravity, \citep[see e.g.][]{Plyatsko2016} and the spacetime dynamics associated with multiple spinning objects revolving around each other \citep{Obukhov2009, Ambrosi2015}. All existing observations  based on electromagnetic radiation have probed the weak-field regime, where $\epsilon \lesssim 10^{-6}, \xi_c \lesssim 10^{-28} $.  The exploration of a more extreme parameter space is achievable through gravitational wave observations  of binary black hole and binary BH-neutron stars mergers, or pulsar timing of extreme (and still undetected) systems, as those considered in this work. In particular, the expected precision attainable with pulsar timing potentially allows for system parameters to be determined with a remarkable accuracy: for a pulsar in a 0.3 year period orbit around Sgr A$^*$, determination of the post-Keplerian parameters allows the mass and spin to be determined to a precision of $10^{-5}$ and $10^{-3}$, respectively \citep[see][]{Liu2014}. At shorter orbital radii, relativistic effects would become more pronounced, potentially allowing for greater precision tests and probing deeper into the strong-field regime. 
	\newline 
	If PSRs in an EMRB will be discovered, their precision timing with the next generation of radio telescopes such as SKA \citep{Shao2015} or FAST \citep{Smits2009} would then allow for an unrivaled probing of physics in an extreme gravity environment and precision measurements of the black hole parameters. This in turn may open a vast array of opportunities: with the determination of the mass and spin of the black hole, the spacetime is wholly determined and key questions of GR can then be investigated; the Cosmic Censorship Conjecture and the No Hair Theorem can then be possibly tested to $\le 1 \%$ precision \citep[see e.g.][]{Kramer2004, Liu2012, Liu2014, Wex1999,Eatough2015}. By extension, the nature of the central massive compact object can also be investigated, i.e. whether the Kerr solution is a true astrophysical solution or if parametrized deviations from the Kerr solution exist, so called `bumpy' black holes \citep{Yagi2016}, or even whether the massive central compact object is some more exotic object like a boson star \citep{Kleihaus2012}. Pulsar timing of EMRBs can also be used to test alternative theories of gravity such as scalar-tensor theories \citep{Liu2014}, search for quantum gravitational effects \citep{Yagi2016, Estes2017} and constrain the cosmological constant \citep{Iorio2018}. Moreover, EMRB systems are also of great interest in their own right, as they are progenitors of the extreme-mass-ratio-inspiral (EMRI) gravitational-wave sources, a major class of targets for the future LISA observations \citep{Gair2010, Gair2017, Babak2017}. Furthermore, such a system can be used for astrophysical purposes, e.g. precisely determining the mass of black holes at the centre of Local Group galaxies and globular clusters in order to constrain the low end of the $M-\sigma$ relation \citep{Ferrarese2000} and establishing the existence of otherwise of intermediate mass black holes \citep{Singh2014}. \newline

	\noindent In this work we compute the signal emitted by a spinning PSR in orbit around a Kerr black hole, accounting for the interaction between the radiation and the material close to the black hole, and by using a ray tracing method. The orbital dynamics of the spinning pulsar is computed considering the coupling between the pulsar spin and the spacetime curvature. We derive the theoretical time-frequency signal from a PSR in an EMRB, and discuss the key factors which influence the signal. \newline
	
	\noindent This paper is structured as follows. In Section \ref{section:raytracing} we present the equations for ray tracing a photon that propagates through a non-gravitating plasma in the Kerr spacetime. In Section \ref{section:MPD} we review how to determine the orbital dynamics of an extended spinning body (i.e. a PSR) in the extreme mass ratio limit. In Section \ref{sec:algorithm} we then combine the ray tracing and orbital dynamics calculations, and present a method for constructing the complete time-frequency signal from the pulsar. In Section \ref{section:application} we apply this method to investigate the various relativistic factors which influence the photon time of arrival. Discussion and conclusions follow in Section \ref{sec:discussion}.
	
	\noindent We adopt the natural units, with $ c=G=\hbar = 1$, and a $(-,+,+,+)$ metric signature.  Unless otherwise stated, a c.g.s. Gaussian unit system is used 
	in the expressions for the electromagnetic properties of matter. The gravitational radius of the black hole is $r_{\rm g} = M$ and the corresponding Schwarzschild radius is $r_{\rm s} = 2M$, where $M$ is the black-hole mass. A comma denotes a partial derivative (e.g.$\;\! x_{,r}$), and a semicolon denotes the covariant derivative (e.g.$\;\! x_{;r}$).

\section{Ray Tracing of light under gravity and influence of plasma}
\label{section:raytracing}
\subsection{Formulation}
The Hamiltonian for the photon propagation in vacuo is  
\begin{eqnarray}
H(x^\mu,k^\mu) = \frac{1}{2} g_{\mu \nu} k^{\mu} k^{\nu }   \ , 
\end{eqnarray} 
and the corresponding equations of motion, Hamilton's equations are 
\begin{eqnarray}
\dot{x^{\mu}} = \frac{\partial H}{\partial k_{\mu}} \, , \hspace*{0.2cm} \dot{k}_{\mu} = -\frac{\partial H}{\partial x^{\mu}} \ , 
\end{eqnarray}
where $g_{\mu \nu}$ is the spacetime metric, $k^\mu$ the contravariant 4-momenta, $x^{\mu}$ the spacetime coordinates, and an overdot denotes the derivative with respect to some affine parameter.

\noindent For photon propagation through a cold, non-magnetized electron-proton plasma, 
the Hamiltonian has an additional terms, expressed in terms of the 
electron plasma frequency $\omega_{\rm{p}}(x^\mu)$:  
\begin{eqnarray}
H(x^\mu,k^\mu) = \frac{1}{2} \left[ g_{\mu \nu} k^{\mu} k^{\nu } + \omega_{\rm{p}}(x^\nu)^2\right] \ ,
\label{eq:PlasmicHamiltonian}
\end{eqnarray}  
in the geometrical optics approximation \citep{Synge1960}. 
The plasma frequency of a cold non-magnetised plasma is $\omega^2_{\rm{p}} = C\;\! n$,         
where $C = 4\pi e^2/m_e$ with $m_e$ the electron mass,  $e$ the electron charge, and $n$ is the electron number density.

\noindent Due to the conservation of angular momentum, astrophysical black holes are thought to be rotating. The vacuo spacetime is then described by the Kerr metric, with spacetime interval,
\begin{eqnarray}
{\rm d}s^2 = -\left(1 - \frac{2Mr}{\Sigma}\right) {\rm d}t^2 
- \frac{4aMr \sin^2 \theta}{\Sigma}\ {\rm d}t \;\! {\rm d}\phi 
+ \frac{\Sigma}{\Delta}{\rm d}r^2 + \Sigma\ {\rm d} \theta^2 \nonumber \\ 
\hspace*{1.2cm} + \frac{\sin^2 \theta}{\Sigma} \left[(r^2+a^2)^2 - \Delta a^2 \sin^2 \theta \right] {\rm d}\phi^2 \ , 
\label{eq:kerr_metric} 
\end{eqnarray}
where $\Sigma = r^2 + a^2 \cos^2 \theta$, $\Delta = r^2 - 2Mr +a^2$, $a$ is the black-hole spin parameter and we work in Boyer-Lindquist coordinates. Going forward, we normalize the black hole mass $M=1$ and so the spacetime interval is then a lengthscale in terms of the gravitational radius $r_{\rm g} (=1)$.  \newline 

\noindent A vector $\xi^{\mu}$ which satisfies the Killing equation,
\begin{eqnarray}
\xi^{(\mu;\nu)} = 0 \ ,
\end{eqnarray}
is described as a Killing vector. Killing vectors are associated with the existence of spacetime symmetries since the inner product of a Killing vector with a tangent vector is conserved along a geodesic, i.e. if $\mathcal{K}_{\mu} =\xi^{\mu}
k_{\mu} $ then $\dot{\mathcal{K}_{\mu}} = 0$. The Kerr spacetime possesses two Killing vectors $\xi^{t}, \xi^{\phi}$. The temporal Killing vector related with the stationarity of the spacetime is associated with the conservation of energy, $k_t = -E$. The azimuthal Killing vector is related with the asymmetry of the spacetime and associated with the angular momentum, $k_{\phi} = L_z$, where $L_z$ is the projection of the particle angular momentum along the black hole spin axis.  In addition to the conservation of $E, L_{z}$, the particle rest mass rest mass is also conserved ($H=0$ for photons). Moreover, the Kerr spacetime possesses an additional rank-2 Killing tensor which in the vacuum case can be associated with Carter's Constant, $Q$, \citep{Carter1968}, related to the separability of the Hamiltonian in $r$ and $\theta$ terms. The exact physical meaning of $Q$ is discussed in \citet{DeFelice1999, Rosquist2009}. 

\begin{figure}
	\includegraphics[width=\columnwidth]{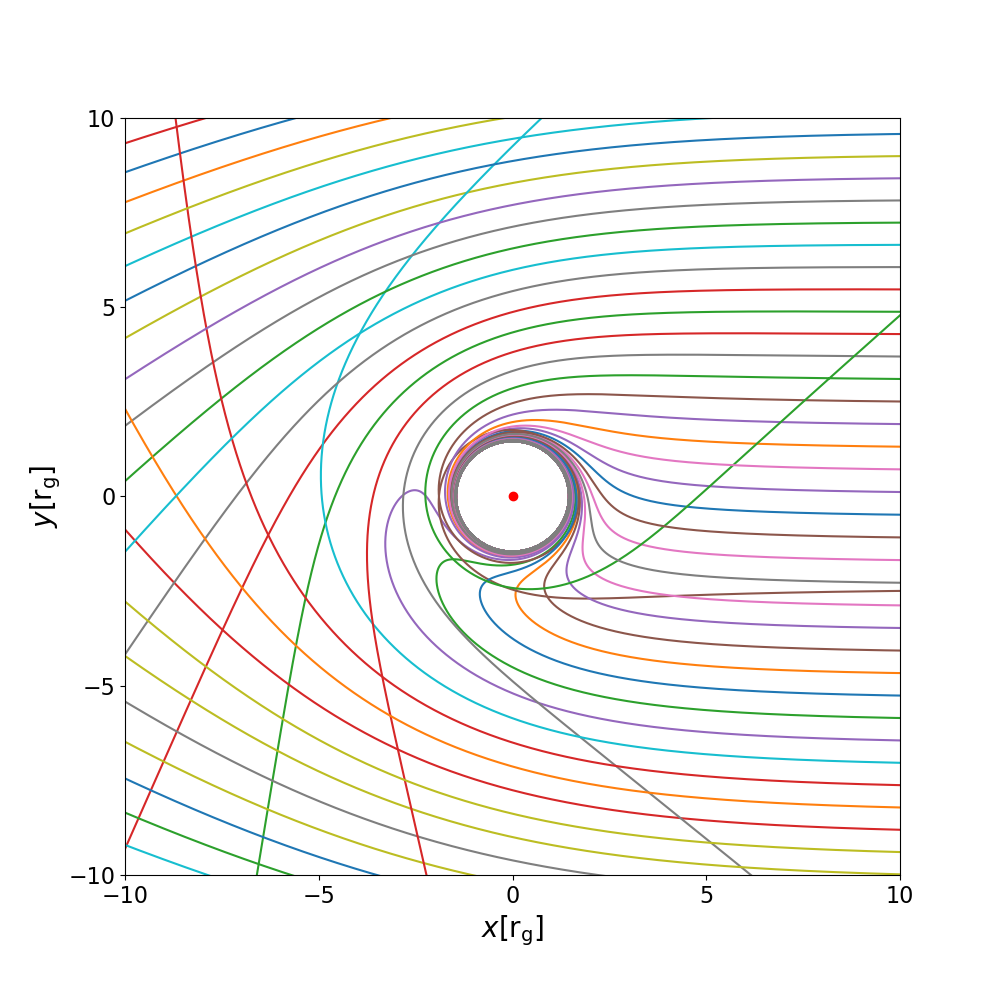}
	\caption{Light ray trajectories through a vacuum around a spinning (Kerr, $a=0.998$) black hole, integrated with a 5th order Runge-Kutta method from a distant observer in the positive $x$-direction.}
	\label{fig:raytracing}
\end{figure}

\noindent For the general plasma density distributions, where the plasma frequency, $\omega_{\rm{p}}(r,\theta)$, has a spatial dependence, the Hamiltonian is not separable in terms of the usual co-ordinate variables and the vacuum Carter constant is no longer a constant along the geodesic, \citep[see e.g.][]{Perlick2017, Kimpson2018}. In order for the  equations of motion to be integrable, it is necessary for $\omega_p$ to take the form,
\begin{eqnarray}
\omega^2_{\rm{p}} = C \ \frac{f(r)+g(\theta)}{\Sigma (r,\theta)}  \  ,
\label{eq:plasma}
\end{eqnarray} 
where $f(r)$ and $g(\theta)$ are two general functions of the coordinate variables $r,\theta$. This represents a plasma density distributions with independent radial and polar terms. From the standpoint of astrophysical modelling, the form of $\omega^2_{\rm{p}}$ has specific advantageous properties, allowing the description - with an appropriate choices of $f(r)$ and $g(\theta)$ - of the characteristic features of an axisymmetric density profile. We recognize that this form of $\omega^2_{\rm{p}}$ will fundamentally be an approximation to the true astrophysical plasma frequency profile. However, we adopt this functional form in our calculations to serve as a decent first-order approximation to the type of plasma density profiles we expect. This then allows for the development of useful insights into the impact of non-vacuum effects on the time-frequency model signal from PSR in EMRB systems and enables the construction of more advanced model calculations that allow us to extract information from observations to test fundamental aspect gravitational physics in the future. The separable form of Eq. \ref{eq:plasma} allows us to define an analogous Carter constant, $Q_p$, for the `plasmic' Hamiltonian (that is, Eq. \ref{eq:PlasmicHamiltonian}).  The equations of motion therefore reduce to a problem of quadratures whereby we have four ordinary differential equations $(\dot{t}, \dot{r}, \dot{\theta}, \dot{\phi})$  and four associated constants of motion $E,$ $L_z$, $Q_p$, and $H$.  The system of equations is integrable.  It follows that the complete set of equations of motion is given, via Hamilton's equations, as \citep{Kimpson2018},
\begin{align}
\dot{t} &= E + \frac{2r(r^2 +a^2)E - 2arL_z}{\Sigma \Delta}  \  ;  \\ 
\dot{r} &= \frac{p_r \Delta}{\Sigma}  \ ;  \\ 
\dot{\theta} &= \frac{p_{\theta}}{\Sigma} \ ;    \\ 
\dot{\phi} &= \frac{2arE + (\Sigma - 2r)L_z\csc^2\theta}{\Sigma \Delta} \ ;   \\ 
\dot{k}_{\theta} &= \frac{1}{2 \Sigma} \left[-C g (\theta)_{, \theta} 
-2a^2 E^2 \sin \theta \cos \theta + 2L_z^2 \cot \theta \csc^2 \theta \right] \ ;  \\ 
\dot{k}_r &= \frac{1}{\Sigma \Delta} \biggl[-\kappa(r-1) +2r(r^2+a^2)E^2 - 2aEL \\ 
&-  \frac{C f(r)_{,r} \Delta}{2}  -C(r-1) f(r)\biggl] - \frac{2 p_r^2(r-1)}{\Sigma}  \nonumber\\ 
\label{eq:pr}
\end{align}
where $\kappa = p_{\theta}^2 +  E^2a^2 \sin^2 \theta + L_z^2 \csc ^2 \theta + a^2 \omega_{\rm p}^2 \cos^2 \theta$.
The two integration constants, $E$ (energy at infinity) and $L_z$ (the azimuthal component of the angular momentum at infinity), 
can be determined by the initial conditions using the following relations:  
\begin{eqnarray}
E^2 = (\Sigma - 2r) \left(\frac{\dot{r}^2}{\Delta} + \dot{\theta}^2 
+ \frac{ \omega_{\rm{p}}^2}{\Sigma}\right) + \Delta \dot{\phi}^2 \sin^2 \theta   \ ;   \\ 
L_z = \frac{(\Sigma \Delta \dot{\phi}-2arE)\sin^2\theta}{\Sigma - 2r}   \ . 
\end{eqnarray}  
Whilst it is possible to express the vacuum equations in terms of elliptic integrals \citep[see e.g.][]{Dexter2009}, for our case we directly integrate the set of equations numerically (e.g. Fig \ref{fig:raytracing}) along the ray using a 5th-order Runge-Kutta integrator \citep{Press1992}. The complete specification of the initial conditions is described in the Appendix.

\section{Spinning object dynamics in GR}
\label{section:MPD}
The most basic approximation in GR is that of the test particle which has no spin, no internal structure and is not subject to self-force effects. Such a particle then directly follows a geodesic of the spacetime metric. However, real astrophysical objects are not in actuality test particles and to obtain an accurate description of their dynamics, higher-order effects must be considered. For a system in which one spinning object (e.g. a PSR) orbits a much more massive one (e.g. Galactic centre Kerr black hole), it is possible to construct two Hamiltonians in the extreme mass ratio and non-relativistic limit, one for the spin-orbit interaction and one for the spin-spin coupling \citep[e.g.][]{Iorio2012}. This approach is sufficient to reproduce the conventional Lense-Thirring precession effects, however it neglects the spin-curvature coupling between the two masses. This coupling occurs because the spin of the small mass itself will modify the Kerr metric of the spacetime. Consequently, in the presence of spin-curvature coupling, the PSR will not follow a geodesic through the Kerr spacetime. Instead we must model the PSR motion through an alternative means, known as the Mathisson-Papatrou-Dixon (MPD) formalism \citep{Mathisson1937,Papapetrou1951,Dixon1974}.
\subsection{Spin-Interaction in MPD formalism}
\begin{figure}
	\subfloat[]{%
		\includegraphics[clip,width=\columnwidth]{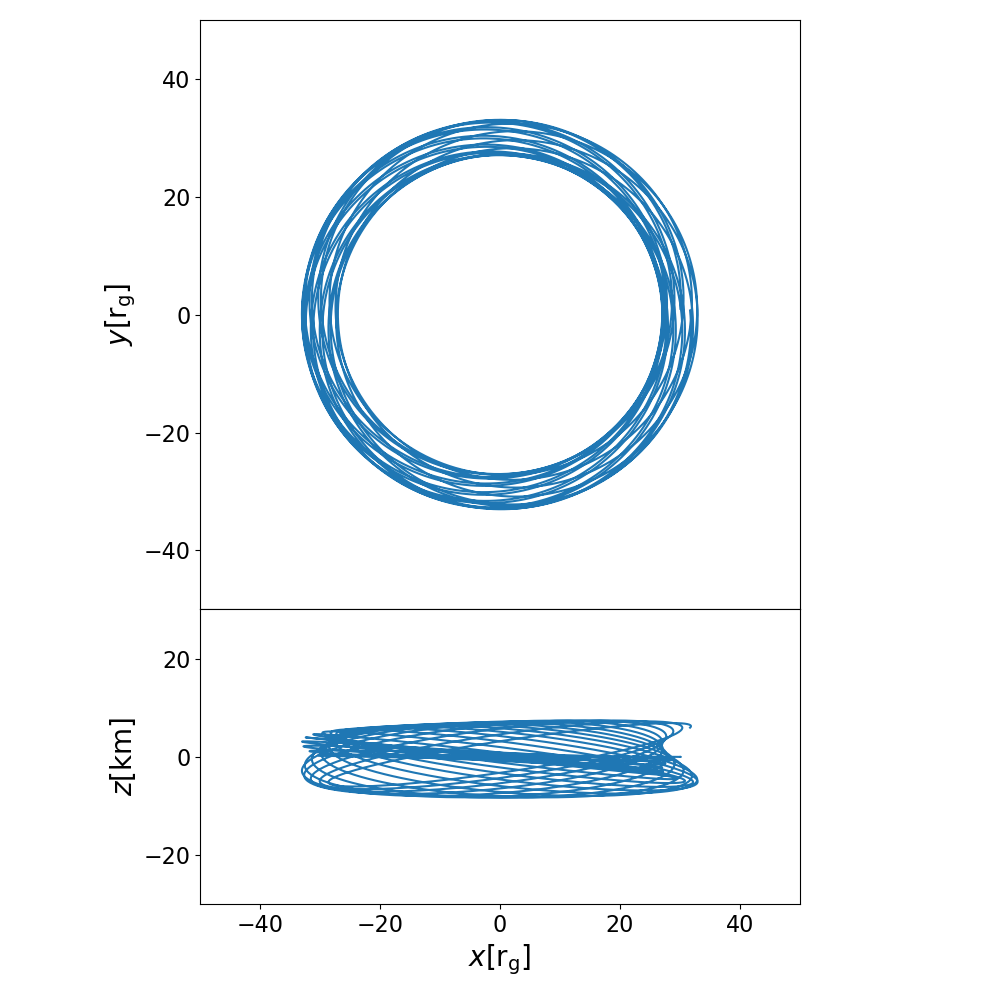}%
	}

	\subfloat[]{%
		\includegraphics[clip,width=\columnwidth]{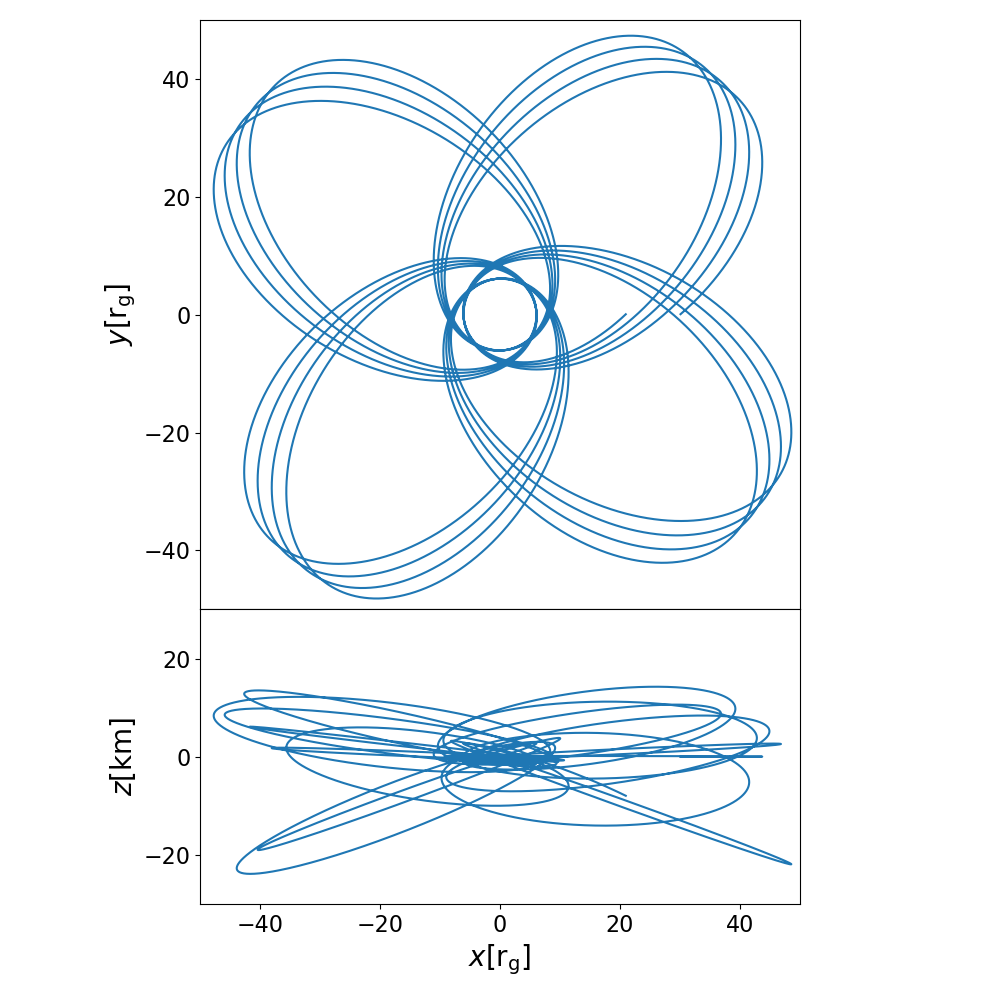}%
	}

	\caption{Trajectories of a spinning particle around a Kerr ($a=0.998$ black hole) over 20 orbits. The particle is initialised in the orbital plane ($\theta = \pi/2$) with initial spin orientation $S_{\theta} = S_{\phi}= \pi/4$, semi-major axis $30 \, r_{\rm g}$ and eccentricity  (a) $e=0.1$, (b) $e=0.8$. Without spin-curvature coupling, a particle initialised in the equatorial plane would remain in the plane ($z = 0$). Relativistic effects such as the precession of periastron and the complex $z$-motion due to spin-curvature coupling will influence the pulse arrival times.}
	\label{fig:spinorbit}
\end{figure}

\begin{figure}
	
	\subfloat[]{%
		\includegraphics[clip,width=\columnwidth]{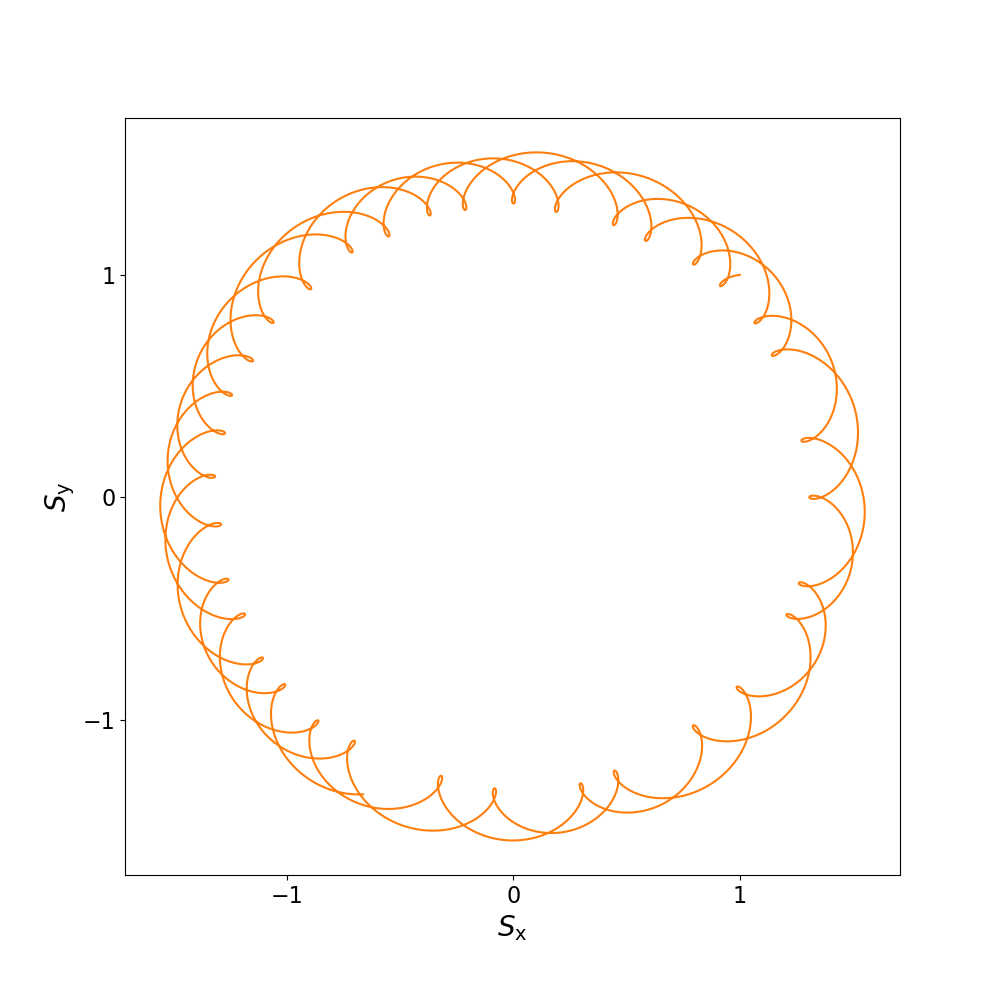}%
	}
	
	\subfloat[]{%
		\includegraphics[clip,width=\columnwidth]{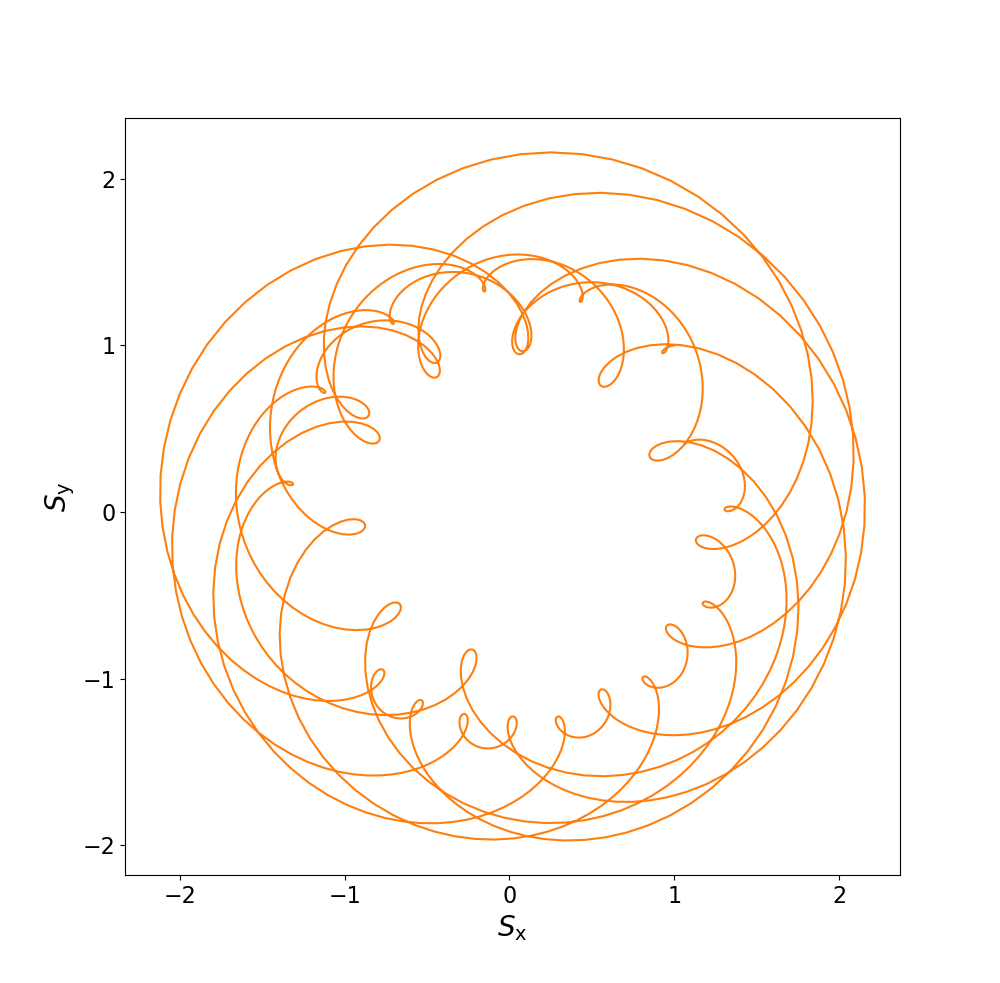}%
	}
	
	\caption{Precession and nutation of the spin axis for a MSP in a prograde orbit around a spinning (Kerr, $a=0.998$) BH with eccentricity (a) $e=0.1$ (b) $e = 0.8$ in the equatorial plane with semi-major axis 10 $r_{\rm g}$. The shift in the spin axis orientation will affect the pulse arrival time and the pulse profile. The $x$ and $y$ components of the spin vector, $S_x$, $S_y$ have been normalised with respect to their initial values.}
	\label{fig:spinprecession}
\end{figure}

\begin{figure*} 
	\subfloat[]{\includegraphics[width=0.333\textwidth]{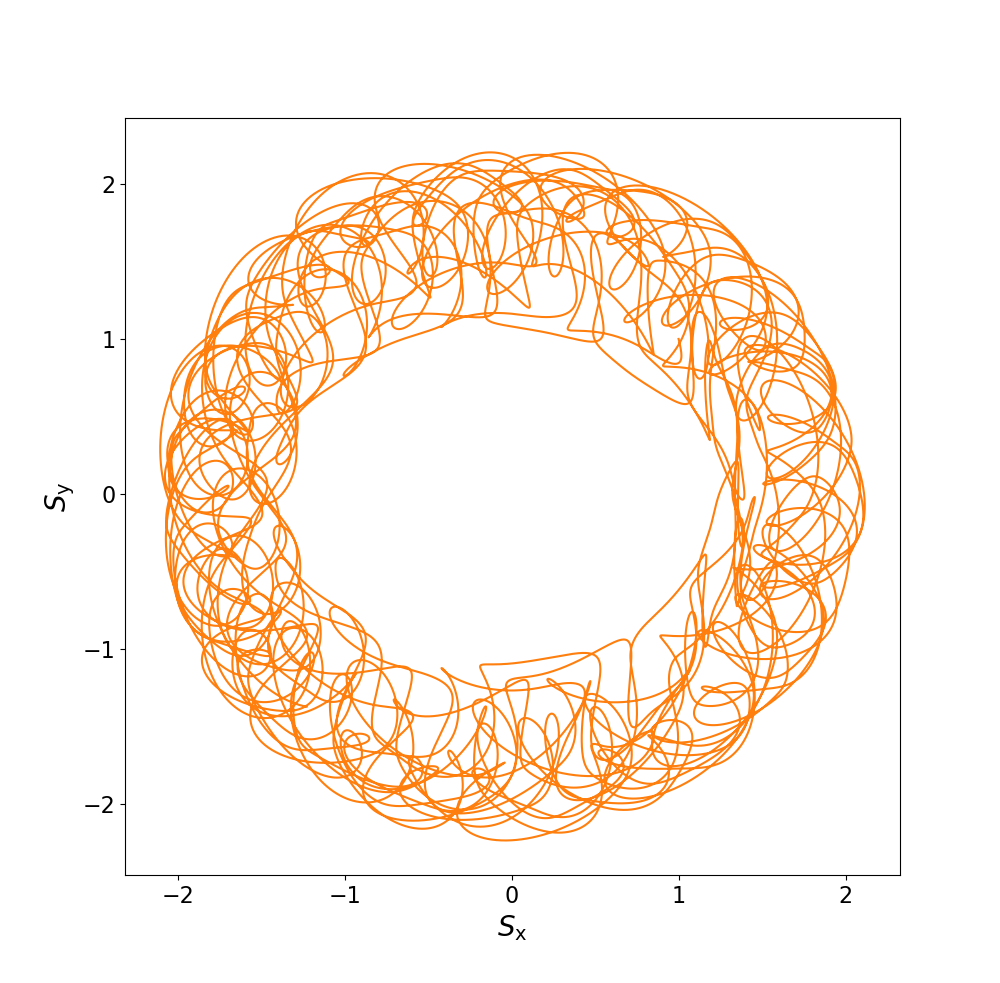}}
	\subfloat[]{\includegraphics[width=0.333\textwidth]{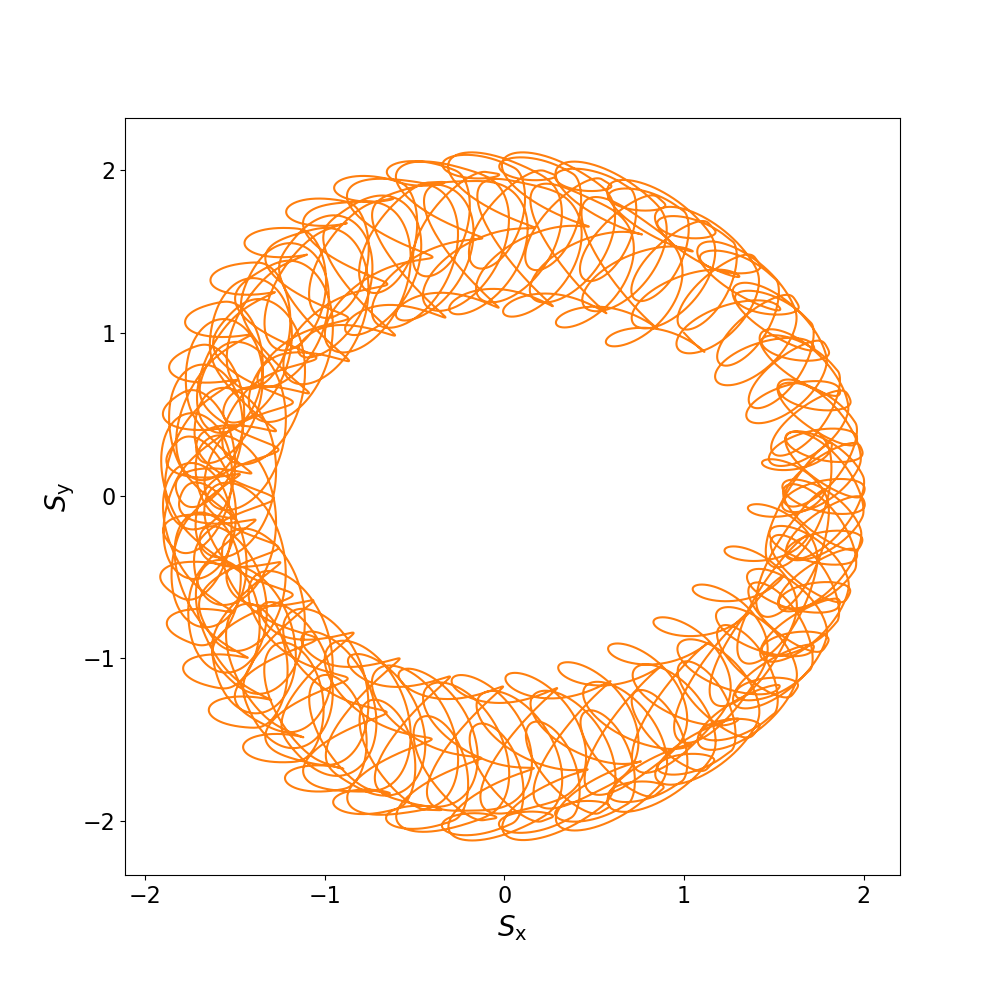}}
	\subfloat[]{\includegraphics[width=0.333\textwidth]{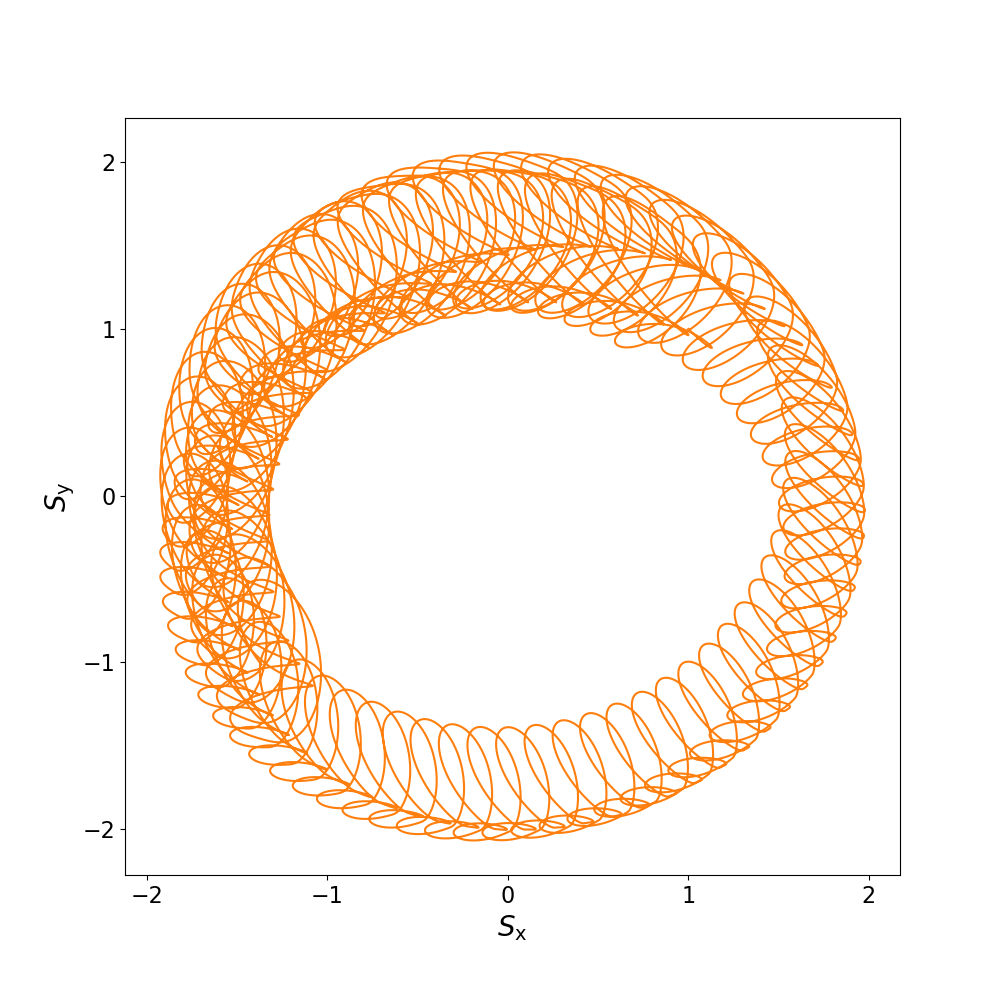}} \\
	
	\subfloat[]{\includegraphics[width=0.333\textwidth]{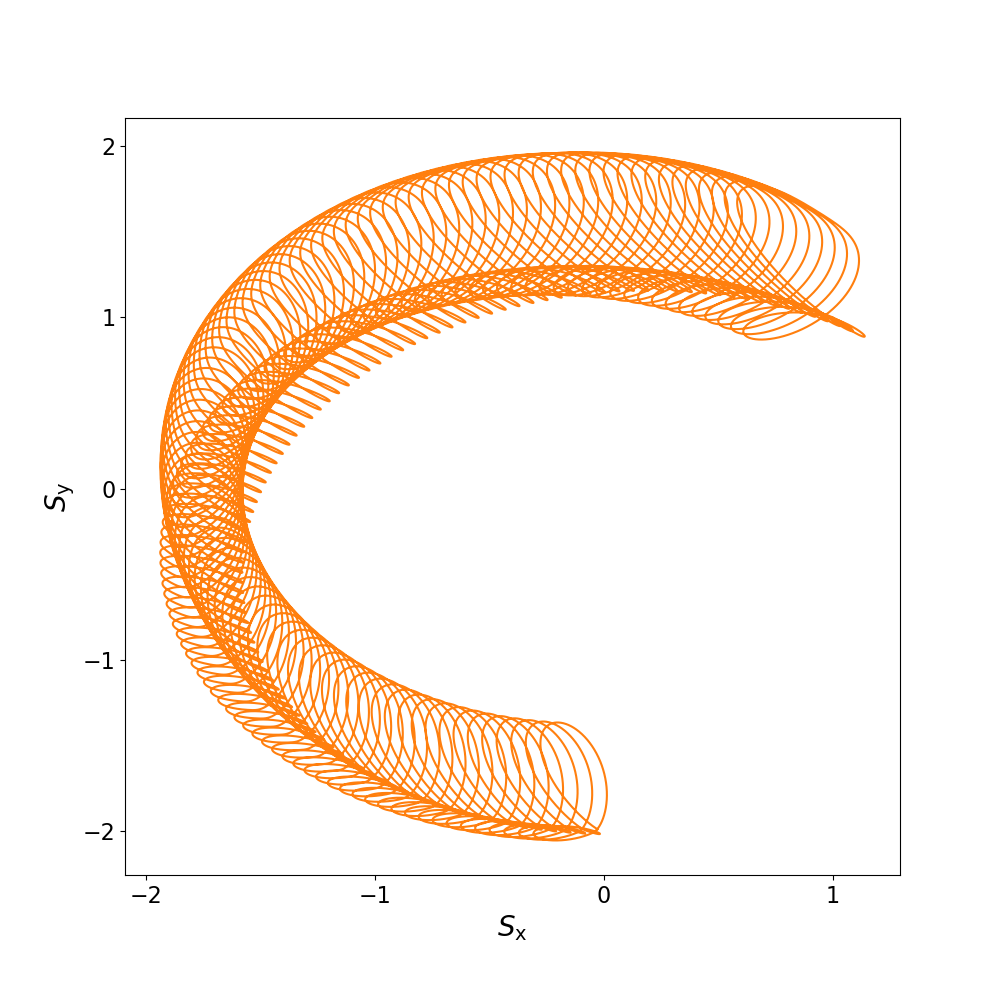}}
	\subfloat[]{\includegraphics[width=0.333\textwidth]{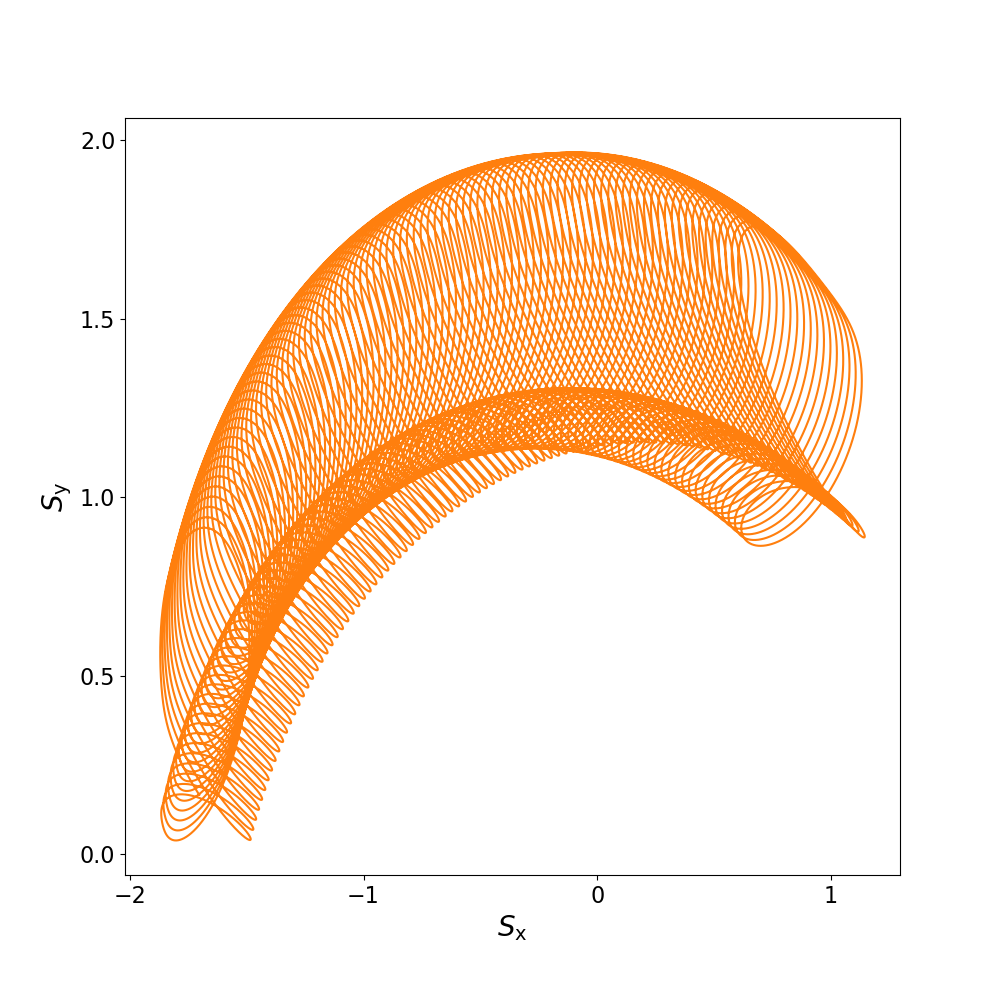}}
	\subfloat[]{\includegraphics[width=0.333\textwidth]{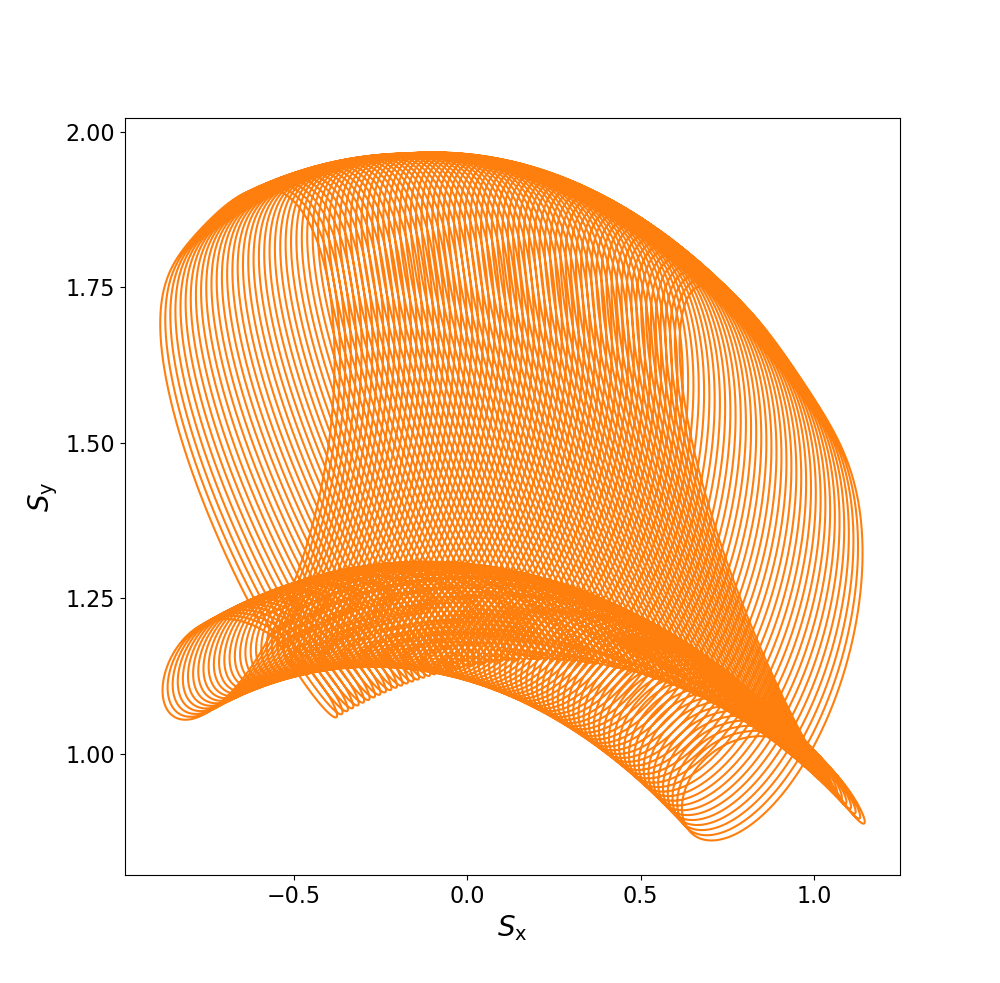}} \\
	
	\medskip

\caption{Complex precession of pulsar spin axis over 100 orbits for some non-equatorial orbital motion with an almost circular eccentricity and semi-major axis (a) 10 $r_{\rm g}$ (b) 50 $r_{\rm g}$ (c) 100 $r_{\rm g}$ (d) 250 $r_{\rm g}$ (e) 500 $r_{\rm g}$ (f) 1000 $r_{\rm g}$. As the semi-major axis shrinks and the pulsar encounters stronger spacetime curvatures the spin precession shifts from periodic to disordered.} \label{fig:spinprecession2}
\end{figure*}

The most general equation of motion of the PSR is given by,
\begin{eqnarray}
T^{\mu \nu} _{;\nu} = 0 \ ,
\end{eqnarray}
where $T^{\mu \nu}$ is the energy-momentum tensor. The `gravitational skeleton' can be constructed via a multipole expansion of the energy-momentum tensor, defined with respect to some reference world line $z^{\alpha} (\lambda)$. Since the black hole mass ($M$)  is much greater than pulsar mass ($m$), the pulsar can be treated as a test mass and its motion entirely determined by the background black hole spacetime and the dynamical spin interaction with this field. In this extreme mass ratio limit and since the typical pulsar radius $R_{ \rm PSR} \ll r_{\rm g}$, moments greater than the quadrupole can be neglected. The first two moments are the 0th mass moment, encoded in the 4-momentum $p^{\mu}$ and the 1st dipole moment, given by the spin tensor $s^{\mu \nu}$. The corresponding equations of motion are \citep{Mathisson1937,Papapetrou1951,Dixon1974},
\begin{eqnarray}
\frac{D p^{\mu}}{d \tau} = - \frac{1}{2} R^{\mu}_{\nu \alpha \beta} u^{\nu} s^{\alpha \beta} \ ,
\label{Eq:mpd1}
\end{eqnarray}
\begin{eqnarray}
\frac{D s^{\mu \nu}}{d \tau} = p^{\mu} u^{\nu} -p^{\nu} u^{\mu} \ ,
\label{Eq:mpd2}
\end{eqnarray}
where $D/d\tau $ denotes a covariant derivative with respect to the proper time along the PSR worldline, $u^{\nu}$ is the PSR 4-velocity and $R^{\mu}_{\nu \alpha \beta}$ the Riemann curvature tensor. This system of equations is not determinate, since there exist more unknowns than equations. This is related to the uncertainty in choosing a reference world line for the multipole expansion. A natural choice of a representative point of the bulk motion of the body is the centre of mass. However in GR the centre of mass of a spinning body is not invariant. It is therefore necessary to specify a spin supplementary condition (SSC) which renders the system of equations determinate. This is equivalent to choosing an observer with respect to which the centre of mass is defined. We adopt the Tulczyjew-Dixon (TD) condition, 
\begin{eqnarray}
s^{\mu \nu} p_{\nu} = 0 \ ,
\label{Eq:ssc}
\end{eqnarray}
\citep{Tulczyjew1959,Dixon1964}. This is equivalent to choosing the centre of mass as measured in the zero 3 momentum frame. This SSC is advantageous since it specifies a unique worldline, whilst other choices of SSC are infinitely degenerate \citep[for discussion see][]{Costa2014}. Now, since $m \ll M$ and the pulsar Moller radius $R_{\rm Moller}$ (the radius of the disk of all possible centroids) is much less than the radius of the pulsar, the pole-dipole terms are much stronger than the dipole-dipole terms. Therefore, to first order the 4-velocity and 4 momentum are parallel, i.e. $p^{\mu} \approx m u^{\mu}$. The equations of motion then become,
\begin{eqnarray}
\frac{D u^{\mu}}{d \tau} = - \frac{1}{2m} R^{\mu}_{\nu \alpha \beta} u^{\nu} s^{\alpha \beta} \ ,
\end{eqnarray}
\begin{eqnarray}
\frac{D s^{\mu \nu}}{d \tau} \approx 0 \ ,
\end{eqnarray} 
\citep{Chicone2005,Mashhoon2006}. 
The ordinary differential equations to then be integrated are \citep{Singh2005,Mashhoon2006}:
\begin{eqnarray}
\frac{dp^{\alpha}}{d\tau} = - \Gamma_{\mu\nu}^{\alpha} p^{\mu}u^{\nu} + \rho \left( \frac{1}{2m} R^{\alpha}_{\beta \rho \sigma} \epsilon^{\rho \sigma}_{\quad \mu \nu} s^{\mu} p^{\nu} u^{\beta}\right) \ ,
\end{eqnarray}

\begin{eqnarray}
\frac{ds^{\alpha}}{d \tau} = - \Gamma^{\alpha}_{\mu \nu} s^{\mu}u^{\nu} + \rho \left(\frac{1}{2m^3}R_{\gamma \beta \rho \sigma} \epsilon^{\rho \sigma}_{\quad \mu \nu} s^{\mu} p^{\nu} s^{\gamma} u^{\beta}\right)p^{\alpha} \ ,
\end{eqnarray}

\begin{eqnarray}
\frac{dx^{\alpha}}{d\tau} = -\frac{p^{\delta}u_{\delta}}{m^2} \left[ p^{\alpha} + \frac{1}{2} \frac{\rho (s^{\alpha \beta} R_{\beta \gamma \mu \nu} p^{\gamma} s^{\mu \nu})}{m^2 + \rho(R_{\mu \nu \rho \sigma} s^{\mu \nu} s^{\beta \sigma}/4)}\right] \ ,
\end{eqnarray}
where $s^{\mu}$ is the spin 4-vector and the dimensionless parameter $\rho$ is used to label the terms which contribute to MPD spin-curvature coupling ($\rho = 1$ includes spin-curvature coupling, for $\rho = 0$ the coupling is omitted). In the $\rho \rightarrow 0$ limit the conventional spin-spin and spin-orbit couplings are recovered. Spin-curvature coupling causes additional complexities in the pulsar's orbital motion; a particle initialized in the orbital plane will remain in the orbital plane without spin-curvature coupling, but lifts in the vertical $z$-direction in the presence of coupling e.g. Fig \ref{fig:spinorbit} \citep[see also][]{Singh2014}.  We integrate this set of ODE's numerically via a standard 4th-order Runge-Kutta algorithm \citep{Press1992}. This completely determines the motion of a pulsar around a spinning black hole. At any integration step we can determine not only the pulsar's spacetime coordinates, but also the tangent momentum 4-vector and the spin 4-vector. The convolution of these vectors leads to complex relativistic dynamics, spin-axis precession (e.g. Figs \ref{fig:spinprecession}, \ref{fig:spinprecession2}), relativistic aberration, relativistic and time dilation and gravitational and relativistic energy shift. The influence on the pulse arrival time, profile and energetics will be discussed in Section \ref{section:application}.

\section{Constructing the PSR timing signal}
\label{sec:algorithm}
\begin{figure*} 
	\subfloat[]{\includegraphics[width=0.48\textwidth]{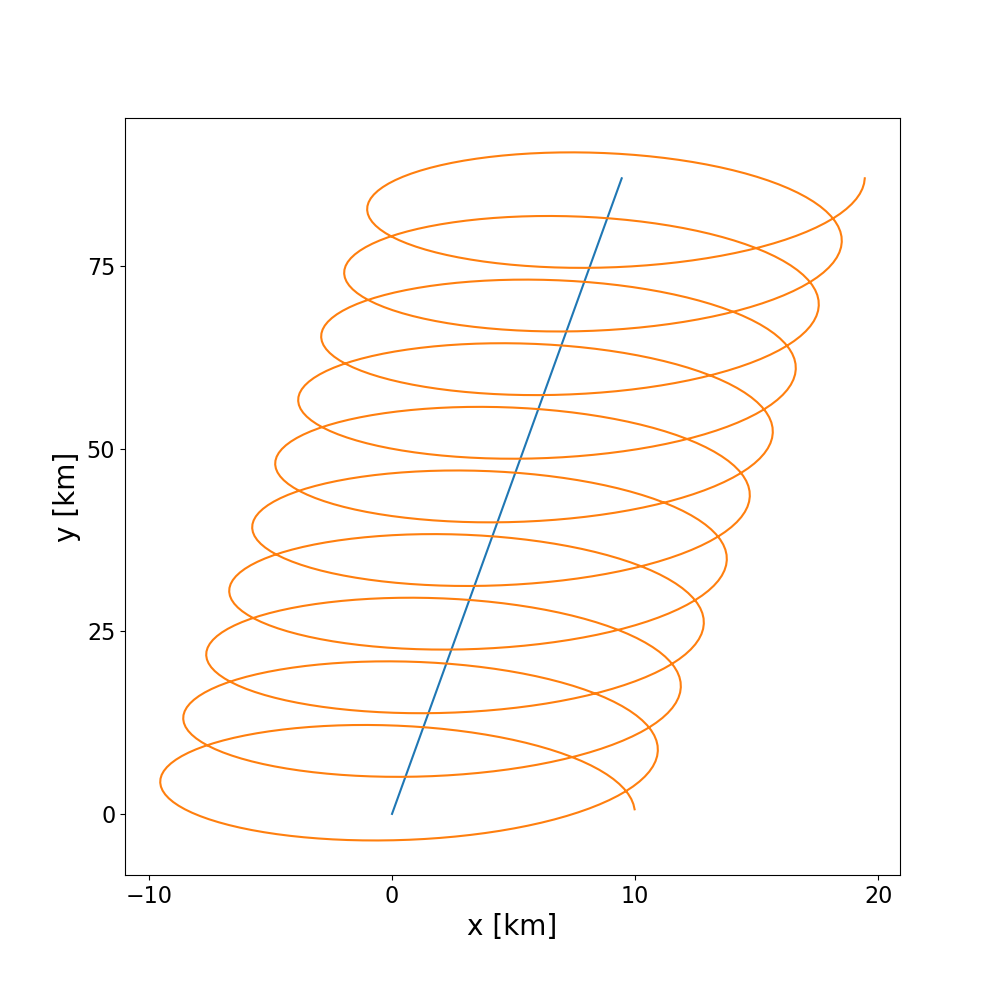}}
	\subfloat[]{\includegraphics[width=0.48\textwidth]{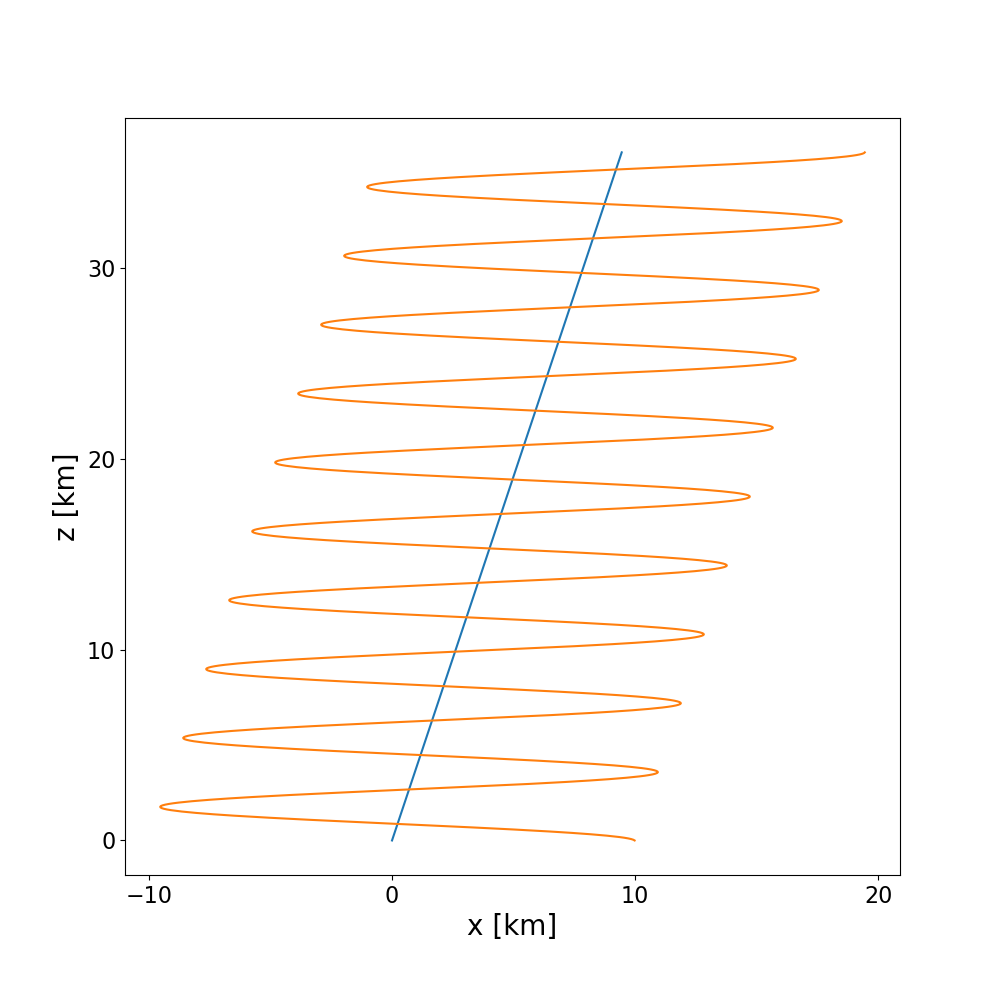}} \\
	
	\subfloat[]{\includegraphics[width=0.48\textwidth]{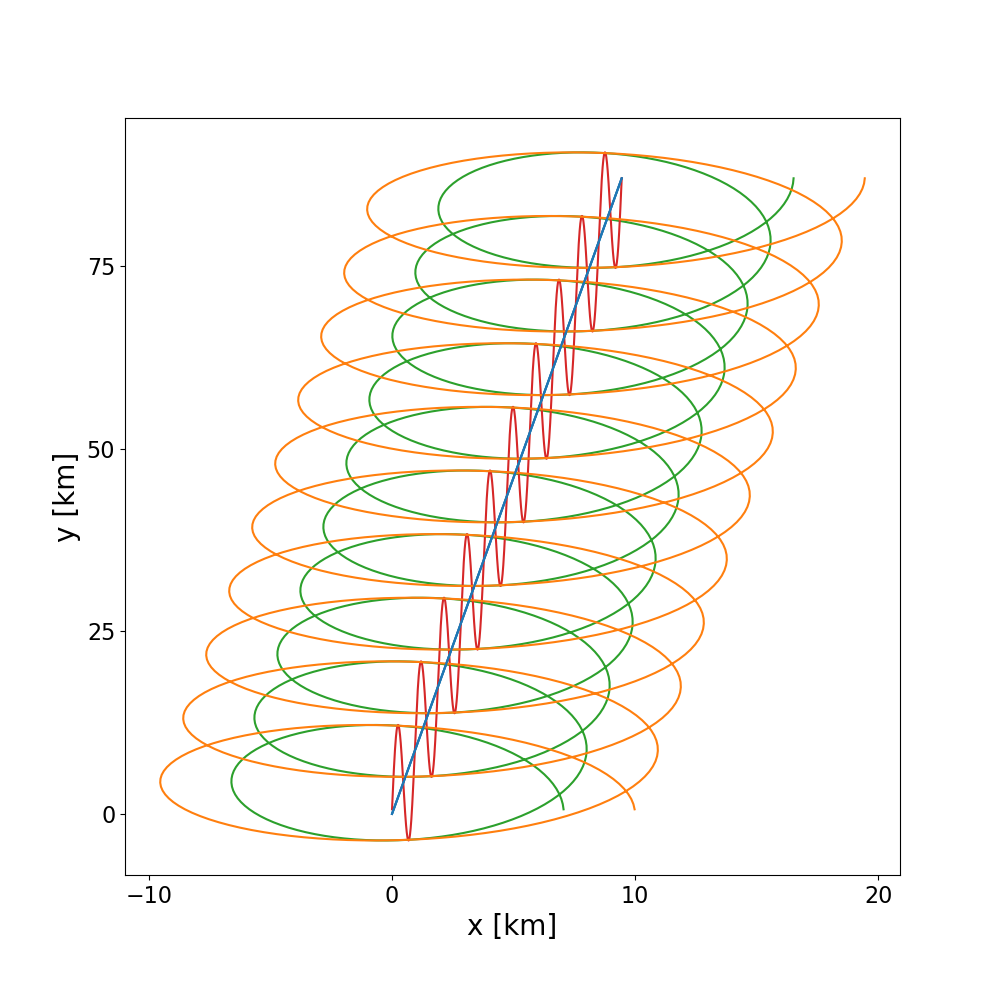}}
	\subfloat[]{\includegraphics[width=0.48\textwidth]{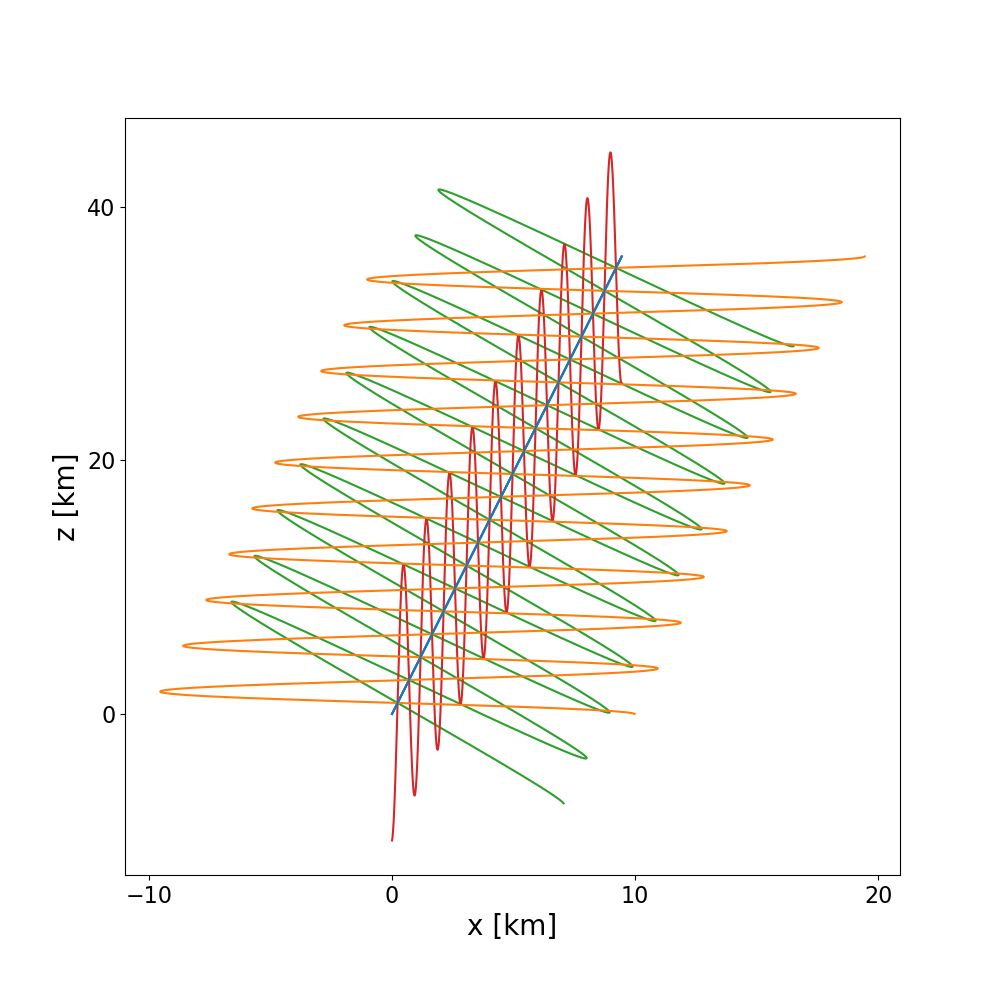}} \\
	
	\medskip
	
	\caption{Motion of the pulsar centre of mass (blue) and the radiation point (orange) in the (a) $x-y$ and (b) $x-z$ planes over 10 pulsar rotations, for pulsar period $P = 1$ ms, $S_{\theta} = 0$, and $\psi = \theta_{\rm obs} = \pi/2$. Coordinates are normalised with respect to initial pulsar position. (c,d) As (a,b) at $S_{\theta} = (0, \pi/4, \pi/2)$ (orange,red,green respectively). As $S_{\theta}$ is increased from 0 to $\pi/2$ the effective area traced by the beam as seen by the observer decreases.} \label{fig:com}
\end{figure*}

We now combine the ray tracing and MPD orbital dynamics tools to create a  theoretical timing model. Whilst it is possible to use a `forwards-in-time' method and integrate a ray from the pulsar outwards, this is computationally wasteful since it involves integrating many geodesics that do not hit the observer's image plane. Instead we use a `backwards-in-time' approach and integrate rays from the observer image plane towards the PSR-BH system, the method outlined in Section \ref{section:raytracing}. For each integration step on the PSR orbit, we want to find the ray which intersects with this location and arrives at the distant observer. Were pulsars isotropic emitters, an intersection (to within some tolerance, e.g. $|x^i_{\rm pulsar} - x^i_{\rm photon}| < R_{\rm PSR}$, such that the ray strikes the PSR surface) would be a sufficient condition to declare that the pulsar is `seen' by some distant observer. However, pulsars clearly do not emit isotropically but instead have characteristic beamed radiation. We therefore instead search for an intersection not with the pulsar's centre of mass, but instead with a `radiation point', $x^i_{\rm rad}$, taken as being the point which lies at a radius $R_{\rm PSR}$ from the PSR centre of mass, in the direction of the radiation beam $\mathbf{n}(\tau)$ (see Fig.  \ref{fig:schematic}) . If the radiation axis is at angles $(\psi, \chi)$ from the pulsar spin axis $\mathbf{S}$, and the spin axis is at angles $S_{\theta}, S_{\phi}$ from the coordinate $z$-axis (i.e. parallel to the black hole spin axis), then we can transform from the pulsar centre of mass $x^i_{\rm pulsar}$ to the radiation point, accounting for the misalignment of the spin and global vertical $z$-axis as, 
\begin{eqnarray}
x^i_{\rm rad} = R_z(S_{\phi}) R_y (S_{\theta}) R_{PSR}
\begin{bmatrix}
\sin(\psi) \cos({\chi}) \\
\sin(\psi) \sin({\chi}) \\
\cos(\psi)
\end{bmatrix} + x^i_{\rm pulsar} \ ,
\end{eqnarray}
where $R_z$ and $R_y$ are 3-dimensional rotation matrices about the coordinate $z$ and $y$ axes respectively. Typically, $\psi$ is set to be some constant angle corresponding to the case where the latitude of the radiation beam does not evolve with respect to the spin axis. The phase $\chi = 2 \pi \tau /P$, for pulsar period $P$. The spin axis angles are time dependent, i.e. $S_{\theta} =S_{\theta}(\tau)$, $S_{\phi} =S_{\phi}(\tau) $ and are related to the precession and nutation of the spin axis with their evolution described by the MPD formalism. The evolution of $x^i_{\rm rad}$ and $x^i_{\rm pulsar}$ is illustrated in Fig. \ref{fig:com}.

\subsection{Finding an intersection}
\begin{figure}
	\subfloat[\label{fig:schematic}]{%
		\includegraphics[clip,width=\columnwidth]{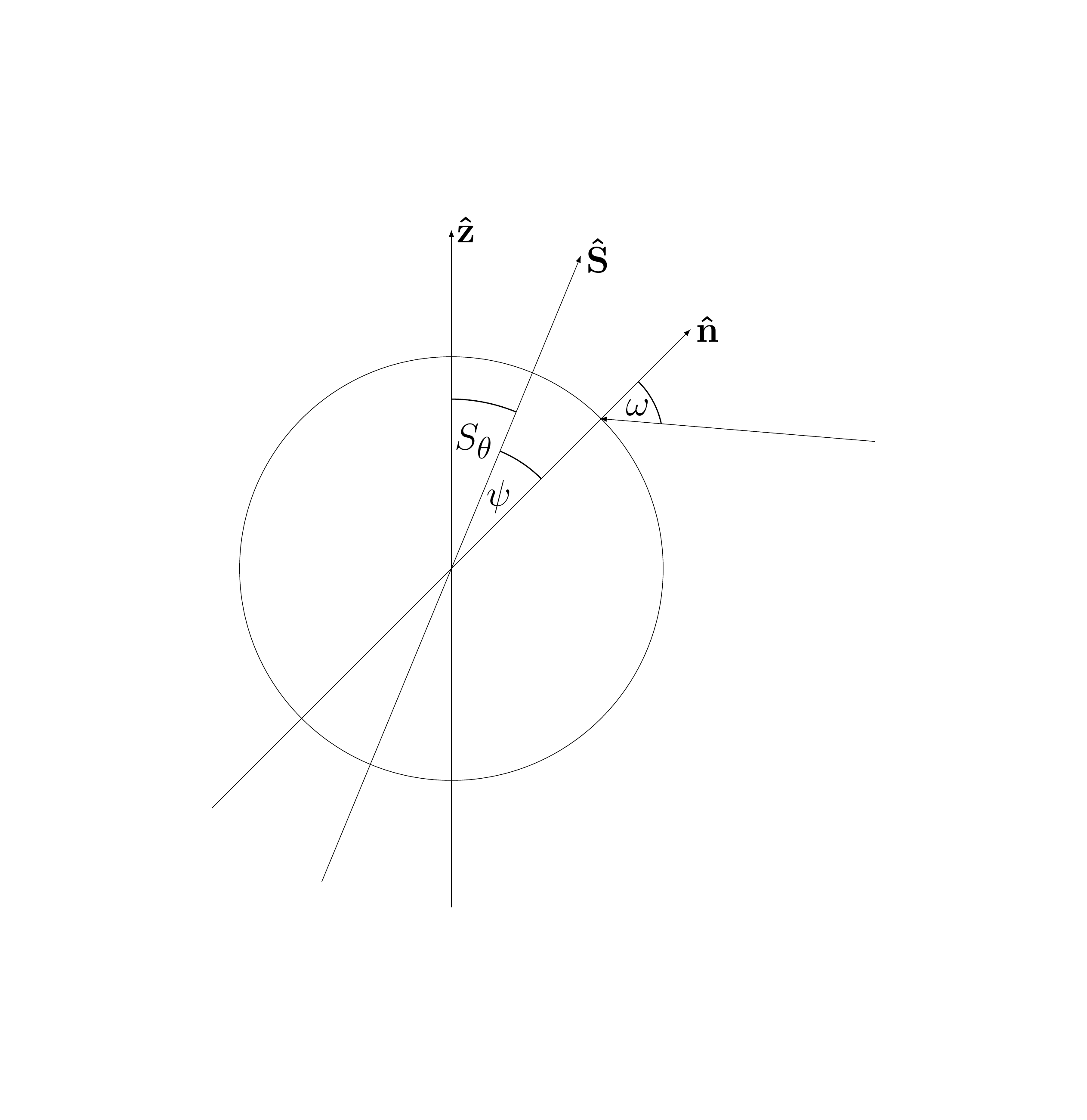}%
	}
	
	\subfloat[\label{fig:finding_intersectionB}]{%
		\includegraphics[clip,width=\columnwidth]{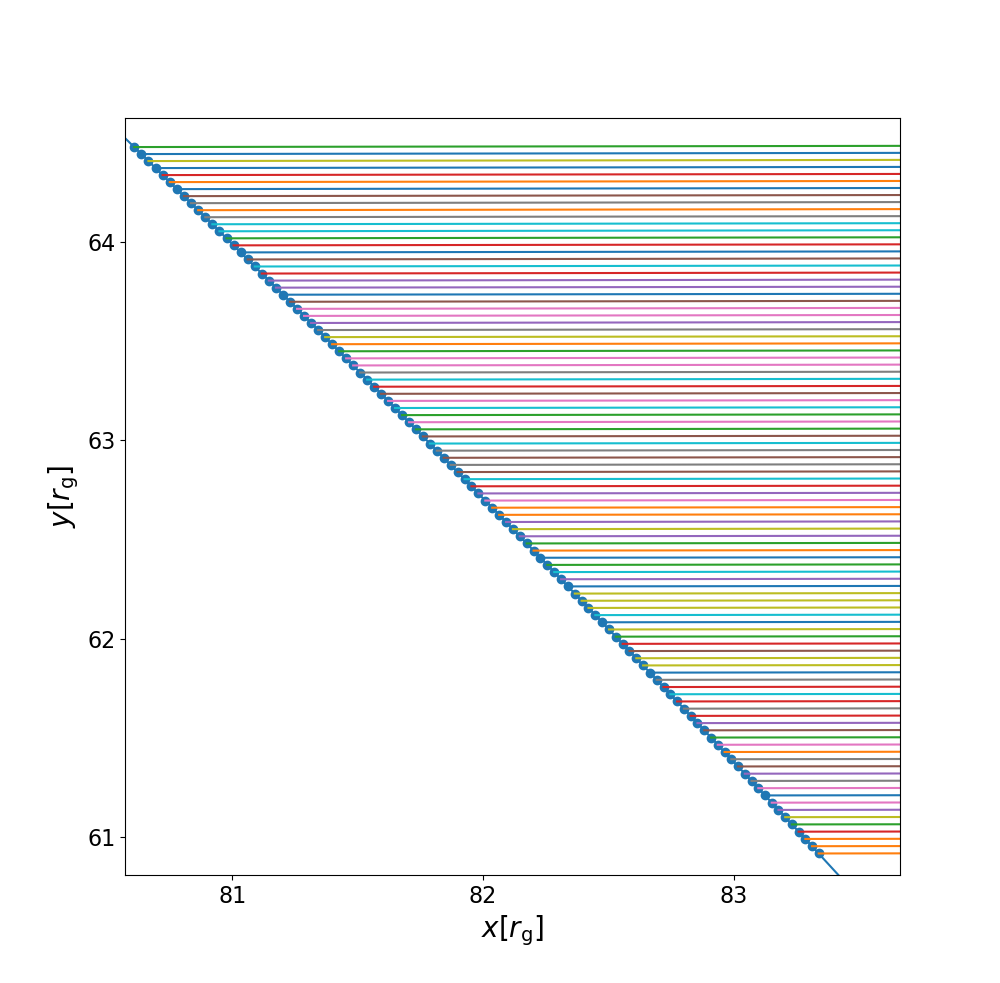}%
	}
	
	\caption{(a) Schematic of the relevant angles used in this work (b) The intersection of photon rays (multicoloured lines) with a section of pulsar orbit (blue line) projected in the $x-y$ plane.}\label{fig:tests}
	
\end{figure}

The problem of finding an intersection is effectively a two point boundary value problem; we know the desired final boundary value (i.e. radiation point $x^i_{\rm rad}$) and we want to find the initial boundary value, i.e. the values of the image plane coordinates $\alpha, \beta$. The ray then traces between these two limits. We solve this problem numerically using a shooting method \citep{Press1992}. The ray tracing is effectively a black box function $f(\alpha, \beta)$ which takes some $\alpha, \beta$ and returns the minimum distance along the ray from the target value $x^i_{\rm rad}$ for each time-step of the PSR orbital integration. We want to minimize $ds = f(\alpha, \beta)$, to within some tolerance $\epsilon$ by varying initial conditions $\alpha, \beta$. The adjustment of $\alpha, \beta$ proceeds via a non-linear conjugate gradient descent algorithm \citep{Fletcher1964, Press1992}, where the new computed gradient is conjugate to previous gradients. The conjugate direction vector $\mathbf{h}_j $,  is updated at each iteration step $j$ as,
\begin{eqnarray}
\mathbf{h}_{j+1} = \mathbf{g}_{j+1} + \gamma_j \mathbf{h}_j \ ,
\end{eqnarray}
where $\mathbf{g}_j = - \nabla f(\alpha_j, \beta_j)$ and
\begin{eqnarray}
\gamma_j = \frac{\mathbf{g}_{j+1} \cdot \mathbf{g}_{j+1}}{\mathbf{g}_{j} \cdot \mathbf{g}_{j}} \ ,
\end{eqnarray}
\citep{Fletcher1964, Press1992}. The variables $\alpha, \beta$ are then updated at each iteration as,
\begin{eqnarray}
\mathbf{x}_{j+1} = \mathbf{x}_j + \delta_j \mathbf{h}_j \ ,
\end{eqnarray}
for vector $\mathbf{x}_j = (\alpha_j , \beta_j)$ and where $\delta_j$ is the variable stepsize, determined via  an inexact line search. This method is advantageous over a vanilla implementation of gradient descent since for certain target points $x^i_{\rm rad}$, the function $f(\alpha, \beta)$ takes the form of an ill-conditioned narrow valley. In such an environment, gradient steepest descent becomes inordinately slow, since the direction of steepest descent is not, in general, in the direction of the minimum and the algorithm instead follows a `criss-cross' pattern, oscillating between the sides of the valley. This issue is avoided by moving in a direction which is conjugate to previous directions. Since the analytical form of $f(\alpha, \beta)$ is unknown, the gradients necessary for the optimization are evaluated numerically via the difference quotient. The algorithm exhibits dependable and fast convergence and 
for this work we declare the algorithm to have converged - i.e. an intersection to be found - when $ds^2 < 10^{-19}$ natural units,working in quadruple precision. Taking the black hole mass to be $4 \times 10^6 M_{\odot}$, then this corresponds  to a spatial distance of $\sim 1.9 m$ or a light travel time of $\sim 6$  ns. The intersection of multiple rays with a section of pulsar orbit is illustrated in Fig \ref{fig:finding_intersectionB}. More complex ray paths and intersections are presented in Section \ref{section:application}.

\subsection{Pitch angle in the comoving local frame}
\begin{figure}
	
	\subfloat[\label{fig:pitcha}]{\includegraphics[clip,width=\columnwidth]{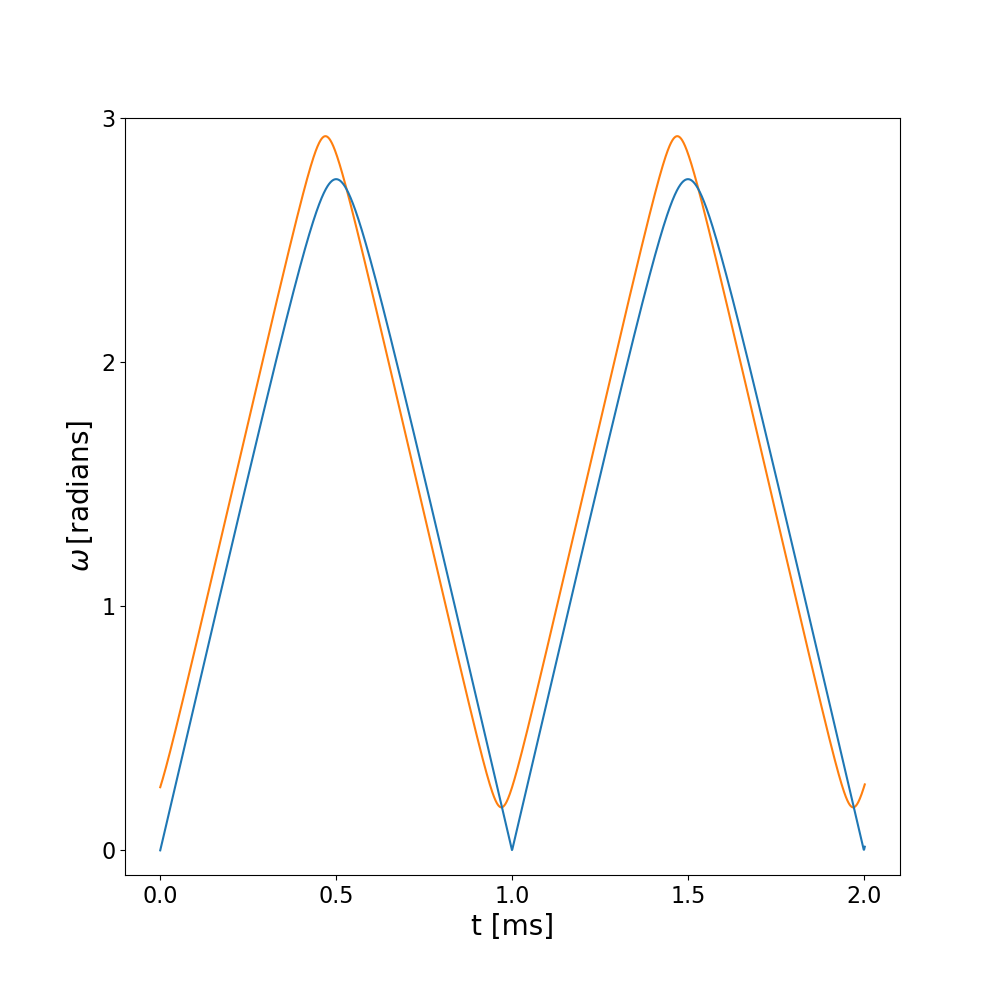}}

	\subfloat[\label{fig:pitchb}]{%
		\includegraphics[clip,width=\columnwidth]{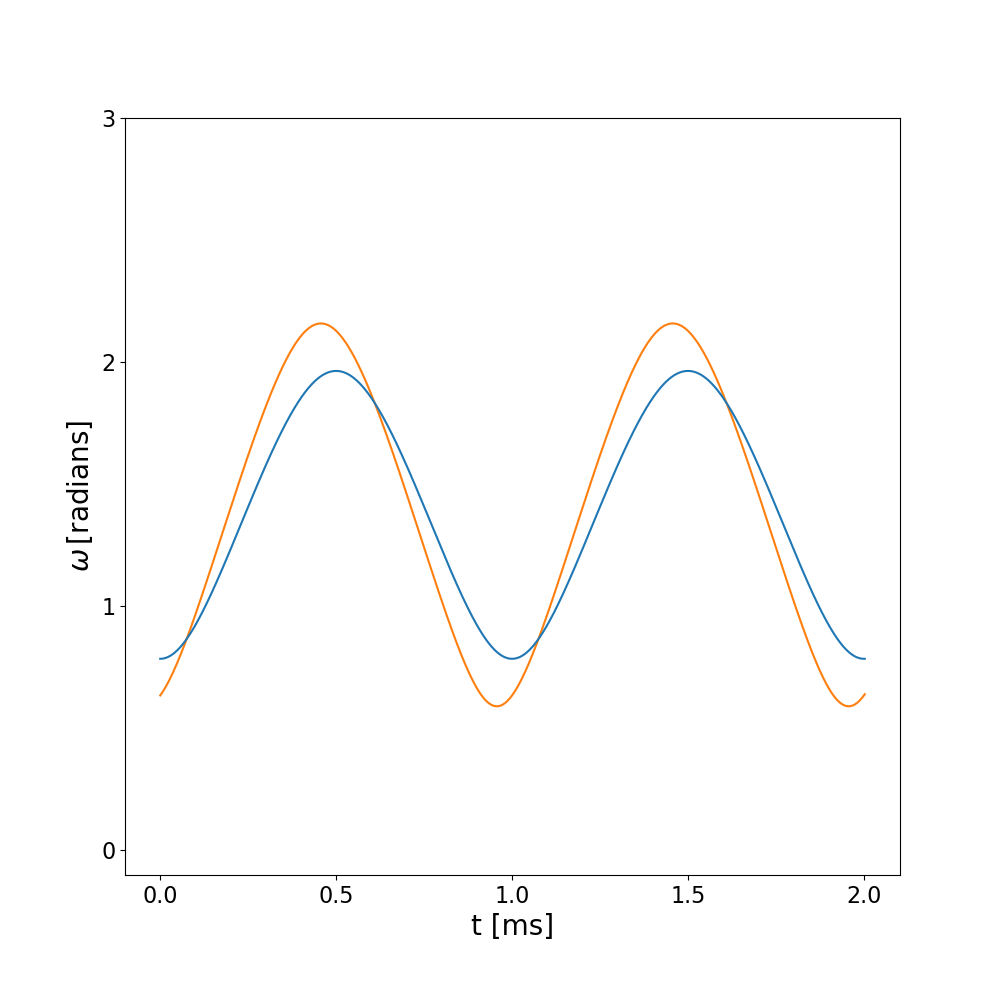}%
	}
	\label{fig:pitch1}
	\caption{(a) Pitch angle in both the coordinate (blue) and comoving (orange) frames for a pulsar with period $P = 1$ms, $S_{\theta} = 0$ and $\psi = 7\pi/16$ as observed by some distant observer at $\theta_{\rm obs} = \psi$. In the coordinate frame, the pitch angle drops to $\sim$ 0 once per period, as expected. However, in the comoving frame the $\omega-t$ profile is shifted due to relativistic aberration. (b) As (a) for $S_{\theta} = \pi/4$. The amplitude of the pitch angle is reduced since the beam direction has been shifted due to the precession of the spin axis.}
	
\end{figure}

Once an intersection with the radiation point is found, in order to determine if the beam is seen by a distant observer we require the pitch angle $\tilde{\omega}$, i.e. the angle the photon ray makes with the normal to the stellar surface. All the MPD and ray tracing calculations take place in the coordinate frame, and so any calculation will result in a pitch angle also measured in the coordinate frame. However the angle that determines whether or not the beam will be visible to an observer is the angle as measured in the orthonormal tetrad frame comoving with the star, $\hat{\omega}$. Due to relativistic aberration these two angles are not the same, $\tilde{\omega} \ne \hat{\omega}$. The pitch angle in the comoving frame is given by,
\begin{eqnarray}
\hat{\omega}  =\arccos \left( \frac{n^i \hat{k}_i}{|n| |k|}\right) \ ,
\label{eq:pitch}
\end{eqnarray}
where $n^i$ is the vector normal to the stellar surface and $\hat{k}_i$ is a vector tangent to the ray path, as measured in the comoving frame. Naturally the the covariant 4-momentum of the ray $k_{\mu}$, is tangent to the ray. We can transform a general vector $A^{\nu}$ from the coordinate basis to the tetrad basis as,
\begin{eqnarray}
A^{(\beta)} = \eta^{(\alpha)(\beta)} g_{\mu \nu} e^{\mu}_{(\alpha)} A^{\nu} \ ,
\end{eqnarray}
where $\eta^{(\alpha)(\beta)}$ is the Minkowski metric, indices with braces (e.g. $A^{(\beta)}$) indicate the comoving basis and the basis 4-vectors are, 
\begin{equation}
e^{\mu}_{(t)} = u^{\mu} \ ,
\end{equation}
\begin{equation}
e^{\mu}_{(r)} = (u_r u^t, -u_tu^t - u_{\phi}u^{\phi},0,u_r u^{\phi})/N_r \ ,
\end{equation}
\begin{equation}
e^{\mu}_{(\theta)} = (u_{\theta}u^t, u_{\theta} u^r, 1+u_{\theta}u^{\theta}, u_{\theta} u^{\phi})/N_{\theta} \ ,
\end{equation}
\begin{equation}
e^{\mu}_{(\phi)} = (u_{\phi},0,0,-u_t)/N_{\phi} \ ,
\end{equation}
\citep{Krolik2005, Kulkarni2011, Shcherbakov2011, Dexter2016} where,
\begin{equation}
N_r^2 = -g_{rr}(u_tu^t + u_{\phi} u^{\phi})(1+u_{\theta} u^{\theta}) \ ,
\end{equation}
\begin{equation}
N_{\theta}^2 = -g_{\theta \theta} (1+u_{\theta} u^{\theta}) \ ,
\end{equation}
\begin{equation}
N_{\phi}^2 = -(u_t u^t + u_{\phi} u^{\phi}) \Delta \sin^2 \theta \ ,
\end{equation}
for transformation to a frame moving with 4-velocity $u^{\mu}$ at position $(r, \theta, \phi)$. We can then transform the ray tangent vector to the comoving frame, $ k_{\mu} \rightarrow k_{(\mu)}$. The comoving frame is locally flat and so indices are raised and lowered with the Minkowski metric $\eta^{(\mu) (\nu)}$. The vector normal is $n^i = x^i_{\rm rad} - x^i_{\rm pulsar}$ and considering only the spatial components of the comoving ray tangent vector, the pitch angle can be calculated via Eq. \ref{eq:pitch}. Once $\hat{\omega}$ has been determined, we declare an observation if $\hat{\omega} < \omega_c$ where $\omega_c$ is some critical angle, i.e. the pulsar jet opening angle. \newline 
\noindent The variation in the pitch angle over 2 rotations of a millisecond pulsar is presented in Fig. \ref{fig:pitcha}, for pitch angle calculated in both the global ($\tilde{\omega}$) and comoving ($\hat{\omega}$) frames. Whilst the periodicity of $\hat{\omega}$ and $\omega$ are the same, $\hat{\omega}$ is shifted in both time and amplitude with respect to $\omega$. If we change the latitude angle of the spin axis from $S_{\theta} =0$ to $S_{\theta} = \pi/4$ (Fig. \ref{fig:pitchb}) the amplitude of the pitch angle reduces in both the global and comoving frames since the radiation beam is shifted further from the observer's line of sight. Furthermore, the amplitude of the pitch angle in the comoving frame is now markedly greater than in the global frame.

\section{Application}
\label{section:application}
We have established the framework for determining the pulsar ray path and orbital dynamics, along with the algorithm to find the intersection of the ray and calculate the pitch angle. We can now apply this framework to investigate the impact of the convolution of relativistic and astrophysical effects on the signal from a PSR in an EMRB, in particular the photon time of arrival and time-frequency profile. 

Clearly, any pulsar timing model is going to depend on both the BH mass and spin. The mass of the BH at the centre of the Milky Way is thought to be $\sim 4 \times 10 ^ 6 M_{\odot}$ \citep{Gillessen2009}. If intermediate mass black holes exist at the centre of globular clusters then the advanced capabilities of next-generation radio telescopes may allow for the detection of an extragalactic EMRB of the Local Group \citep{Keane2015}. In this case the mass would be more intermediary, $\sim 10^3 - 10^5 M_{\odot}$ \citep{Wrobel2018}. In contrast to the mass determination, there remains considerable uncertainty in the spin parameter of astrophysical BHs. Astrophysical measurements of the nearest BH source, Sgr A*, range from $0.44 - 0.996$ \citep{Aschenbach2010, Kato2010a,Dokuchaev2014}. The determination of the spin of BHs other than Sgr A* have been possible via continuum fitting and iron lines measurements, and again cover a wide range from $a <-0.2$ to $a > 0.98$ \citep[see Table 1 of][]{Bambi2016}. Going forward we set $M = 4 \times 10 ^ 6 M_{\odot}$ and have the BH as rapidly spinning such that $a = 0.998$. These parameters are adopted in our demonstrative calculations so as to illustrate the sorts of effects a strong field pulsar timing model must account for. Exploration of the effects of alternative spin parameters and masses and their consequent impact on the photon time of arrival will take place in a later paper. 
\subsection{Effects of relativistic spin orbital dynamics}
\begin{figure}
	\includegraphics[width=\columnwidth]{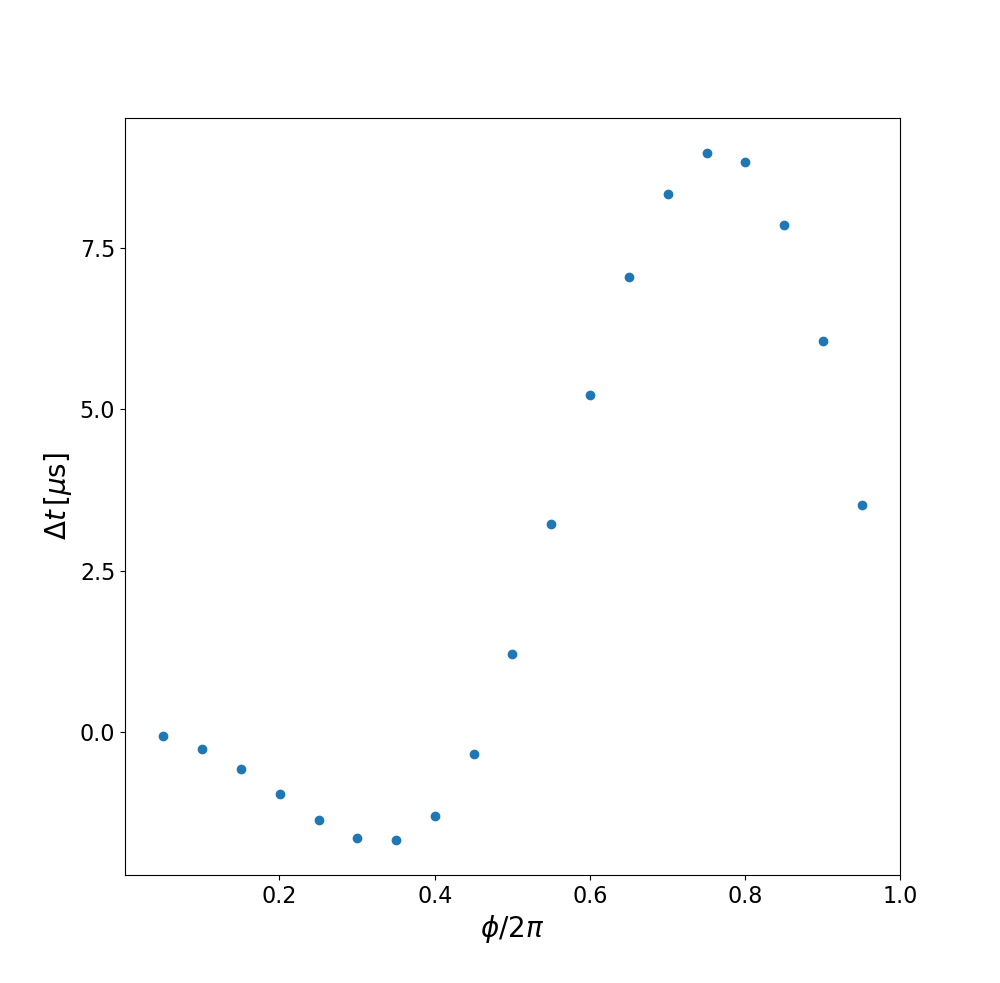}
	\caption{Variation in the photon arrival time between the $\lambda = 0$ and $\lambda=1$ spin curvature couplings over a single orbit. Semi-major axis $30 \, r_{\rm g}$, eccentricity $= 0.1$. We set the observer at some inclination $\theta_{\rm obs} = \pi/4$. The out-of-plane motion induced by the coupling causes $\mu$s order variations in the photon arrival time. Naturally the severity of the time variation depends on the orbital configuration and the observer's line of sight These variations will be further exacerbated in conjunction with other relativistic and astrophysical effects. }
	\label{fig:spin_curvature_timeshift}
\end{figure}

Treating the pulsar as a rigid spinning object and accounting for the associated spin couplings causes substantial variation in the pulsar orbital dynamics. In particular, spin-curvature coupling causes the pulsar to deviate from geodesic motion in a Kerr spacetime and exhibit complex orbital dynamics. In particular there is noticeable motion in the vertical direction, out of the orbital plane, that would not be present in the $\rho = 0$ case. The severity of the effects of spin-curvature coupling become more pronounced for lower mass ratios (e.g. IMBH in globular clusters), for faster spinning objects (e.g. millisecond pulsars) and retrograde orbits \citep[see e.g][]{Singh2014}. These dynamical spin effects will influence the pulse time of arrival. Figure \ref{fig:spin_curvature_timeshift} illustrates the change in the light travel time, in vacuum, between a photon emitted at the same orbital phase  in the $\rho = 0$ and $\rho =1$ cases over a single orbit. We neglect any complications induced by secondary pulses from strong gravitational bending (see Sec \ref{sec:gravbending}). The light travel time varies on the order of $\sim \pm 10 \mu$s \citep[similar values are found in][]{Singh2014} a variation which should be detectable with the next generation of radio telescopes; 10 minute integrations of millisecond pulsars at 1.4 GHz should achieve a ToA precision of the order $100$ ns \citep{Liu2011}. Furthermore, any variation in the light travel time due to the orbital dynamics will be further compounded by the consideration of additional factors (e.g. time dispersion, time dilation, spin axis precession etc. see subsequent sections). 

An accurate description of the PSR orbital dynamics is essential, since the apparent or observed pulse frequency is related to the intrinsic frequency 
\begin{eqnarray}
\nu_{\rm obs} =  \frac{p_{\alpha} u^{\alpha} |_{\rm observed}}{p_{\alpha} u^{\alpha} |_{\rm emitted}}  \nu_{\rm emitted} \ ,
\end{eqnarray}
In order to determine the Doppler shift accurately, one must therefore have an appropriate description of the PSR velocity. Furthermore, for systems with short orbital periods, the line of sight velocity may not be constant over an integration period. The complex orbital dynamics of a spinning pulsar around a spinning black hole will also naturally influence both the acceleration and the change in acceleration (i.e. jerk) of the pulsar, depending on the orbital and observer configuration. Highly accelerated systems - the most scientifically interesting systems from the perspectives of testing GR - are difficult to search for when the integration time is comparable to the orbital period. Assuming the acceleration to be constant, the changing velocity of the pulsar causes the signal to drift into multiple frequency bins, with the number of bins drifted,
\begin{eqnarray}
N_{\rm drift} \propto a_0 T^2 \ ,
\end{eqnarray}
\citep{EatoughThesis}, for constant line-of-sight acceleration $a_0$ and integration time $T$. Search algorithms can correct for this effect (`acceleration searches'), but the correction is typically computationally intensive. Spin-curvature coupling can cause additional acceleration as measured by the observer, depending on the line of sight, and will influence the number of frequency bins drifted. Furthermore, in the presence of spin-curvature dynamics the assumption of constant acceleration may need modification, since in highly relativistic regimes the jerk may be non-negligible. In this case the number of frequency bins drifted becomes time dependent,
\begin{eqnarray}
N_{\rm drift} \propto (a_0+j_0t) T^2 \ ,
\end{eqnarray}
at time $t$ with jerk $j_0$, which may introduce additional complications in searching for highly relativistic systems.

\subsection{Effects of spin axis precession}
Since the motion of the pulsar is inextricably tied to its spin, the geodetic precession of the spin axis works in conjunction with the relativistic aberration to influence the observability of the pulsar signal, the pulse profile \citep[e.g.][]{Roman2006} and the pulse arrival time. Over the course of its orbital motion the pulsar spin axis oscillates (e.g. Fig. \ref{fig:spinprecession}), with the timescale of oscillation shortening for more compact orbital radii where the gravitational curvature is greater. For some orbital configurations, this oscillation will leave the pulsar beam invisible at certain times \citep[e.g.][]{Istomin1991} as $\hat{\omega} > \omega_c$. The pulsar signal would consequently be intermittent, leading to additional complications in search observations and any consistent timing model.\newline 

\begin{figure}
	\includegraphics[width=\columnwidth]{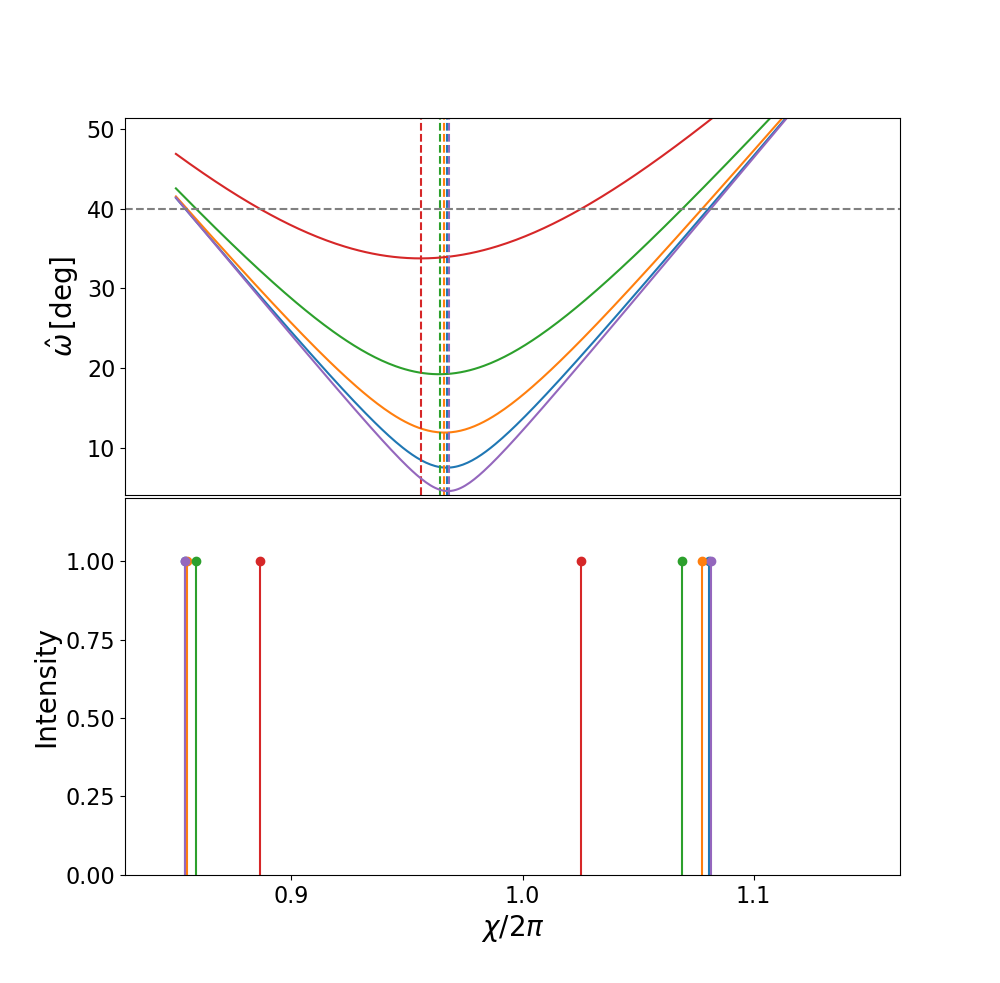}
	\caption{Change in the pitch angle (top panel) as the pulsar beam rotates with phase $\chi$ and (bottom panel) phase angle at the edge of the radiation beam cone. As $S_{\theta}$ varies through $\pi/12, \pi/10, \pi/8, \pi/6, \pi/4$ (purple through to red, respectively) the minimum of $\omega'$ increases, since the line of sight now intersects more sharply with the beam. The vertical coloured lines of the top panel denote the location of the minimum of the curve.}
	\label{fig:aberration_shift}
\end{figure}

\noindent To inspect the effects of spin axis precession, we consider variations in $S_{\theta}$ whilst holding $S_{\phi}$ constant at  $S_{\phi}= 0$. (Fig \ref{fig:aberration_shift}). We arbitrarily define the pulse width $W_{40}$ as the phase subtended when $\hat{\omega}$ varies from $40^{\circ}$, to the minimum, and back to $40^{\circ}$. Since the orientation of the spin axis affects the effective latitude on the radiation cone with which the line of sight intersects, the pulse width exhibits a clear variation with the orientation of the spin axis. The pulse width can be considered as a proxy for the observed pulse duration, and so precession of the spin axis over the course of the pulsar's orbit will directly influence the time over which a pulse is viewable. Whilst the variations in the pulse width is due to how close the line of sight is to the edge of the emission cone, the asymmetry evident in the broadening is a result of the relativistic aberration. Consequently, due to spin effects the pulse profile will vary over the course of the orbit, and also vary between different kinds of orbit due to the difference in the 4-velocity and 4-spin vectors. \newline 

\begin{figure}
	\includegraphics[width=\columnwidth]{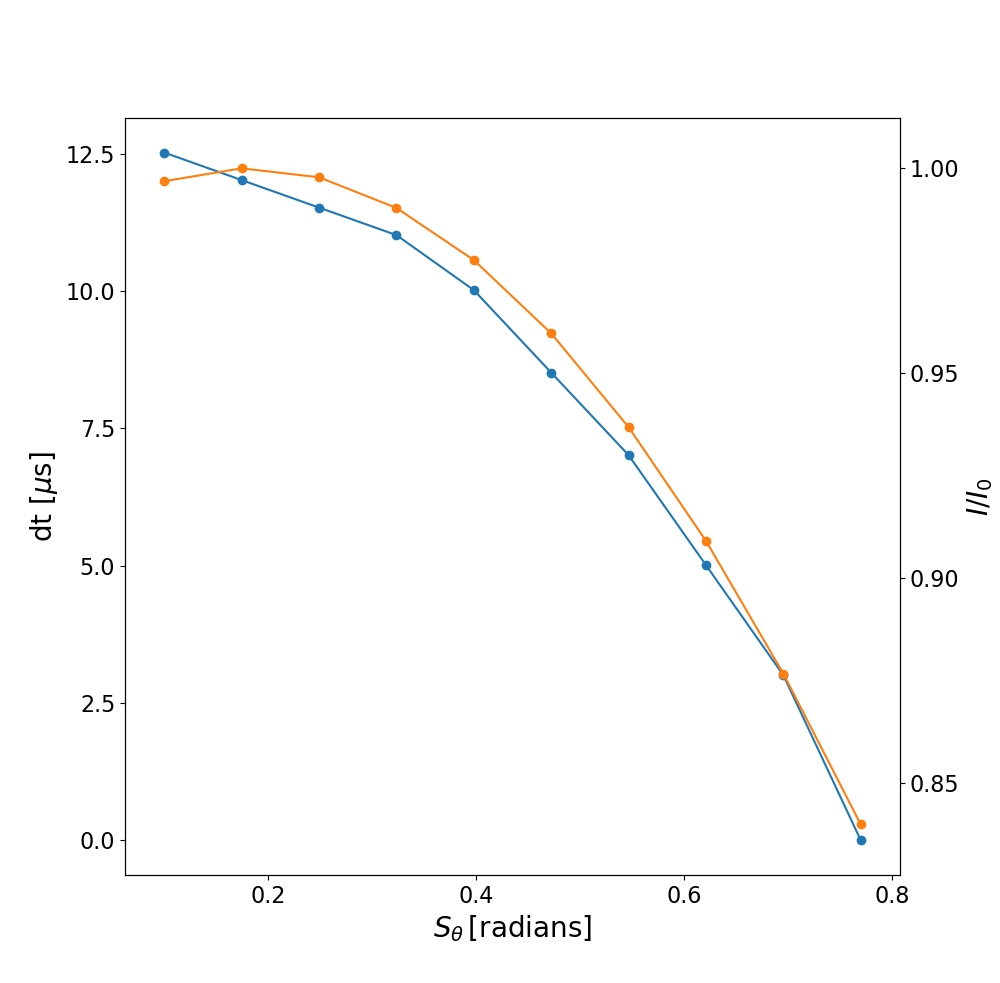}
	\caption{Time delay (blue, left axis) and relative shift in the intensity (orange, right axis) due to the shift in the centroid location, assuming a pulsar with a period of $1$ ms. A ray originating from a pulsar with $S_{\theta} = \pi/12$ is delayed by over $10 \, \mu$s compared to a ray which originates from a pulsar with $S_{\theta} = \pi/4$, whilst, approximating the pulsar cone as a Lambertian surface, the peak intensity can be reduced on the order $\sim 0.1-0.2$.}
	\label{fig:time_width}
\end{figure}

\noindent The asymmetry in the $\hat{\omega}-t$ profiles can be quantified by considering the phase angle $\chi$ at which the minimum of $\hat{\omega}$ occurs, which corresponds to the centroid of the arc subtended by the radiation beam across our line of sight (denoted by vertical dashed lines in Fig \ref{fig:aberration_shift}). Variation in $S_{\theta}$ in conjunction with relativistic aberration causes a shift in the centroid location $\Delta \chi$. This centroid shift in turn causes the pulse arrival times from a pulsar at a given location to vary with $S_{\theta}$ (Fig. \ref{fig:time_width}), due to the extra time elapsed traversing the angle $\Delta \chi$. For a pulsar with a 1 ms spin period, the centroid of the profile from a pulsar with $S_{\theta} = \pi/12$ is delayed on the order of $\sim 10 \mu \rm s$ compared to a pulsar with $S_{\theta} = \pi/4$ and typical delays are of the order $\sim (1 - 10)\mu $s, i.e. $0.1 - 1$ \% of a MSP spin period. Since the centroid shift will occur smoothly over orbital timescales, during observations this effect should be traceable and measurable and used to inform the evolution of the pulsar spin axis.  Furthermore, at steeper $S_{\theta}$ angles, the amplitude of $\hat{\omega}$ is also reduced since the radiation beam is shifted further from the observer's line of sight. Approximating the pulsar beam as a Lambertian surface, precession of the spin axis will cause variations in the pulse intensity $dI$ as,
\begin{eqnarray}
dI = \cos (d \hat{\omega}).
\end{eqnarray}
Consequently, the precession of the spin axis can directly affect the observed intensity. Clearly this is an oversimple approximation and to correctly determine the pulse intensity also accounting for scattering, absorption and Doppler boosting would require covariant general relativistic radiative transfer \citep[e.g.][]{Younsi2012} along the geodesics determined via the framework presented in this paper. However, the Lambertian approximation serves as a adequate first order estimate to inspect the influence of spin precession on the pulse intensity. The peak intensity of a beam that originates from a pulsar with $S_{\theta} = \pi/4$ is $\lesssim 85 \% $ that of a pulsar with $S_{\theta} = \pi/12$ for the observer at the same location (Fig. \ref{fig:time_width}). \newline 

\noindent In addition to the orbital dynamics, spin precession can then also affect the pulse width (and hence duration), the pulse ToA due to the relativistic centroid shift and the observed pulse intensity. The precision that can be achieved with pulsar timing is a function of both the pulse intensity and the sharpness of the pulse profile, and so an accurate description of these quantities is key for realistic assessment of the prospects of the detection of a PSR-EMRB. Variations in the pulse profile width may provide a method to determine the PSR spin behaviour \citep[e.g.][]{Roman2006}, and hence the orbital parameters of the system. For very compact orbits or periastron passages, the timescale of spin orientation variation can severely shorten, causing rapid changes in the pulse profile (width, intensity, ToA). Corrections for this change in the overall pulse profile shape will be necessary for both long term timing and - if the timescale of spin axis oscillation is comparable to the observation integration time -  initial detection. This analysis has involved just varying $S_{\theta}$. Changes in $S_{\phi}$ over the course of the orbit adds an additional degree of freedom to influence the pulse profile, which are naturally included within our framework. 

\subsection{Gravitational Light Bending}
\label{sec:gravbending}
The notion of a geodesic, i.e. a straight line in curved spacetime, means that the spatial path of the ray emitted from the pulsar can be bent due to the spacetime curvature induced by the central massive black hole. That is, the signal from a pulsar may not follow a purely radial path. A strongly deflected pulsar beam may propagate directly though the strong-field regime, considered in the analytical case for a Schwarzchild black hole in \citet{Wang2008, Wang2009}. Gravitational lensing by Sgr A* has been considered for emission for S-stars close to the Galactic centre \citep{Bozza2009, BinNun2010, Bozza2012} as observed by the next generation of Very Large Telescope Interferometer instrumentation \citep[e.g. GRAVITY,][]{Gravity2017}. The magnitude of the astrometric shift that results from gravitational bending is naturally dependent on the configuration of the system \citep[i.e. relative alignment of source, lens, observer. See e.g.][for a description of the relevant angles and geometry]{Bozza2012}. For the S-stars, at alignment angles of $\sim 35  ^{\circ}$, the expected astrometric shift is of order $ 30 \mu$as, whilst the contribution of post-Newtonian and spin corrections to gravitational bending are expected to appear at $\sim 5 \mu $as, below current interferometric instrumental sensitivity \citep{Bozza2012}.

\begin{figure*} 
	\subfloat[]{\includegraphics[width=0.48\textwidth]{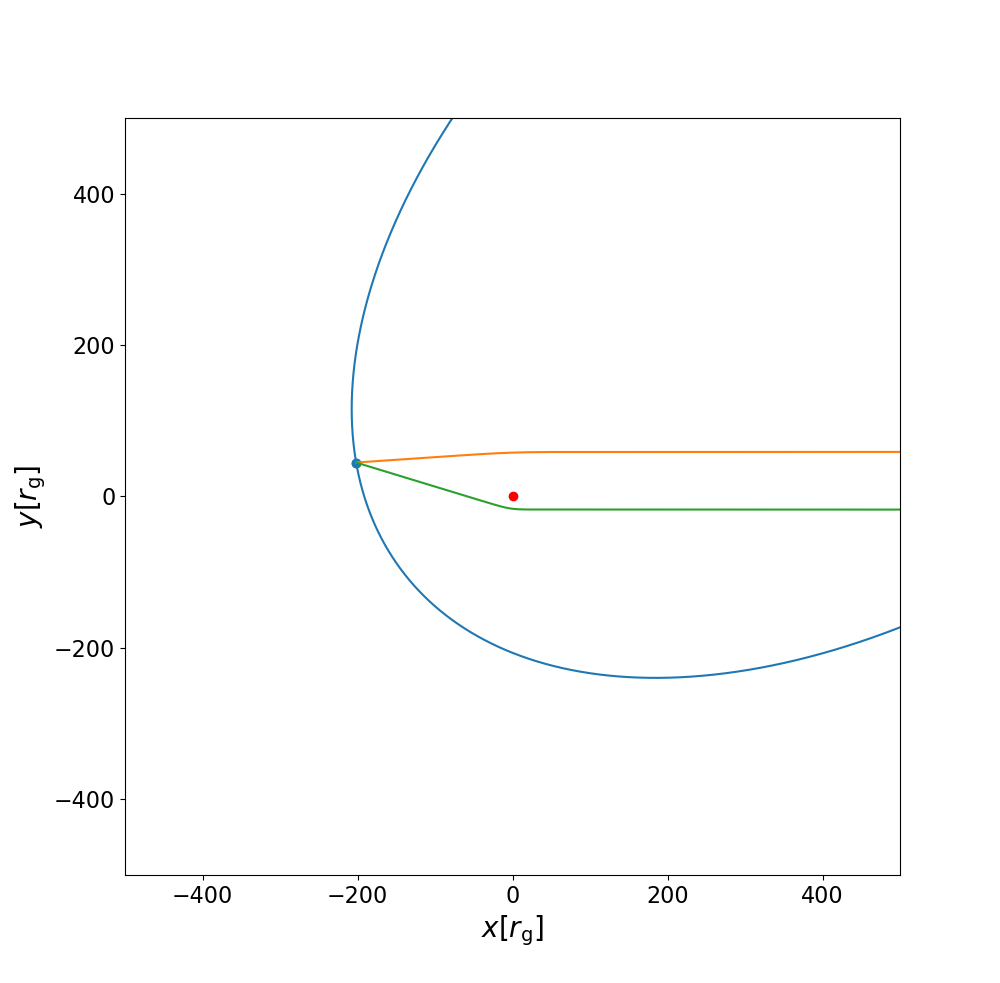}}
	\subfloat[]{\includegraphics[width=0.48\textwidth]{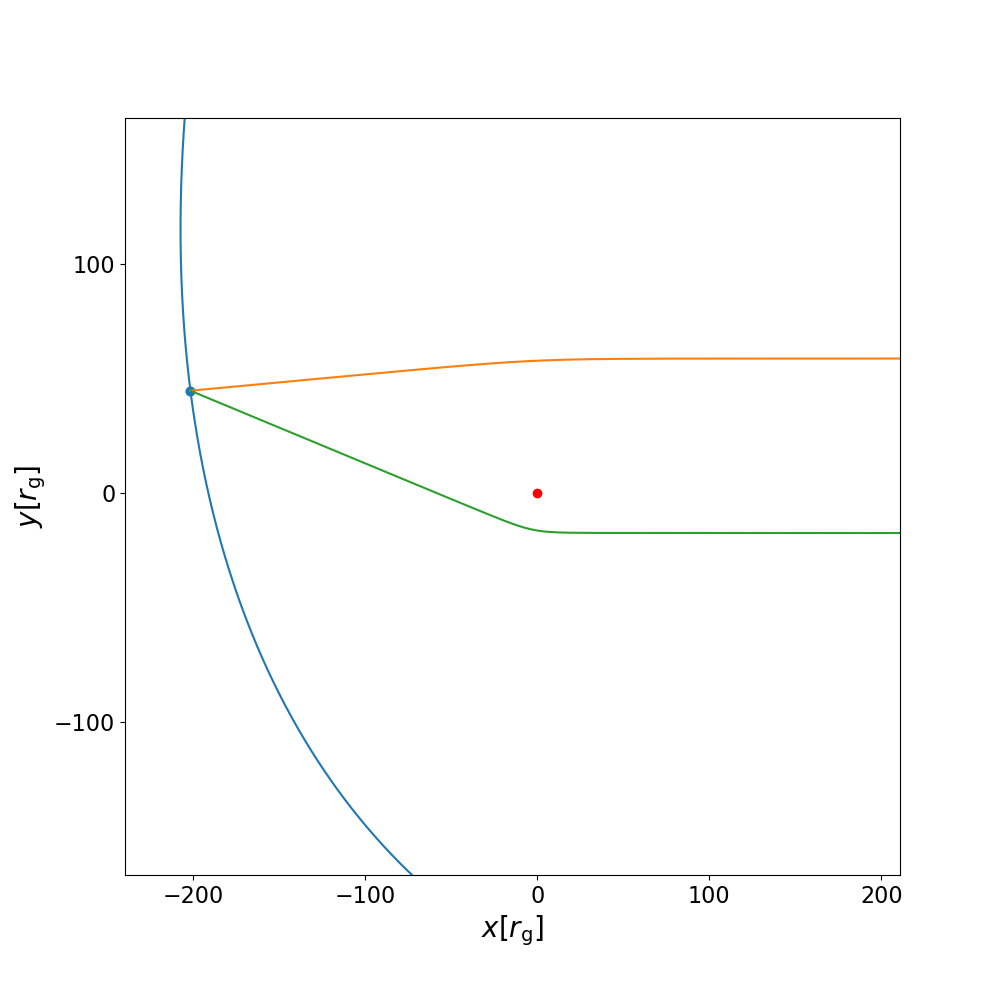}} \\
	
	\medskip
	
	\caption{(a) Weak gravitational bending of the primary ray (orange) and strong gravitational bending of a secondary ray (green) due to the curvature of the spacetime induced by the central massive black hole (red point). The blue dot represents a pulsar on an example eccentric orbit (blue line). The red dot is the location of the massive BH. The observer will receive two pulses as the pulsar rotates. (b) Zoomed in view of (a).} \label{fig:gravbending}
\end{figure*}

Gravitational bending in a Kerr spacetime is naturally included in the ray tracing set up described in Sec. \ref{section:raytracing}. In the weak deflection case, gravitational bending directly influences the photon ToA since the spatial trajectory of the ray is modified from flat Minkowski spacetime. Furthermore, in addition to the primary pulses received by the observer that suffer weak or no gravitational bending, there also exist secondary pulses that are associated with the primary pulses, but which are  strongly bent \citep[see e.g. Fig \ref{fig:gravbending} and also ][]{Wang2008}. Whilst higher order pulses are possible, for the purposes of this paper we only consider the primary and secondary pulses, since the intensity of higher order pulses is naturally low. Our framework and algorithm as described in Sec. \ref{sec:algorithm} is able to find both the primary and secondary pulses. Naturally these strong bending effects are most prominent when the pulsar is on the far side of the black hole. For certain orbital configurations, the pulsar `primary' pulses may be invisible, and instead the pulsar is only visible via the strongly bent `secondary' pulses. Whether strong bending of pulsar rays will occur astrophysically is highly dependent on both the system orbital configuration and the observer viewing angle, with the maximal bending occurring when the pulsar lies close to the central black hole and on the far side with respect to the observer. Analysis from \citet{Stovall2012} suggests that with current radio observation facilities, the probability of detecting strong bent pulsar beams is small but non-negligible. Regarding future radio facilities, optimistic estimates suggest that SKA should be able to detected strongly bent beams from multiple pulsars. The analysis of \citet{Stovall2012} is focused towards the Galactic centre where there are complications due to scattering which decreases the detection probability. Observations in globular clusters of the Local Group might provide more favourable hunting grounds. Furthermore, in estimating a probability \citet{Stovall2012} take as a prototypical model a pulsar at $r = 10^4 r_{\rm g}$. Taking this as the semi-major axis, this gives a Keplerian period of $\sim 4 $ years. At such radii, as noted in \citet{Stovall2012}, the degree of strong deflection is rather small and if an observer receives the primary beam it will also receive the secondary beams. However, for pulsars with smaller orbital periods, or at small orbital radii (e.g. at periastron) the probability of observing a strongly deflected beam would be increased. Extending the considered pulsar population to include not just pulsars of the Milky Way, but also nearby globular clusters would also increase this rate. Moreover, the large baselines offered by future radio interferometers like SKA will enable $\sim \mu$as astrometry \citep{Fomalont2004,Smits2011}. This provides a complementary pathway to timing for identifying secondary rays.

Both strong and weak gravitational bending have important implications for not only the photon ToA due to differing spatial paths, but also due to gravitational and relativistic time dilation, relativistic energy shift (see Sec. \ref{sec:timedilation}) and time-frequency behaviour due to temporal and spatial dispersion (Sec. \ref{sec:timefrequency}).

\subsection{Gravitational and relativistic time dilation }
\label{sec:timedilation}
\begin{figure*} 
	\subfloat[]{\includegraphics[width=0.333\textwidth]{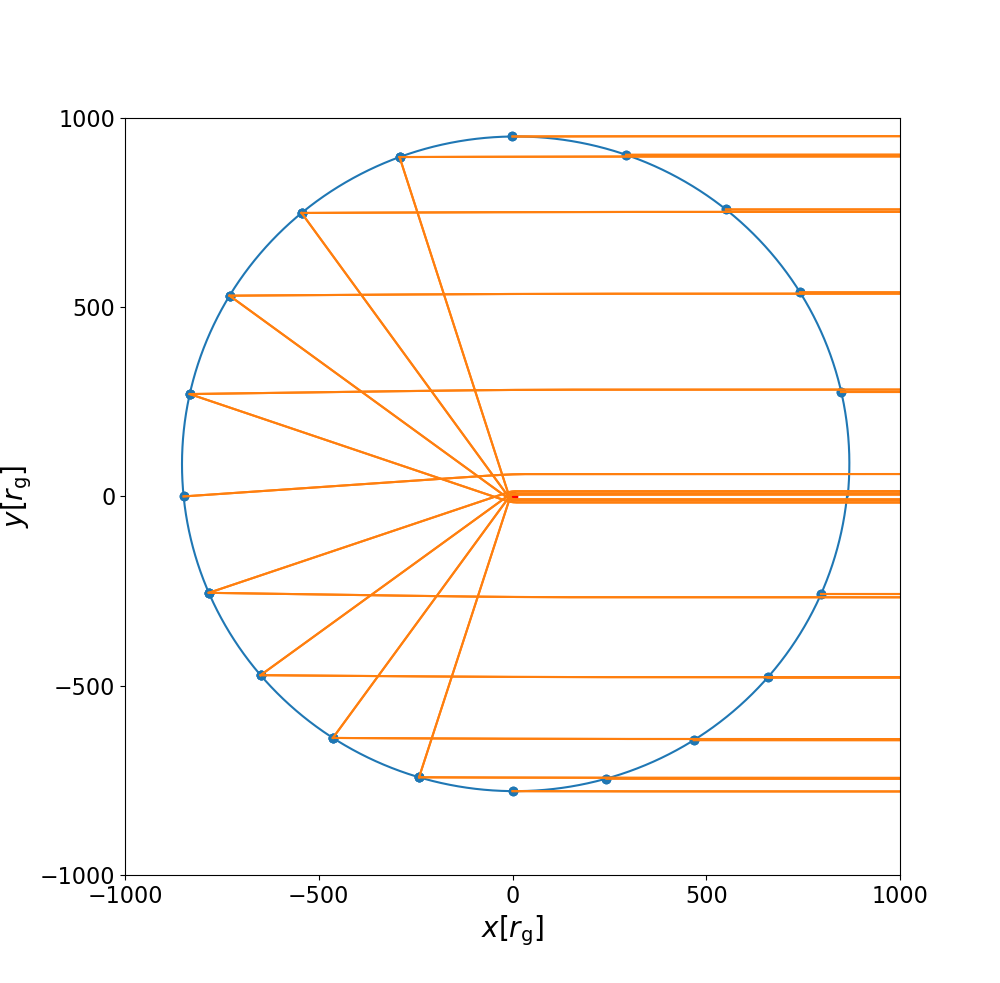}}
	\subfloat[]{\includegraphics[width=0.333\textwidth]{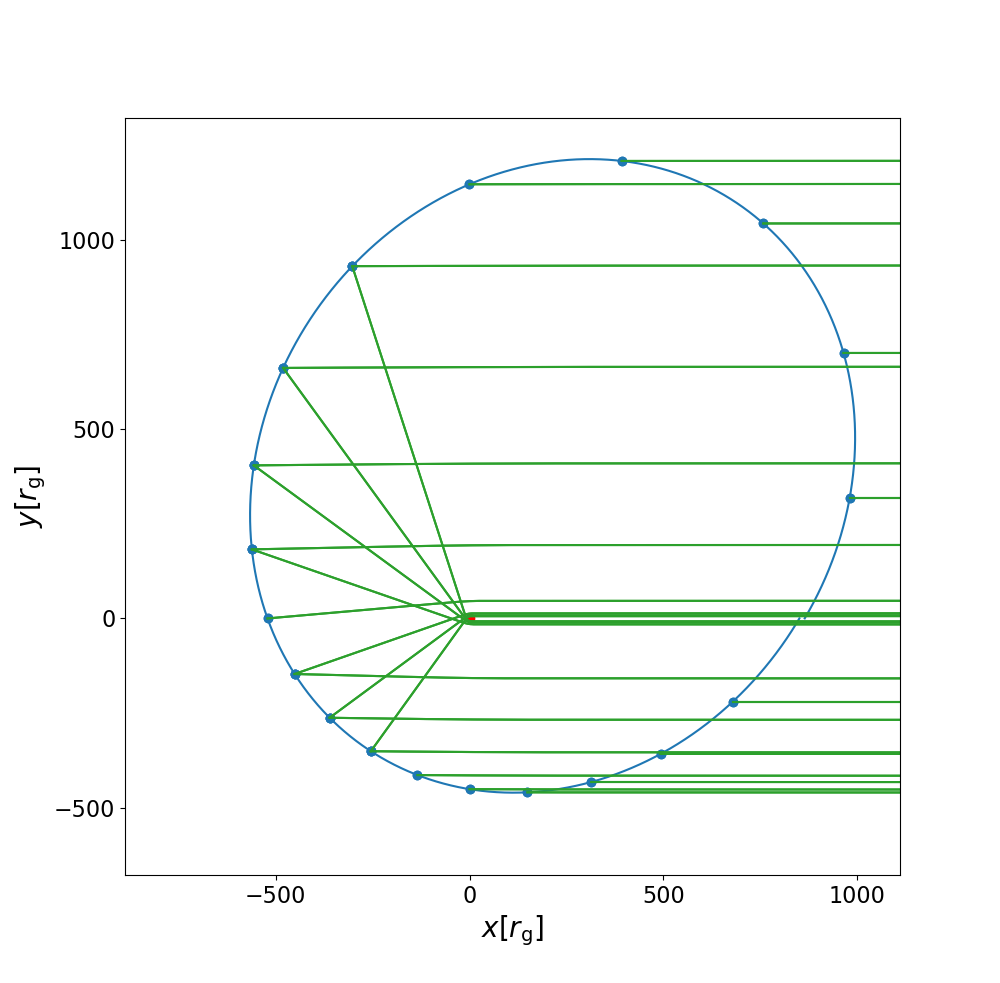}}
	\subfloat[]{\includegraphics[width=0.333\textwidth]{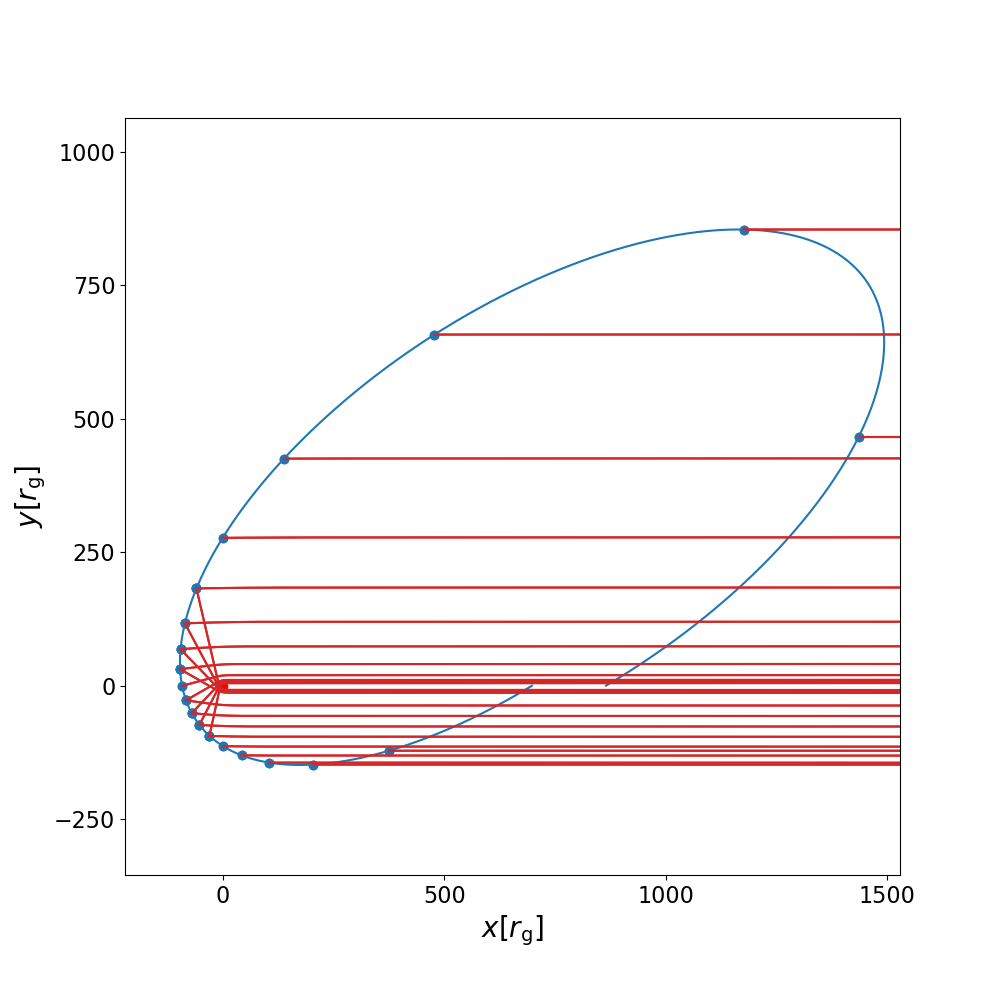}} \\	
	\medskip
	\caption{$P = 0.1$ year orbits for (a) $e =0.1 $, (b) $e = 0.5$, (c) $e=0.9$ as used to determine frequency variations as quantified by $\gamma$ and illustrated in Fig. \ref{fig:gamma}. The orbit lies mostly in the equatorial plane, with some variation due to spin curvature coupling $\rho = 1$.} \label{fig:example_orbits}
\end{figure*}

\begin{figure}
	\includegraphics[width=\columnwidth]{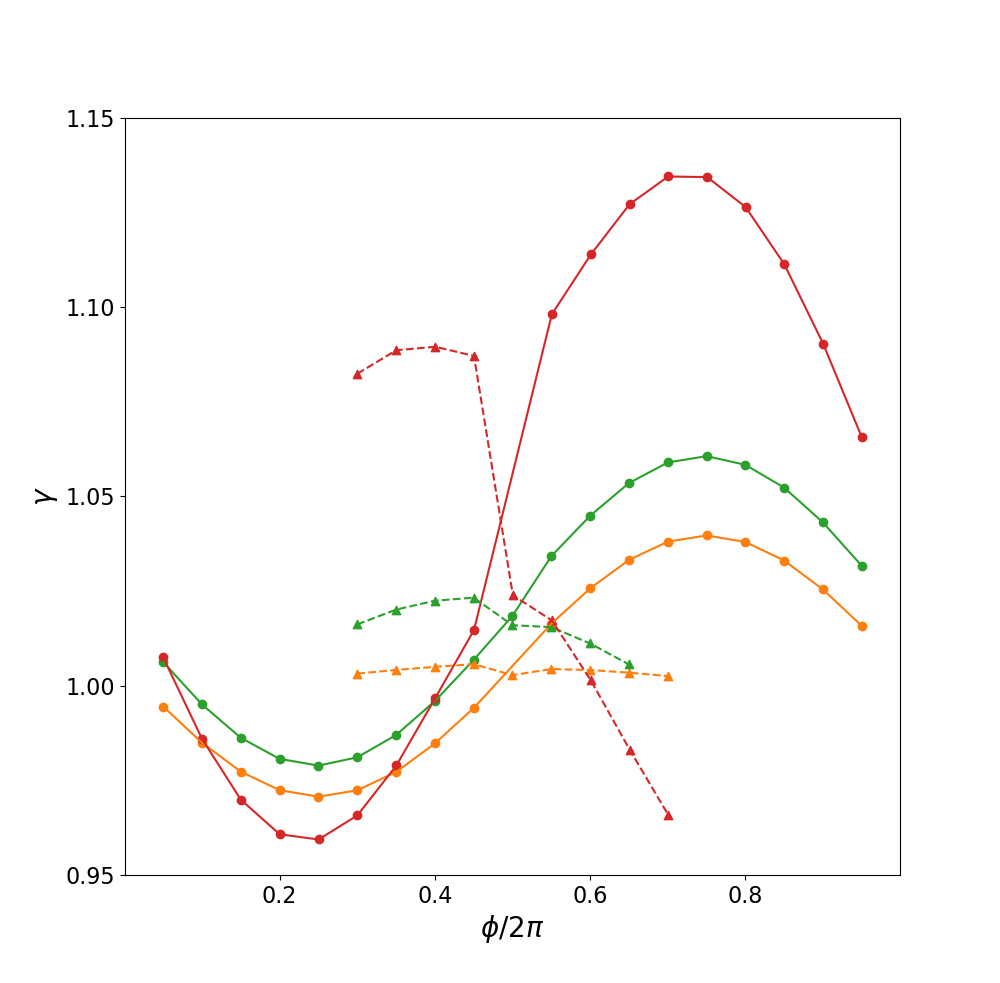}
	\caption{Fractional variation in the pulse frequency due to gravitational time dilation and relativistic Doppler shift over the course of a single $P = 0.1$ year orbit for $e =0.1 , e = 0.5, e=0.9$ (orange, green, red respectively). Solid circular points denote $\gamma$ as calculated from a primary ray whilst the triangular points are due to secondary rays. More eccentric orbits exhibit large amplitude variations due to their greater velocities.}
	\label{fig:gamma}
\end{figure}

For a pulsar in a strong-field environment, the interplay of both gravitational time dilation (clocks run slower in potential wells) and the relativistic Doppler shift induced by the motion of the pulsar will have an impact on the observed radio signal; an apparent modulation in the observed pulsar period where the intrinsic pulsar rotation frequency is different to that recorded by some distant observer. This effect can be quantified as,
\begin{eqnarray}
\gamma = \frac{\nu_{\rm emitted}}{\nu_{ \rm observed}} = \frac{p_{\alpha} u^{\alpha} |_{\rm emitted}}{p_{\alpha} u^{\alpha} |_{\rm observed}} \ ,
\label{eq:timefrequency}
\end{eqnarray}
where $\nu$ can be either the apparent pulsar rotation frequency or the photon frequency. This quantity is frame-invariant and we choose to evaluate it in the global coordinate frame. Since $\gamma$ is a function of both the gravitational time dilation and the relativistic motion of the pulsar, the net shift varies over the pulsar orbit (Figs \ref{fig:example_orbits}, \ref{fig:gamma}). Due to the gravitational bending of light rays, for some sections of the orbit the observer will also receive a secondary beam. When the pulsar is directly behind the black hole the primary ray cannot reach the observer. Gravitational light bending allows for secondary rays to reach the observer from sections of the orbit that would otherwise exist in the black hole shadow. Since this bent ray follows a different geodesic to the primary ray, it is subject to a different general relativistic time dilation and so will also exhibit a different value of $\gamma$. 

The modulation of the observed pulsar rotation frequency must be considered on both long and short timescales. For the former, the variation in the observed rotation frequency will cause a pulse smearing, akin to that observed in relativistic binaries \citep[e.g.][]{Jouteux2002}, but with the added contribution due to strong gravitational time dilation. This would be especially pronounced in the detection and timing of pulsars close to periastron and may be further complicated by the existence of secondary rays which in addition to having different light travel times, will have different values for $\gamma$. Over longer timescales the modulation in the observed pulsar frequency will need to be considered for any accurate timing model, especially one which is then used to test fundamental aspects of GR. 

 Taking into account these relativistic effects on the apparent pulsar rotation period is essential to accurately model the signal from a pulsar in an extreme gravity environment. The dependence of the net time dilation on the motion of the emitter also emphasizes the necessity to have an accurate description of the pulsar orbital dynamics - accounting for all general relativistic effects e.g. spin-curvature coupling - in modeling the pulsar signal.

\subsection{Time-frequency signal}
\label{sec:timefrequency}
\begin{figure}
	\subfloat[]{%
		\includegraphics[clip,width=\columnwidth]{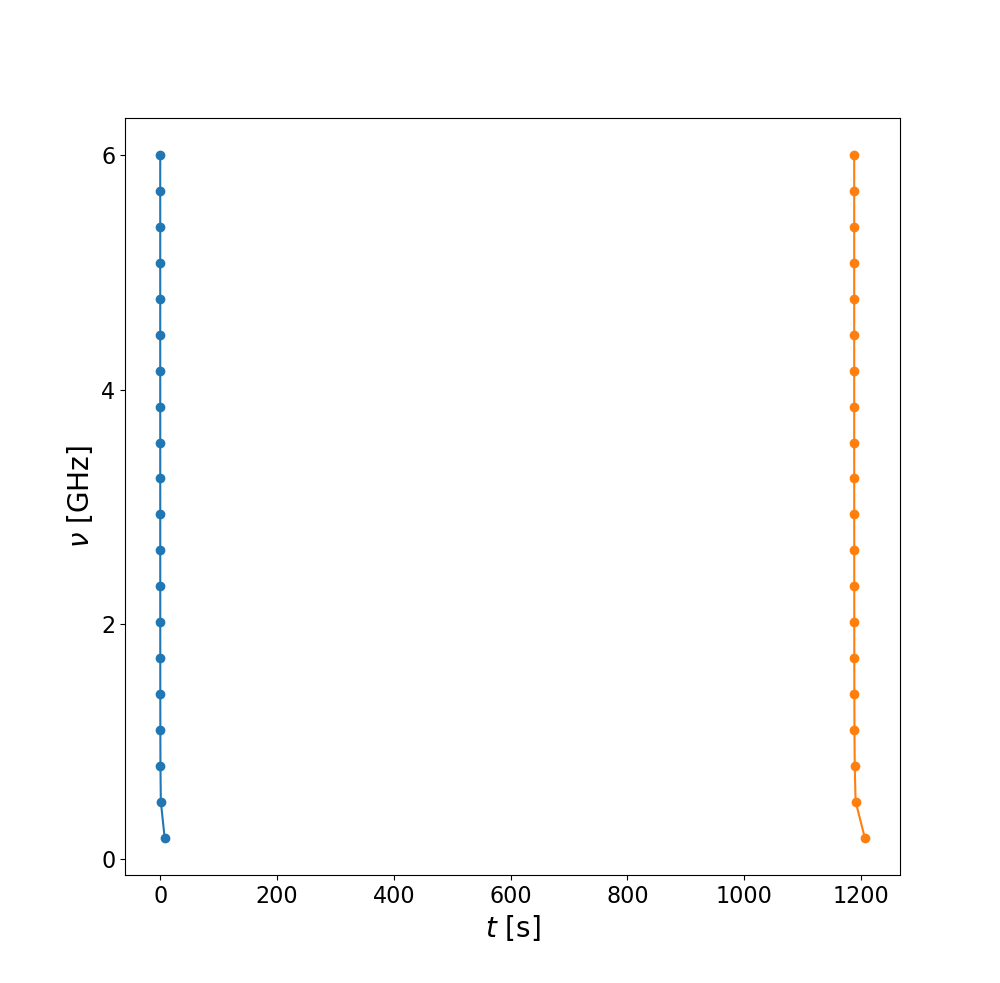}%
	}

	\subfloat[]{%
		\includegraphics[clip,width=\columnwidth]{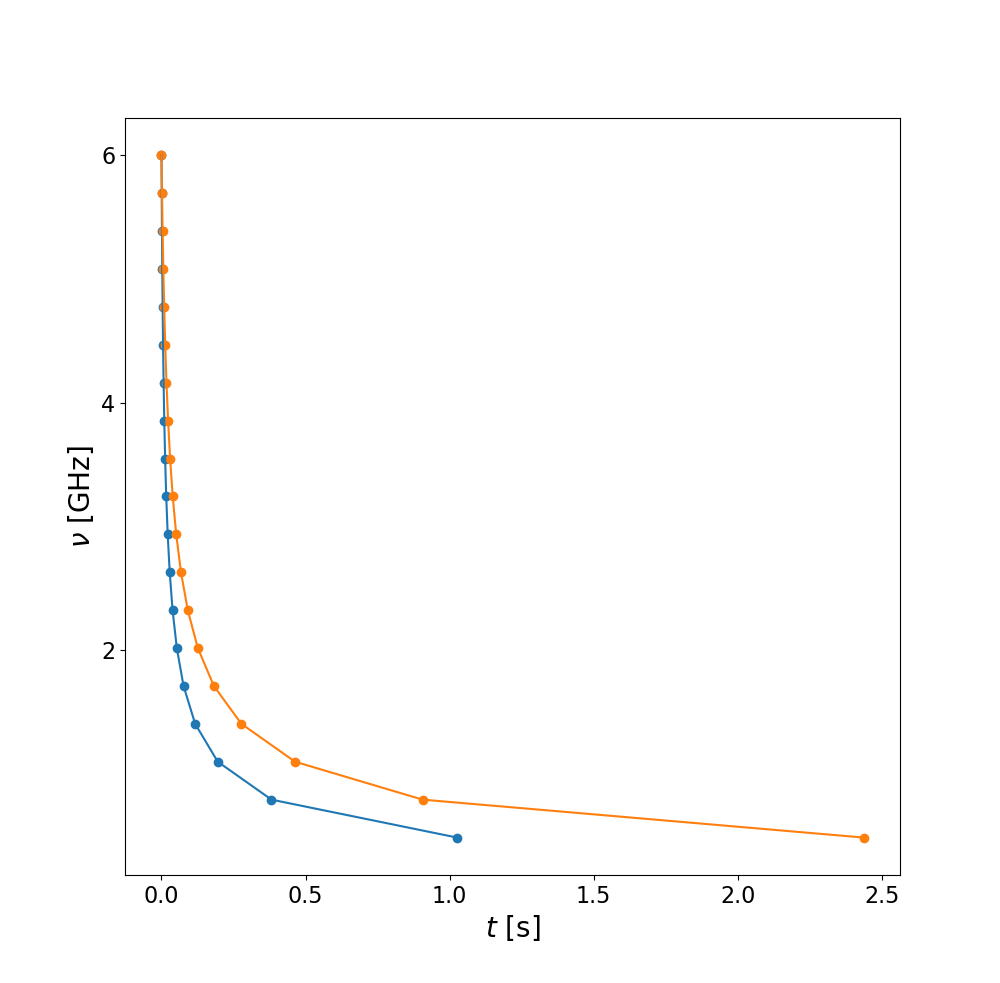}%
	}
	\caption{(a) Time-frequency signal of the primary and secondary rays received from a pulsar at approximately the same location. Two pulses are received due to gravitational light bending, with the bent pulse (orange) delayed with respect to the first (blue). (b) as (a) with $t$ normalized to $t_0$. The different $t - \nu$ profile of the primary and secondary pulses occurs due to the alternative ray path and DM. Due to spatial dispersion, each ray frequency also follows a different spatial path and so is subject to a different DM. We consider frequencies in the range $0.18 - 6$ GHz, and use the electron number density model described in \citet{Psaltis2012, Kimpson2018}.}
	\label{fig:time-freq}
\end{figure}

Typically, pulsar signals are dispersed in time due to the interstellar medium (ISM) and any signal needs correcting to account for this dispersion \citep[e.g.][]{Cordes2016}. The time delay induced by the presence of plasma along the line of sight,
\begin{eqnarray}
\Delta t \propto DM \nu^{-2} \ ,
\end{eqnarray}
for frequency $\nu$ and dispersion measure,
\begin{eqnarray}
DM = \int_{0}^{L} n_e (l) dl \ ,
\end{eqnarray}
where $n_e$ is the free electron number density. During pulsar searches, attempts are made to de-disperse the signal at a number of trial DMs, since the true DM is unknown, before searching for periodicities in Fourier space. Once an accurate DM is determined, long-term precision timing measurements can then take place. Dispersion can also be induced due to relativistic fluids close to the black hole \citep[e.g. accretion flow,][]{Psaltis2012}. Dispersive effects due to the interaction with plasma along the ray path are fully accounted for within our framework, as all electromagnetic effects are included within the $\omega_{\rm p}^2$ term of the Hamiltonian. As discussed, due to strong gravitational bending we receive two radio signals from a PSR; the primary minimally bent ray and the ray which undergoes strong bending due to the curvature of the space time (illustrated in Fig \ref{fig:gravbending}). The secondary ray is typically retarded with respect to the first ray, since it follows a longer spatial path. The alternative trajectory ensures that the secondary ray is subject to a different gravitational time dilation and also the DM along the two rays will be different. Higher order time variations also occur since the pulsar must rotate by a certain angle between the primary and secondary emission, during which time the pulsar is also traveling, but these variations are a fraction of the pulsar period and dominated by the difference in the ray path. All these factors conspire to affect the pulse arrival time and the degree of dispersion across the frequency bandwidth within that pulse (Fig \ref{fig:time-freq}). 

Gravitational light bending means that the apparent position of the pulsar - the position which lies on the asymptote of the tangent line to the ray that converges at the observer - is distinct from the true pulsar position. This introduces complexities in timing observations since two pulses which arrive at a similar time could originate from different times of the orbit and so be are subject to different time dilations, energy shifts, spin and orbital effects and have different dispersion measures and pulse profiles. Since the DM and $t-\nu$ profile will be different for primary and secondary rays this may provide a mechanism to distinguish between the two in any timing solution. 
 
 In addition to temporal dispersion, the combination of strong-field curvature and electron plasma induces a spatial dispersion such that the trajectory followed by the ray after being deflected by the black hole is frequency dependent \citep{Kimpson2018}. This has several implications for rays which are gravitationally bent. Firstly, some rays may no longer be visible in specific frequency bins since if the spatial dispersion is sufficiently severe the ray path is bent such that is does not hit the observer's image plane. Moreover, each ray which does reach the observer has followed a different spatial trajectory and so suffers from a time delay due to the alternative ray path, in addition to different DM's along that path and encounters a different spacetime. Any corrections for gravitational light bending are frequency dependent; the apparent position of the pulsar can be related to multiple true positions, depending on the ray frequency, each necessitating a different DM correction to be applied in each frequency bin. The total received signal is then not some function that varies smoothly with the pulsar orbital phase, but instead the convolution of different energy rays emitted at different orbital phases and consequently subject to differing relativistic and line-of-sight effects. This may, depending on the orbital configuration, result in additional difficulties in detecting signals from PSR-EMRB systems.
\begin{figure}
	\subfloat[]{%
		\includegraphics[clip,width=\columnwidth]{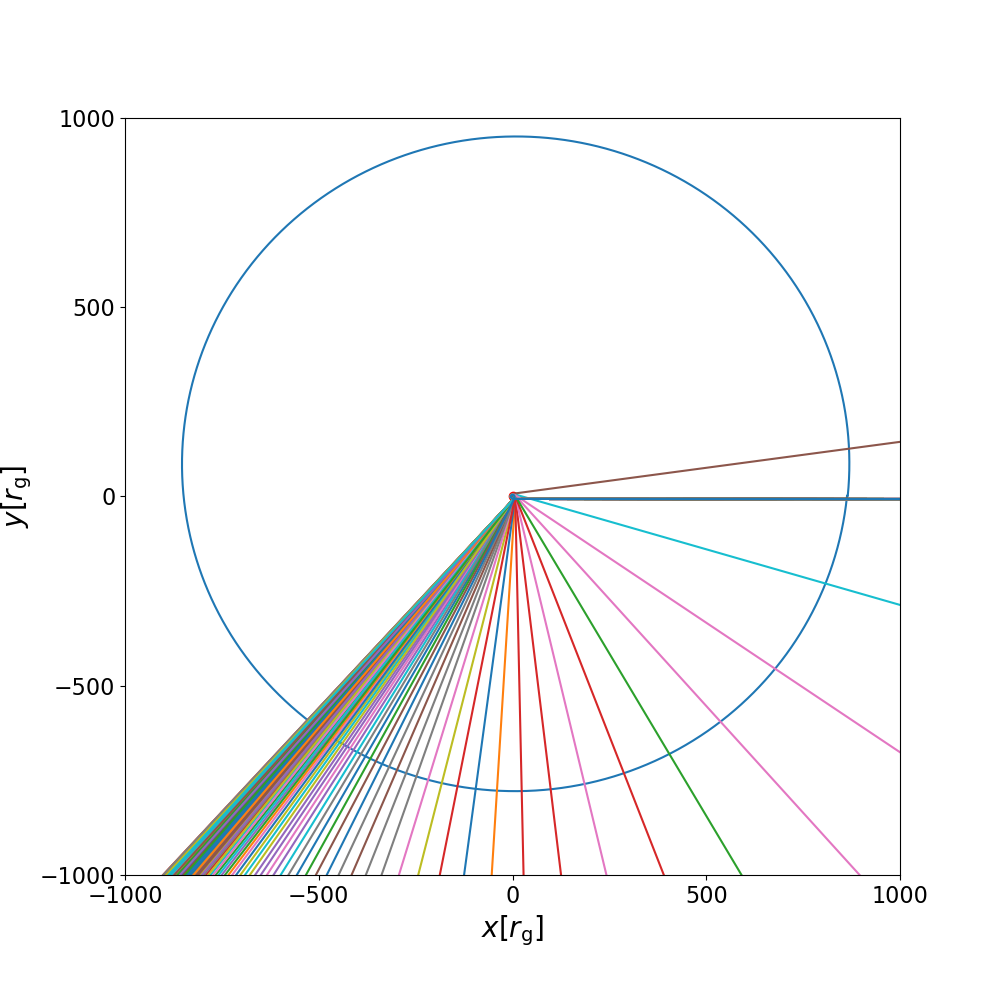}%
	}
	\label{fig:spatdisp}
	\caption{Spatial dispersion of a bundle of light rays with frequencies 0.18 - 6 GHz that originate from the same location on the observer image plane. Spatial dispersion means that the DM correction applied is now frequency dependent. The blue line is some illustrative circular $P=0.1$ year orbit. The model for the electron number density is the same as that used in \citet{Psaltis2012, Kimpson2018}}
\end{figure}
\section{Discussion and Conclusions}
\label{sec:discussion}
We have presented the principles and framework for calculating the radio signal from a PSR in an EMRB. 
We restrict our study to the extreme mass ratio of EMRB systems and so do not consider PSRs in stellar-mass black hole binaries with finite mass ratios \citep[e.g.][]{Blanchet2014,Liu2014}. 
We account for both relativistic and astrophysical effects and the convolution between the two. This includes gravitational and relativistic time dilation and energy shift, gravitational light bending, complex orbital dynamics induced by spin couplings, temporal variation and distortion of the pulse profile due to spin axis precession and relativistic aberration, 2nd order pulses due to gravitational bending, and dispersions (temporal and spatial) induced by the material along the line of sight. We have demonstrated that within our framework we are able to determine the time-frequency behaviour accounting for all these effects. The framework also applies for any orbital configuration, e.g. we are not restricted to orbital motion in the equatorial plane or beaming confined to the orbital plane. The methods used are entirely covariant and general relativistic, rather than working under any post-Newtonian approximation and so are inherently more accurate. Indeed, the post-Newtonian method is an explicitly weak-field method, and the validity of its application to strong-field dynamical regimes is unclear \citep{Will5938}. Whilst working explicitly in the Kerr metric means that we are unable to independently probe either alternative gravitational theories or extensions to Kerr \citep[e.g. Kerr spacetime with an arbitrary mass quadrupole, see ][]{Bini2009}, our framework provides the basis for a theoretical timing model which can then be compared with observations for tests of strong field GR. We approximate the PSR body as a perfect sphere. However due to the spin of the PSR the true shape is more oblate. This will ultimately influence the pitch angle of the ray with the neutron star surface. This effect is considered to be minor, but the method could easily be extended to account for this oblateness \citep[see][]{Nattil2018}. We neglect the effects of hydrodynamic drag due to the plasma that surrounds that black hole, since at compact radii ( $ \lesssim 10^4 r_g$) the gravitational and relativistic effects dominate \citep{Psaltis2012}. We also do not take account of any potential Newtonian perturbations on the motion of the pulsar \citep[e.g.][]{Merritt2011} due to the presence of other masses (e.g. stars, other compact objects etc.) since these factors are likely negligible for the orbital periods considered in this work \citep[$\lesssim 0.3$ years,][]{Liu2012}. Indeed, the potential for external perturbations to hamper tests of strong-field GR necessitates that an ideal PSR-EMRB systems should have orbital periods on the order of $0.1$ years (or better), or else observations should be taken close to periapsis \citep[see discussion in][]{Psaltis2016}. These are precisely the regions where the spacetime curvature and orbital acceleration is greatest, further stressing the importance of a strong-field timing model. We also neglect any influence of gravitational radiation on the orbit or the ray trajectory. The neglect of gravitational radiation is justified since in the extreme mass ratio limit, the timescale for orbital decay due to gravitational wave emission is \citep{MTW},
\begin{eqnarray}
\tau_{GW} \sim \frac{5 r^4}{96mM(m+M)} f(e)^{-1} \ ,
\end{eqnarray}
where $M$ is the mass of the black hole, $m$ the pulsar mass and $r$ the orbital separation. The eccentricity function is,
\begin{eqnarray}
f(e) = (1-e^2)^{-7/2} \left( 1 + \frac{73}{24} e^2 + \frac{37}{96} e^4\right) \ .
\end{eqnarray}
If we take the PSR orbital period $P$ to be Keplerian, then for a pulsar with mass $1.4 M_{\odot}$ on an eccentric ($e = 0.8$), $P=0.1$ year  orbit around a BH with mass $4.3 \times 10^6 M_{\odot}$ , 
\begin{eqnarray}
\frac{\tau_{GW}}{P}  \sim 10^9 >> 1 \ ,
\end{eqnarray}
and so the effects of gravitational radiation  can be neglected. Even for smaller radii and more eccentric orbits the spacetime is well approximated as stationary (e.g. $\tau_{GW}/P \ \sim 10^5$ for $e=0.9$, $r = 100 M$ ). Whilst the effects of gravitational radiation are then not important for a single orbit, for observations over longer periods of time the effect of gravitational emission on the orbit and hence the timing solution will need to be considered. The PSR may also emit a gravitational wave burst during passage through periastron \citep{Berry2013MNRAS, Berry2013}. The influence of this gravitational radiation on both the PSR trajectory and the photon ToA is highly non-trivial and not considered here.

	In this work we do not address the task of how to use our calculation to perform mock data analysis and extract  orbital parameters from simulated PSR-EMRB timing data. This is achievable, for instance,  by using  a software package such as TEMPO \citep{Hobbs2006} to determine the parameters of the timing model from the simulated data \citep[e.g.][]{Liu2014},  or/and by performing a Markov Chain Monte Carlo fitting to investigate 
	constraints on the orbital parameters \citep[e.g.][]{Zhang2017}. In principle, the method outlined in this work can be used to generate time-frequency  data which could then be analysed by the aforementioned methods, but this is beyond the scope of this paper. We also caveat that, due to the high stellar density of the Galactic centre, timing data can be influenced by external Newtonian perturbations \citep[from e.g. stars, stellar mass compact objects][]{Merritt2011}. Any consistent timing solution should therefore  provide a method to correct for these gravitational foreground disturbances \citep[e.g.][]{Angelil2014,Zhang2017}. The handling of such perturbations is not explicitly considered in this work, but we restrict our analysis to pulsars on orbits of $P \lesssim 0.1 $ years where such perturbations are more likely to be negligible \citep{Liu2012}. The analysis of timing data taken from orbital systems with longer periods would need a method to remove these effects (another possibility is to analyse only data taken close to periapsis, where the magnitude of these perturbations is expected to be less severe). \newline

Whilst we have started the theoretical basis for timing observations of a strong-field PSR, there are a number of further potential developments of this work. With the ray tracing solution, we can then perform general relativistic radiative transfer along the rays \citep[e.g.][]{Fuerst2004, Fuerst2007, Younsi2012} so as to determine the effects of line-of-sight material on the beam intensity. Understanding the received intensity and consequent S/N ratio is essential for accurately exploring the prospect of using PSR as probes of strong-field GR. As mentioned, it would also be of interest to investigate how well the weak-field post-Keplerian parametrization can describe strong-field effects. This is important both for determining the types of PSR we require to test GR  and for creating an accurate model to then compare with observations. A coherent $t-\nu$ model which accounts for all relativistic and line of sight effects could then also be potentially used to inform detections; e.g. are acceleration searches necessary for the detection of MSP close to the Galactic centre? As noted by \citet{Faucher2011}, the high stellar density in the Galactic Centre may allow for the creation of some rare binaries (e.g. triple systems). The subsequent dynamics \citep[e.g.][]{Remmen2013} and impact on the PSR signal would be another interesting pursuit. \newline

\noindent To summarize, accurately modelling the time-frequency behaviour from a radio PSR in the strong-field regime leads to a number of higher-order effects which will influence the photon ToA. These include:
\begin{itemize}
	\item Consideration of spin curvature coupling can lead to variations of order $\pm 10\mu$s  in the photon ToA compared to when spin-curvature coupling is neglected. Lower order estimates to the orbital motion (e.g. pure Keplerian dynamics) will further exacerbate the discrepancy, whilst accurate determination of the orbital motion (i.e. including spin-spin, spin-orbit, spin-curvature couplings) is essential for accurately modeling the frequency modulation.
	\item Precession of the spin axis in conjunction with relativistic aberration influences both the pulse profile, pulse duration, the pulse arrival time and the pulse intensity. Aperiodicity in the spin precession may introduce additional complications in the detection of pulsars, whilst severe precession could leave the pulsar signal intermittent.
	\item Gravitational bending causes deviation from a simple Minkowski geodesic which naturally influences the photon ToA. Strong gravitational bending can cause multiple (primary/secondary) pulses to be received by the observer, emitted from a pulsar at approximately the same location. Each ray follows a distinct spacetime path and so suffers different transfer effects, both gravitational (e.g. time dilation) and those due to interaction with material along the line of sight (i.e. temporal dispersion with the DM different for the primary and secondary rays). The interplay of gravitational bending with spatial dispersion further compounds these effects.
	\item Gravitational and relativistic time dilation causes substantial shift in the observed pulse frequency over the orbit. The magnitude of the dilation varies over the orbit and is more pronounced for more eccentric orbits. Secondary, highly bent rays traverse a different curvature of spacetime and so the frequency modulation is different from that of the primary rays. 
	\item The presence of material along the ray path causes a temporal dispersion in the photon ToA, whilst a coupling with the strong spacetime curvature causes a spatial dispersion of the rays. Since each ray of a given frequency follows a different path -  and so is subject to different gravitational and line of sight effects - the ToA will vary.
\end{itemize}

\noindent To conclude, in order to undertake precision tests of GR it is necessary to have a coherent, accurate theoretical model with which to compare observations. In this work we present a framework for calculating the time-frequency behaviour from a pulsar signal in an entirely general relativistic context, including the effects of spin axis precession and nutation, relativistic aberration, relativistic and gravitational time shift, relativistic energy shifts, spatial and temporal dispersion and gravitational light bending. The convolution of these effects will all ultimately influence the photon arrival time. Such a framework is {\bf the first step toward the creation} of an accurate timing model of a PSR signal in the strong-field regime.

\section*{Acknowledgments}
We thank the reviewer Prasenjit Saha for useful critique and suggestion which helped to improve the paper, and in particular his recommendations towards making explicit what the timing model does and does not include. We also thank Ellis Owen and Ziri Younsi for careful reading of the manuscript and useful helpful suggestions towards its improvement. TK acknowledges support from a UK Science and Technology Facilities Council studentship. This research has made use of NASA’s Astrophysics Data Systems. 




\bibliographystyle{mnras}
\bibliography{Paper2} 

\begin{thebibliography}{}
\makeatletter
\relax
\def\mn@urlcharsother{\let\do\@makeother \do\$\do\&\do\#\do\^\do\_\do\%\do\~}
\def\mn@doi{\begingroup\mn@urlcharsother \@ifnextchar [ {\mn@doi@}
  {\mn@doi@[]}}
\def\mn@doi@[#1]#2{\def\@tempa{#1}\ifx\@tempa\@empty \href
  {http://dx.doi.org/#2} {doi:#2}\else \href {http://dx.doi.org/#2} {#1}\fi
  \endgroup}
\def\mn@eprint#1#2{\mn@eprint@#1:#2::\@nil}
\def\mn@eprint@arXiv#1{\href {http://arxiv.org/abs/#1} {{\tt arXiv:#1}}}
\def\mn@eprint@dblp#1{\href {http://dblp.uni-trier.de/rec/bibtex/#1.xml}
  {dblp:#1}}
\def\mn@eprint@#1:#2:#3:#4\@nil{\def\@tempa {#1}\def\@tempb {#2}\def\@tempc
  {#3}\ifx \@tempc \@empty \let \@tempc \@tempb \let \@tempb \@tempa \fi \ifx
  \@tempb \@empty \def\@tempb {arXiv}\fi \@ifundefined
  {mn@eprint@\@tempb}{\@tempb:\@tempc}{\expandafter \expandafter \csname
  mn@eprint@\@tempb\endcsname \expandafter{\@tempc}}}

\bibitem[\protect\citeauthoryear{Abbott et~al.,}{Abbott
  et~al.}{2017a}]{AbbotBHb}
Abbott B.~P.,  et~al., 2017a, \mn@doi [Phys. Rev. Lett.]
  {10.1103/PhysRevLett.118.221101}, 118, 221101

\bibitem[\protect\citeauthoryear{Abbott et~al.,}{Abbott
  et~al.}{2017b}]{AbbottNS}
Abbott B.~P.,  et~al., 2017b, \mn@doi [Phys. Rev. Lett.]
  {10.1103/PhysRevLett.119.161101}, 119, 161101

\bibitem[\protect\citeauthoryear{{Ang{\'e}lil} \& {Saha}}{{Ang{\'e}lil} \&
  {Saha}}{2014}]{Angelil2014}
{Ang{\'e}lil} R.,  {Saha} P.,  2014, \mn@doi [\mnras] {10.1093/mnras/stu1686},
  \href {http://adsabs.harvard.edu/abs/2014MNRAS.444.3780A} {444, 3780}

\bibitem[\protect\citeauthoryear{{Aschenbach}}{{Aschenbach}}{2010}]{Aschenbach2010}
{Aschenbach} B.,  2010, MmSAI, \href
  {http://adsabs.harvard.edu/abs/2010MmSAI..81..319A} {81, 319}

\bibitem[\protect\citeauthoryear{{Babak} et~al.,}{{Babak}
  et~al.}{2017}]{Babak2017}
{Babak} S.,  et~al., 2017, \mn@doi [\prd] {10.1103/PhysRevD.95.103012}, \href
  {http://adsabs.harvard.edu/abs/2017PhRvD..95j3012B} {95, 103012}

\bibitem[\protect\citeauthoryear{{Bambi}, {Jiang}  \& {Steiner}}{{Bambi}
  et~al.}{2016}]{Bambi2016}
{Bambi} C.,  {Jiang} J.,   {Steiner} J.~F.,  2016, \mn@doi [Classical and
  Quantum Gravity] {10.1088/0264-9381/33/6/064001}, \href
  {https://ui.adsabs.harvard.edu/\#abs/2016CQGra..33f4001B} {33, 064001}

\bibitem[\protect\citeauthoryear{{Barausse}, {Hughes}  \&
  {Rezzolla}}{{Barausse} et~al.}{2007}]{Barausse2007}
{Barausse} E.,  {Hughes} S.~A.,   {Rezzolla} L.,  2007, \mn@doi [\prd]
  {10.1103/PhysRevD.76.044007}, \href
  {http://adsabs.harvard.edu/abs/2007PhRvD..76d4007B} {76, 044007}

\bibitem[\protect\citeauthoryear{{Berry} \& {Gair}}{{Berry} \&
  {Gair}}{2013a}]{Berry2013MNRAS}
{Berry} C.~P.~L.,  {Gair} J.~R.,  2013a, \mn@doi [\mnras]
  {10.1093/mnras/stt990}, \href
  {http://adsabs.harvard.edu/abs/2013MNRAS.433.3572B} {433, 3572}

\bibitem[\protect\citeauthoryear{{Berry} \& {Gair}}{{Berry} \&
  {Gair}}{2013b}]{Berry2013}
{Berry} C.~P.~L.,  {Gair} J.~R.,  2013b, in {Auger} G.,  {Bin{\'e}truy} P.,
  {Plagnol} E.,  eds,  Astronomical Society of the Pacific Conference Series
  Vol. 467, 9th LISA Symposium. p.~185 (\mn@eprint {arXiv} {1209.5731})

\bibitem[\protect\citeauthoryear{Bin-Nun}{Bin-Nun}{2010}]{BinNun2010}
Bin-Nun A.~Y.,  2010, \mn@doi [Phys. Rev. D] {10.1103/PhysRevD.82.064009}, 82,
  064009

\bibitem[\protect\citeauthoryear{{Bini}, {Geralico}, {Luongo}  \&
  {Quevedo}}{{Bini} et~al.}{2009}]{Bini2009}
{Bini} D.,  {Geralico} A.,  {Luongo} O.,   {Quevedo} H.,  2009, \mn@doi
  [Classical and Quantum Gravity] {10.1088/0264-9381/26/22/225006}, \href
  {http://adsabs.harvard.edu/abs/2009CQGra..26v5006B} {26, 225006}

\bibitem[\protect\citeauthoryear{Blanchet}{Blanchet}{2014}]{Blanchet2014}
Blanchet L.,  2014, \mn@doi [Living Reviews in Relativity]
  {10.12942/lrr-2014-2}, 17, 2

\bibitem[\protect\citeauthoryear{Bozza \& Mancini}{Bozza \&
  Mancini}{2009}]{Bozza2009}
Bozza V.,  Mancini L.,  2009, The Astrophysical Journal, 696, 701

\bibitem[\protect\citeauthoryear{{Bozza} \& {Mancini}}{{Bozza} \&
  {Mancini}}{2012}]{Bozza2012}
{Bozza} V.,  {Mancini} L.,  2012, \mn@doi [\apj] {10.1088/0004-637X/753/1/56},
  \href {https://ui.adsabs.harvard.edu/\#abs/2012ApJ...753...56B} {753, 56}

\bibitem[\protect\citeauthoryear{Carter}{Carter}{1968}]{Carter1968}
Carter B.,  1968, \mn@doi [Phys. Rev.] {10.1103/PhysRev.174.1559}, 174, 1559

\bibitem[\protect\citeauthoryear{{Chicone}, {Mashhoon}  \& {Punsly}}{{Chicone}
  et~al.}{2005}]{Chicone2005}
{Chicone} C.,  {Mashhoon} B.,   {Punsly} B.,  2005, \mn@doi [Physics Letters A]
  {10.1016/j.physleta.2005.05.072}, \href
  {http://adsabs.harvard.edu/abs/2005PhLA..343....1C} {343, 1}

\bibitem[\protect\citeauthoryear{{Cordes}, {Shannon}  \& {Stinebring}}{{Cordes}
  et~al.}{2016}]{Cordes2016}
{Cordes} J.~M.,  {Shannon} R.~M.,   {Stinebring} D.~R.,  2016, \mn@doi [\apj]
  {10.3847/0004-637X/817/1/16}, \href
  {http://adsabs.harvard.edu/abs/2016ApJ...817...16C} {817, 16}

\bibitem[\protect\citeauthoryear{{De Felice} \& Preti}{{De Felice} \&
  Preti}{1999}]{DeFelice1999}
{De Felice} F.,  Preti G.,  1999, Class. Quantum Grav., 16, 2929

\bibitem[\protect\citeauthoryear{{Desvignes} et~al.,}{{Desvignes}
  et~al.}{2016}]{Desvignes2016}
{Desvignes} G.,  et~al., 2016, \mn@doi [\mnras] {10.1093/mnras/stw483}, \href
  {http://adsabs.harvard.edu/abs/2016MNRAS.458.3341D} {458, 3341}

\bibitem[\protect\citeauthoryear{{Dexter}}{{Dexter}}{2016}]{Dexter2016}
{Dexter} J.,  2016, \mn@doi [\mnras] {10.1093/mnras/stw1526}, \href
  {http://adsabs.harvard.edu/abs/2016MNRAS.462..115D} {462, 115}

\bibitem[\protect\citeauthoryear{{Dexter} \& {Agol}}{{Dexter} \&
  {Agol}}{2009}]{Dexter2009}
{Dexter} J.,  {Agol} E.,  2009, \mn@doi [\apj] {10.1088/0004-637X/696/2/1616},
  \href {http://adsabs.harvard.edu/abs/2009ApJ...696.1616D} {696, 1616}

\bibitem[\protect\citeauthoryear{Dixon}{Dixon}{1964}]{Dixon1964}
Dixon W.~G.,  1964, \mn@doi [Il Nuovo Cimento (1955-1965)]
  {10.1007/BF02734579}, 34, 317

\bibitem[\protect\citeauthoryear{{Dixon}}{{Dixon}}{1974}]{Dixon1974}
{Dixon} W.~G.,  1974, \mn@doi [Philosophical Transactions of the Royal Society
  of London A: Mathematical, Physical and Engineering Sciences]
  {10.1098/rsta.1974.0046}, 277, 59

\bibitem[\protect\citeauthoryear{{Dokuchaev}}{{Dokuchaev}}{2014}]{Dokuchaev2014}
{Dokuchaev} V.~I.,  2014, \mn@doi [General Relativity and Gravitation]
  {10.1007/s10714-014-1832-x}, \href
  {http://adsabs.harvard.edu/abs/2014GReGr..46.1832D} {46, 1832}

\bibitem[\protect\citeauthoryear{Eatough}{Eatough}{2009}]{EatoughThesis}
Eatough R.,  2009, PhD thesis, University of Manchester

\bibitem[\protect\citeauthoryear{{Eatough} et~al.,}{{Eatough}
  et~al.}{2015}]{Eatough2015}
{Eatough} R.,  et~al., 2015, Advancing Astrophysics with the Square Kilometre
  Array (AASKA14), \href {http://adsabs.harvard.edu/abs/2015aska.confE..45E}
  {p.~45}

\bibitem[\protect\citeauthoryear{{Estes}, {Kavic}, {Lippert}  \&
  {Simonetti}}{{Estes} et~al.}{2017}]{Estes2017}
{Estes} J.,  {Kavic} M.,  {Lippert} M.,   {Simonetti} J.~H.,  2017, \mn@doi
  [Astrophys. J.] {10.3847/1538-4357/aa610e}, \href
  {http://adsabs.harvard.edu/abs/2017ApJ...837...87E} {837, 87}

\bibitem[\protect\citeauthoryear{{Faucher-Gigu{\`e}re} \&
  {Loeb}}{{Faucher-Gigu{\`e}re} \& {Loeb}}{2011}]{Faucher2011}
{Faucher-Gigu{\`e}re} C.-A.,  {Loeb} A.,  2011, \mn@doi [\mnras]
  {10.1111/j.1365-2966.2011.19019.x}, \href
  {http://adsabs.harvard.edu/abs/2011MNRAS.415.3951F} {415, 3951}

\bibitem[\protect\citeauthoryear{{Feng} \& {Soria}}{{Feng} \&
  {Soria}}{2011}]{Feng2011}
{Feng} H.,  {Soria} R.,  2011, \mn@doi [New Astron. Rev.]
  {10.1016/j.newar.2011.08.002}, \href
  {http://cdsads.u-strasbg.fr/abs/2011NewAR..55..166F} {55, 166}

\bibitem[\protect\citeauthoryear{{Ferrarese} \& {Merritt}}{{Ferrarese} \&
  {Merritt}}{2000}]{Ferrarese2000}
{Ferrarese} L.,  {Merritt} D.,  2000, \mn@doi [Astrophys. J.] {10.1086/312838},
  \href {http://adsabs.harvard.edu/abs/2000ApJ...539L...9F} {539, L9}

\bibitem[\protect\citeauthoryear{{Filipe Costa} \& {Nat{\'a}rio}}{{Filipe
  Costa} \& {Nat{\'a}rio}}{2014}]{Costa2014}
{Filipe Costa} L.,  {Nat{\'a}rio} J.,  2014, preprint, \href
  {http://adsabs.harvard.edu/abs/2014arXiv1410.6443F} {} (\mn@eprint {arXiv}
  {1410.6443})

\bibitem[\protect\citeauthoryear{Fletcher \& Reeves}{Fletcher \&
  Reeves}{1964}]{Fletcher1964}
Fletcher R.,  Reeves C.~M.,  1964, \mn@doi [The Computer Journal]
  {10.1093/comjnl/7.2.149}, 7, 149

\bibitem[\protect\citeauthoryear{{Fomalont} \& {Reid}}{{Fomalont} \&
  {Reid}}{2004}]{Fomalont2004}
{Fomalont} E.,  {Reid} M.,  2004, \mn@doi [New Astronomy Reviews]
  {10.1016/j.newar.2004.09.037}, \href
  {https://ui.adsabs.harvard.edu/\#abs/2004NewAR..48.1473F} {48, 1473}

\bibitem[\protect\citeauthoryear{{Freire}}{{Freire}}{2013}]{Freire2013}
{Freire} P.~C.~C.,  2013, in {van Leeuwen} J.,  ed.,  IAU Symposium Vol. 291,
  Neutron Stars and Pulsars: Challenges and Opportunities after 80 years. pp
  243--250 (\mn@eprint {arXiv} {1210.3984}), \mn@doi{10.1017/S1743921312023770}

\bibitem[\protect\citeauthoryear{{Fuerst} \& {Wu}}{{Fuerst} \&
  {Wu}}{2004}]{Fuerst2004}
{Fuerst} S.~V.,  {Wu} K.,  2004, \mn@doi [Astron. Astrophys.]
  {10.1051/0004-6361:20035814}, \href
  {http://ukads.nottingham.ac.uk/abs/2004A%26A...424..733F} {424, 733}

\bibitem[\protect\citeauthoryear{{Fuerst} \& {Wu}}{{Fuerst} \&
  {Wu}}{2007}]{Fuerst2007}
{Fuerst} S.~V.,  {Wu} K.,  2007, \mn@doi [Astron. Astrophys.]
  {10.1051/0004-6361:20066008}, \href
  {http://ukads.nottingham.ac.uk/abs/2007A%26A...474...55F} {474, 55}

\bibitem[\protect\citeauthoryear{{Gair}, {Tang}  \& {Volonteri}}{{Gair}
  et~al.}{2010}]{Gair2010}
{Gair} J.~R.,  {Tang} C.,   {Volonteri} M.,  2010, \mn@doi [\prd]
  {10.1103/PhysRevD.81.104014}, \href
  {http://adsabs.harvard.edu/abs/2010PhRvD..81j4014G} {81, 104014}

\bibitem[\protect\citeauthoryear{{Gair}, {Babak}, {Sesana}, {Amaro-Seoane},
  {Barausse}, {Berry}, {Berti}  \& {Sopuerta}}{{Gair} et~al.}{2017}]{Gair2017}
{Gair} J.~R.,  {Babak} S.,  {Sesana} A.,  {Amaro-Seoane} P.,  {Barausse} E.,
  {Berry} C.~P.~L.,  {Berti} E.,   {Sopuerta} C.,  2017, in Journal of Physics
  Conference Series. p. 012021 (\mn@eprint {arXiv} {1704.00009}),
  \mn@doi{10.1088/1742-6596/840/1/012021}

\bibitem[\protect\citeauthoryear{{Gillessen}, {Eisenhauer}, {Trippe},
  {Alexander}, {Genzel}, {Martins}  \& {Ott}}{{Gillessen}
  et~al.}{2009}]{Gillessen2009}
{Gillessen} S.,  {Eisenhauer} F.,  {Trippe} S.,  {Alexander} T.,  {Genzel} R.,
  {Martins} F.,   {Ott} T.,  2009, \mn@doi [ApJ]
  {10.1088/0004-637X/692/2/1075}, \href
  {http://adsabs.harvard.edu/abs/2009ApJ...692.1075G} {692, 1075}

\bibitem[\protect\citeauthoryear{{Gravity Collaboration} et~al.,}{{Gravity
  Collaboration} et~al.}{2017}]{Gravity2017}
{Gravity Collaboration} et~al., 2017, \mn@doi [\aap]
  {10.1051/0004-6361/201730838}, \href
  {https://ui.adsabs.harvard.edu/\#abs/2017A&A...602A..94G} {602, A94}

\bibitem[\protect\citeauthoryear{{Hobbs}, {Edwards}  \& {Manchester}}{{Hobbs}
  et~al.}{2006}]{Hobbs2006}
{Hobbs} G.~B.,  {Edwards} R.~T.,   {Manchester} R.~N.,  2006, \mn@doi [\mnras]
  {10.1111/j.1365-2966.2006.10302.x}, \href
  {https://ui.adsabs.harvard.edu/\#abs/2006MNRAS.369..655H} {369, 655}

\bibitem[\protect\citeauthoryear{{Iorio}}{{Iorio}}{2012}]{Iorio2012}
{Iorio} L.,  2012, \mn@doi [General Relativity and Gravitation]
  {10.1007/s10714-011-1302-7}, \href
  {http://adsabs.harvard.edu/abs/2012GReGr..44..719I} {44, 719}

\bibitem[\protect\citeauthoryear{{Iorio}}{{Iorio}}{2018}]{Iorio2018}
{Iorio} L.,  2018, \mn@doi [Universe] {10.3390/universe4040059}, \href
  {http://adsabs.harvard.edu/abs/2018Univ....4...59I} {4, 59}

\bibitem[\protect\citeauthoryear{{Istomin}}{{Istomin}}{1991}]{Istomin1991}
{Istomin} Y.~N.,  1991, Soviet Astronomy Letters, \href
  {http://adsabs.harvard.edu/abs/1991SvAL...17..301I} {17, 301}

\bibitem[\protect\citeauthoryear{{Jouteux}, {Ramachandran}, {Stappers},
  {Jonker}  \& {van der Klis}}{{Jouteux} et~al.}{2002}]{Jouteux2002}
{Jouteux} S.,  {Ramachandran} R.,  {Stappers} B.~W.,  {Jonker} P.~G.,   {van
  der Klis} M.,  2002, \mn@doi [\aap] {10.1051/0004-6361:20020052}, \href
  {http://adsabs.harvard.edu/abs/2002A%26A...384..532J} {384, 532}

\bibitem[\protect\citeauthoryear{{Kato}, {Miyoshi}, {Takahashi}, {Negoro}  \&
  {Matsumoto}}{{Kato} et~al.}{2010}]{Kato2010a}
{Kato} Y.,  {Miyoshi} M.,  {Takahashi} R.,  {Negoro} H.,   {Matsumoto} R.,
  2010, \mn@doi [MNRAS] {10.1111/j.1745-3933.2010.00818.x}, \href
  {http://adsabs.harvard.edu/abs/2010MNRAS.403L..74K} {403, L74}

\bibitem[\protect\citeauthoryear{{Keane} et~al.,}{{Keane}
  et~al.}{2015}]{Keane2015}
{Keane} E.,  et~al., 2015, Advancing Astrophysics with the Square Kilometre
  Array (AASKA14), \href {http://adsabs.harvard.edu/abs/2015aska.confE..40K}
  {p.~40}

\bibitem[\protect\citeauthoryear{{Kimpson}, {Wu}  \& {Zane}}{{Kimpson}
  et~al.}{2019}]{Kimpson2018}
{Kimpson} T.,  {Wu} K.,   {Zane} S.,  2019, \mn@doi [\mnras]
  {10.1093/mnras/stz138}, \href
  {https://ui.adsabs.harvard.edu/\#abs/2019MNRAS.484.2411K} {484, 2411}

\bibitem[\protect\citeauthoryear{{Kleihaus}, {Kunz}  \& {Schneider}}{{Kleihaus}
  et~al.}{2012}]{Kleihaus2012}
{Kleihaus} B.,  {Kunz} J.,   {Schneider} S.,  2012, \mn@doi [Phys. Rev. D]
  {10.1103/PhysRevD.85.024045}, \href
  {http://adsabs.harvard.edu/abs/2012PhRvD..85b4045K} {85, 024045}

\bibitem[\protect\citeauthoryear{{Kramer}}{{Kramer}}{2016}]{Kramer2016}
{Kramer} M.,  2016, \mn@doi [International Journal of Modern Physics D]
  {10.1142/S0218271816300299}, \href
  {http://adsabs.harvard.edu/abs/2016IJMPD..2530029K} {25, 1630029}

\bibitem[\protect\citeauthoryear{{Kramer}, {Backer}, {Cordes}, {Lazio},
  {Stappers}  \& {Johnston}}{{Kramer} et~al.}{2004}]{Kramer2004}
{Kramer} M.,  {Backer} D.~C.,  {Cordes} J.~M.,  {Lazio} T.~J.~W.,  {Stappers}
  B.~W.,   {Johnston} S.,  2004, \mn@doi [New Astron. Rev.]
  {10.1016/j.newar.2004.09.020}, \href
  {http://ukads.nottingham.ac.uk/abs/2004NewAR..48..993K} {48, 993}

\bibitem[\protect\citeauthoryear{{Krolik}, {Hawley}  \& {Hirose}}{{Krolik}
  et~al.}{2005}]{Krolik2005}
{Krolik} J.~H.,  {Hawley} J.~F.,   {Hirose} S.,  2005, \mn@doi [\apj]
  {10.1086/427932}, \href {http://adsabs.harvard.edu/abs/2005ApJ...622.1008K}
  {622, 1008}

\bibitem[\protect\citeauthoryear{{Kulkarni} et~al.,}{{Kulkarni}
  et~al.}{2011}]{Kulkarni2011}
{Kulkarni} A.~K.,  et~al., 2011, \mn@doi [\mnras]
  {10.1111/j.1365-2966.2011.18446.x}, \href
  {http://adsabs.harvard.edu/abs/2011MNRAS.414.1183K} {414, 1183}

\bibitem[\protect\citeauthoryear{{Lazarus}, {Karuppusamy}, {Graikou},
  {Caballero}, {Champion}, {Lee}, {Verbiest}  \& {Kramer}}{{Lazarus}
  et~al.}{2016}]{Lazarus2016}
{Lazarus} P.,  {Karuppusamy} R.,  {Graikou} E.,  {Caballero} R.~N.,  {Champion}
  D.~J.,  {Lee} K.~J.,  {Verbiest} J.~P.~W.,   {Kramer} M.,  2016, \mn@doi
  [\mnras] {10.1093/mnras/stw189}, \href
  {http://adsabs.harvard.edu/abs/2016MNRAS.458..868L} {458, 868}

\bibitem[\protect\citeauthoryear{{Li}, {Wu}  \& {Singh}}{{Li}
  et~al.}{2018}]{Li2018}
{Li} K.~J.,  {Wu} K.,   {Singh} D.,  2018, MNRAS, Submitted

\bibitem[\protect\citeauthoryear{{Liu}, {Verbiest}, {Kramer}, {Stappers}, {van
  Straten}  \& {Cordes}}{{Liu} et~al.}{2011}]{Liu2011}
{Liu} K.,  {Verbiest} J.~P.~W.,  {Kramer} M.,  {Stappers} B.~W.,  {van Straten}
  W.,   {Cordes} J.~M.,  2011, \mn@doi [\mnras]
  {10.1111/j.1365-2966.2011.19452.x}, \href
  {http://adsabs.harvard.edu/abs/2011MNRAS.417.2916L} {417, 2916}

\bibitem[\protect\citeauthoryear{Liu, Wex, Kramer, Cordes  \& Lazio}{Liu
  et~al.}{2012}]{Liu2012}
Liu K.,  Wex N.,  Kramer M.,  Cordes J.~M.,   Lazio T. J.~W.,  2012, Astrophys.
  J., 747

\bibitem[\protect\citeauthoryear{{Liu}, {Eatough}, {Wex}  \& {Kramer}}{{Liu}
  et~al.}{2014}]{Liu2014}
{Liu} K.,  {Eatough} R.~P.,  {Wex} N.,   {Kramer} M.,  2014, \mn@doi [Mon. Not.
  R. Astron. Soc.] {10.1093/mnras/stu1913}, \href
  {http://adsabs.harvard.edu/abs/2014MNRAS.445.3115L} {445, 3115}

\bibitem[\protect\citeauthoryear{{Liu}, {Bassa}  \& {Stappers}}{{Liu}
  et~al.}{2018}]{Liu2018}
{Liu} X.~J.,  {Bassa} C.~G.,   {Stappers} B.~W.,  2018, \mn@doi [\mnras]
  {10.1093/mnras/sty1202}, \href
  {http://adsabs.harvard.edu/abs/2018MNRAS.478.2359L} {478, 2359}

\bibitem[\protect\citeauthoryear{{Lorimer}}{{Lorimer}}{2008}]{Lorimer2008}
{Lorimer} D.~R.,  2008, \mn@doi [Living Rev. Relativ.] {10.12942/lrr-2008-8},
  \href {http://ukads.nottingham.ac.uk/abs/2008LRR....11....8L} {11, 8}

\bibitem[\protect\citeauthoryear{{Macquart}, {Kanekar}, {Frail}  \&
  {Ransom}}{{Macquart} et~al.}{2010}]{Macquart2010}
{Macquart} J.-P.,  {Kanekar} N.,  {Frail} D.~A.,   {Ransom} S.~M.,  2010,
  \mn@doi [\apj] {10.1088/0004-637X/715/2/939}, \href
  {http://adsabs.harvard.edu/abs/2010ApJ...715..939M} {715, 939}

\bibitem[\protect\citeauthoryear{{Mashhoon} \& {Singh}}{{Mashhoon} \&
  {Singh}}{2006}]{Mashhoon2006}
{Mashhoon} B.,  {Singh} D.,  2006, \mn@doi [\prd] {10.1103/PhysRevD.74.124006},
  \href {http://adsabs.harvard.edu/abs/2006PhRvD..74l4006M} {74, 124006}

\bibitem[\protect\citeauthoryear{{Mathisson}}{{Mathisson}}{1937}]{Mathisson1937}
{Mathisson} A.,  1937, Acta Phys. Pol., 6, 163

\bibitem[\protect\citeauthoryear{{Merritt}, {Alexander}, {Mikkola}  \&
  {Will}}{{Merritt} et~al.}{2011}]{Merritt2011}
{Merritt} D.,  {Alexander} T.,  {Mikkola} S.,   {Will} C.~M.,  2011, \mn@doi
  [\prd] {10.1103/PhysRevD.84.044024}, \href
  {http://adsabs.harvard.edu/abs/2011PhRvD..84d4024M} {84, 044024}

\bibitem[\protect\citeauthoryear{Misner, Thorne  \& Wheeler}{Misner
  et~al.}{1973}]{MTW}
Misner C.,  Thorne K.,   Wheeler J.,  1973, Gravitation.
No. pt. 3 in Gravitation, W. H. Freeman, \url
  {https://books.google.co.uk/books?id=w4Gigq3tY1kC}

\bibitem[\protect\citeauthoryear{Nampalliwar, Price, Creighton  \&
  Jenet}{Nampalliwar et~al.}{2013}]{Nampalliwar2013}
Nampalliwar S.,  Price R.~H.,  Creighton T.,   Jenet F.~A.,  2013, Astrophys.
  J., 778, 145

\bibitem[\protect\citeauthoryear{{N{\"a}ttil{\"a}} \&
  {Pihajoki}}{{N{\"a}ttil{\"a}} \& {Pihajoki}}{2018}]{Nattil2018}
{N{\"a}ttil{\"a}} J.,  {Pihajoki} P.,  2018, \mn@doi [\aap]
  {10.1051/0004-6361/201630261}, \href
  {http://adsabs.harvard.edu/abs/2018A%26A...615A..50N} {615, A50}

\bibitem[\protect\citeauthoryear{{Norris}}{{Norris}}{2011}]{Norris2011}
{Norris} R.~P.,  2011, in Sixth IEEE International Conference on eScience, p.
  21-24. pp 21--24 (\mn@eprint {arXiv} {1101.1355}),
  \mn@doi{10.1109/eScienceW.2010.13}

\bibitem[\protect\citeauthoryear{{Obukhov}, {Silenko}  \& {Teryaev}}{{Obukhov}
  et~al.}{2009}]{Obukhov2009}
{Obukhov} Y.~N.,  {Silenko} A.~J.,   {Teryaev} O.~V.,  2009, \mn@doi [\prd]
  {10.1103/PhysRevD.80.064044}, \href
  {http://adsabs.harvard.edu/abs/2009PhRvD..80f4044O} {80, 064044}

\bibitem[\protect\citeauthoryear{{Pachner}}{{Pachner}}{1970}]{Pachner1971}
{Pachner} J.,  1970, \mn@doi [Canadian Journal of Physics] {10.1139/p70-126},
  \href {http://adsabs.harvard.edu/abs/1970CaJPh..48..970P} {48, 970}

\bibitem[\protect\citeauthoryear{Papapetrou}{Papapetrou}{1951}]{Papapetrou1951}
Papapetrou A.,  1951, \mn@doi [Proceedings of the Royal Society of London A:
  Mathematical, Physical and Engineering Sciences] {10.1098/rspa.1951.0200},
  209, 248

\bibitem[\protect\citeauthoryear{{Perera} et~al.,}{{Perera}
  et~al.}{2017}]{Perera2017MNRAS}
{Perera} B.~B.~P.,  et~al., 2017, \mn@doi [Mon. Not. R. Astron. Soc.]
  {10.1093/mnras/stx501}, \href
  {http://ukads.nottingham.ac.uk/abs/2017MNRAS.468.2114P} {468, 2114}

\bibitem[\protect\citeauthoryear{{Perlick} \& {Tsupko}}{{Perlick} \&
  {Tsupko}}{2017}]{Perlick2017}
{Perlick} V.,  {Tsupko} O.~Y.,  2017, \mn@doi [\prd]
  {10.1103/PhysRevD.95.104003}, \href
  {http://adsabs.harvard.edu/abs/2017PhRvD..95j4003P} {95, 104003}

\bibitem[\protect\citeauthoryear{{Plyatsko} \& {Fenyk}}{{Plyatsko} \&
  {Fenyk}}{2016}]{Plyatsko2016}
{Plyatsko} R.,  {Fenyk} M.,  2016, \mn@doi [\prd] {10.1103/PhysRevD.94.044047},
  \href {http://adsabs.harvard.edu/abs/2016PhRvD..94d4047P} {94, 044047}

\bibitem[\protect\citeauthoryear{Press, Teukolsky, Vetterling  \&
  Flannery}{Press et~al.}{1996}]{Press1992}
Press W.~H.,  Teukolsky S.~A.,  Vetterling W.~T.,   Flannery B.~P.,  1996,
  Numerical Recipes in Fortran 90 (2Nd Ed.): The Art of Parallel Scientific
  Computing.
Cambridge University Press, New York, NY, USA

\bibitem[\protect\citeauthoryear{{Psaltis}}{{Psaltis}}{2008}]{Psaltis2008}
{Psaltis} D.,  2008, \mn@doi [Living Reviews in Relativity]
  {10.12942/lrr-2008-9}, \href
  {http://adsabs.harvard.edu/abs/2008LRR....11....9P} {11, 9}

\bibitem[\protect\citeauthoryear{{Psaltis}}{{Psaltis}}{2012}]{Psaltis2012}
{Psaltis} D.,  2012, \mn@doi [\apj] {10.1088/0004-637X/759/2/130}, \href
  {http://adsabs.harvard.edu/abs/2012ApJ...759..130P} {759, 130}

\bibitem[\protect\citeauthoryear{{Psaltis}, {Wex}  \& {Kramer}}{{Psaltis}
  et~al.}{2016}]{Psaltis2016}
{Psaltis} D.,  {Wex} N.,   {Kramer} M.,  2016, \mn@doi [\apj]
  {10.3847/0004-637X/818/2/121}, \href
  {http://adsabs.harvard.edu/abs/2016ApJ...818..121P} {818, 121}

\bibitem[\protect\citeauthoryear{{Pu}, {Yun}, {Younsi}  \& {Yoon}}{{Pu}
  et~al.}{2016}]{Pu2016}
{Pu} H.-Y.,  {Yun} K.,  {Younsi} Z.,   {Yoon} S.-J.,  2016, \mn@doi [\apj]
  {10.3847/0004-637X/820/2/105}, \href
  {http://adsabs.harvard.edu/abs/2016ApJ...820..105P} {820, 105}

\bibitem[\protect\citeauthoryear{Rafikov \& Lai}{Rafikov \&
  Lai}{2006}]{Roman2006}
Rafikov R.~R.,  Lai D.,  2006, The Astrophysical Journal, 641, 438

\bibitem[\protect\citeauthoryear{{Rajwade}, {Lorimer}  \& {Anderson}}{{Rajwade}
  et~al.}{2017}]{Rajwade2017}
{Rajwade} K.~M.,  {Lorimer} D.~R.,   {Anderson} L.~D.,  2017, \mn@doi [Mon.
  Not. R. Astron. Soc.] {10.1093/mnras/stx1661}, \href
  {http://ukads.nottingham.ac.uk/abs/2017MNRAS.471..730R} {471, 730}

\bibitem[\protect\citeauthoryear{{Remmen} \& {Wu}}{{Remmen} \&
  {Wu}}{2013}]{Remmen2013}
{Remmen} G.~N.,  {Wu} K.,  2013, \mn@doi [Mon. Not. R. Astron. Soc.]
  {10.1093/mnras/stt023}, \href
  {http://ukads.nottingham.ac.uk/abs/2013MNRAS.430.1940R} {430, 1940}

\bibitem[\protect\citeauthoryear{{Rosquist}, {Bylund}  \&
  {Samuelsson}}{{Rosquist} et~al.}{2009}]{Rosquist2009}
{Rosquist} K.,  {Bylund} T.,   {Samuelsson} L.,  2009, \mn@doi [International
  Journal of Modern Physics D] {10.1142/S0218271809014546}, \href
  {http://ukads.nottingham.ac.uk/abs/2009IJMPD..18..429R} {18, 429}

\bibitem[\protect\citeauthoryear{{Saxton}, {Younsi}  \& {Wu}}{{Saxton}
  et~al.}{2016}]{Saxton2016}
{Saxton} C.~J.,  {Younsi} Z.,   {Wu} K.,  2016, \mn@doi [Mon. Not. R. Astron.
  Soc.] {10.1093/mnras/stw1626}, \href
  {http://ukads.nottingham.ac.uk/abs/2016MNRAS.461.4295S} {461, 4295}

\bibitem[\protect\citeauthoryear{{Shao} et~al.,}{{Shao}
  et~al.}{2015}]{Shao2015}
{Shao} L.,  et~al., 2015, Advancing Astrophysics with the Square Kilometre
  Array (AASKA14), \href {http://adsabs.harvard.edu/abs/2015aska.confE..42S}
  {p.~42}

\bibitem[\protect\citeauthoryear{{Shcherbakov} \& {Huang}}{{Shcherbakov} \&
  {Huang}}{2011}]{Shcherbakov2011}
{Shcherbakov} R.~V.,  {Huang} L.,  2011, \mn@doi [\mnras]
  {10.1111/j.1365-2966.2010.17502.x}, \href
  {http://adsabs.harvard.edu/abs/2011MNRAS.410.1052S} {410, 1052}

\bibitem[\protect\citeauthoryear{{Singh}}{{Singh}}{2005}]{Singh2005}
{Singh} D.,  2005, \mn@doi [\prd] {10.1103/PhysRevD.72.084033}, \href
  {http://adsabs.harvard.edu/abs/2005PhRvD..72h4033S} {72, 084033}

\bibitem[\protect\citeauthoryear{{Singh}, {Wu}  \& {Sarty}}{{Singh}
  et~al.}{2014}]{Singh2014}
{Singh} D.,  {Wu} K.,   {Sarty} G.~E.,  2014, \mn@doi [Mon. Not. R. Astron.
  Soc.] {10.1093/mnras/stu614}, \href
  {http://ukads.nottingham.ac.uk/abs/2014MNRAS.441..800S} {441, 800}

\bibitem[\protect\citeauthoryear{{Smits}, {Lorimer}, {Kramer}, {Manchester},
  {Stappers}, {Jin}, {Nan}  \& {Li}}{{Smits} et~al.}{2009}]{Smits2009}
{Smits} R.,  {Lorimer} D.~R.,  {Kramer} M.,  {Manchester} R.,  {Stappers} B.,
  {Jin} C.~J.,  {Nan} R.~D.,   {Li} D.,  2009, \mn@doi [\aap]
  {10.1051/0004-6361/200911939}, \href
  {http://adsabs.harvard.edu/abs/2009A%26A...505..919S} {505, 919}

\bibitem[\protect\citeauthoryear{{Smits}, {Tingay}, {Wex}, {Kramer}  \&
  {Stappers}}{{Smits} et~al.}{2011}]{Smits2011}
{Smits} R.,  {Tingay} S.~J.,  {Wex} N.,  {Kramer} M.,   {Stappers} B.,  2011,
  \mn@doi [\aap] {10.1051/0004-6361/201016141}, \href
  {https://ui.adsabs.harvard.edu/\#abs/2011A&A...528A.108S} {528, A108}

\bibitem[\protect\citeauthoryear{{Stovall}, {Creighton}, {Price}  \&
  {Jenet}}{{Stovall} et~al.}{2012}]{Stovall2012}
{Stovall} K.,  {Creighton} T.,  {Price} R.~H.,   {Jenet} F.~A.,  2012, \mn@doi
  [\apj] {10.1088/0004-637X/744/2/143}, \href
  {http://adsabs.harvard.edu/abs/2012ApJ...744..143S} {744, 143}

\bibitem[\protect\citeauthoryear{Synge}{Synge}{1960}]{Synge1960}
Synge J.,  1960, {Relativity: The General Theory}.
North-Holland Publishing Company

\bibitem[\protect\citeauthoryear{{Tulczyjew}}{{Tulczyjew}}{1959}]{Tulczyjew1959}
{Tulczyjew} W.,  1959, \mn@doi [Acta Phys. Pol.] {10.1093/mnras/stu614}, 18,
  393

\bibitem[\protect\citeauthoryear{{Verbiest} et~al.,}{{Verbiest}
  et~al.}{2009}]{Verbiest2009}
{Verbiest} J.~P.~W.,  et~al., 2009, \mn@doi [\mnras]
  {10.1111/j.1365-2966.2009.15508.x}, \href
  {http://adsabs.harvard.edu/abs/2009MNRAS.400..951V} {400, 951}

\bibitem[\protect\citeauthoryear{Wang, Jenet, Creighton  \& Price}{Wang
  et~al.}{2009a}]{Wang2008}
Wang Y.,  Jenet F.~A.,  Creighton T.,   Price R.~H.,  2009a, Astrophys. J.,
  697, 237

\bibitem[\protect\citeauthoryear{Wang, Creighton, Price  \& Jenet}{Wang
  et~al.}{2009b}]{Wang2009}
Wang Y.,  Creighton T.,  Price R.~H.,   Jenet F.~A.,  2009b, Astrophys. J.,
  705, 1252

\bibitem[\protect\citeauthoryear{{Wex} \& {Kopeikin}}{{Wex} \&
  {Kopeikin}}{1999}]{Wex1999}
{Wex} N.,  {Kopeikin} S.~M.,  1999, \mn@doi [Astrophys. J.] {10.1086/306933},
  \href {http://adsabs.harvard.edu/abs/1999ApJ...514..388W} {514, 388}

\bibitem[\protect\citeauthoryear{Wharton, Chatterjee, Cordes, Deneva  \&
  Lazio}{Wharton et~al.}{2012}]{Wharton2012}
Wharton R.~S.,  Chatterjee S.,  Cordes J.~M.,  Deneva J.~S.,   Lazio T. J.~W.,
  2012, \mn@doi [Astrophys. J.] {10.1088/0004-637X/753/2/108}, 753

\bibitem[\protect\citeauthoryear{Will}{Will}{2011}]{Will5938}
Will C.~M.,  2011, \mn@doi [Proceedings of the National Academy of Sciences]
  {10.1073/pnas.1103127108}, 108, 5938

\bibitem[\protect\citeauthoryear{{Will}}{{Will}}{2014}]{Will2014}
{Will} C.~M.,  2014, \mn@doi [Living Rev. Relativ.] {10.12942/lrr-2014-4},
  \href {http://ukads.nottingham.ac.uk/abs/2014LRR....17....4W} {17, 4}

\bibitem[\protect\citeauthoryear{{Wrobel}, {Miller-Jones}, {Nyland}  \&
  {Maccarone}}{{Wrobel} et~al.}{2018}]{Wrobel2018}
{Wrobel} J.~M.,  {Miller-Jones} J.~C.~A.,  {Nyland} K.~E.,   {Maccarone} T.~J.,
   2018, preprint, \href {http://adsabs.harvard.edu/abs/2018arXiv180606052W} {}
  (\mn@eprint {arXiv} {1806.06052})

\bibitem[\protect\citeauthoryear{{Yagi} \& {Stein}}{{Yagi} \&
  {Stein}}{2016}]{Yagi2016}
{Yagi} K.,  {Stein} L.~C.,  2016, \mn@doi [Classical and Quantum Gravity]
  {10.1088/0264-9381/33/5/054001}, \href
  {http://adsabs.harvard.edu/abs/2016CQGra..33e4001Y} {33, 054001}

\bibitem[\protect\citeauthoryear{Younsi}{Younsi}{2013}]{YounsiThesis}
Younsi Z.,  2013, PhD thesis, Mullard Space Science Laboratory, UCL

\bibitem[\protect\citeauthoryear{Younsi, Wu  \& Fuerst}{Younsi
  et~al.}{2012}]{Younsi2012}
Younsi Z.,  Wu K.,   Fuerst S.~V.,  2012, Astron. Astrophys., 545

\bibitem[\protect\citeauthoryear{{Zhang} \& {Saha}}{{Zhang} \&
  {Saha}}{2017}]{Zhang2017}
{Zhang} F.,  {Saha} P.,  2017, \mn@doi [\apj] {10.3847/1538-4357/aa8f47}, \href
  {http://adsabs.harvard.edu/abs/2017ApJ...849...33Z} {849, 33}

\bibitem[\protect\citeauthoryear{{d'Ambrosi}, {Satish Kumar}  \& {van
  Holten}}{{d'Ambrosi} et~al.}{2015}]{Ambrosi2015}
{d'Ambrosi} G.,  {Satish Kumar} S.,   {van Holten} J.~W.,  2015, \mn@doi
  [Physics Letters B] {10.1016/j.physletb.2015.03.007}, \href
  {http://adsabs.harvard.edu/abs/2015PhLB..743..478D} {743, 478}

\makeatother
\end{thebibliography}



\appendix

\section{Ray Tracing Initial Conditions}

The differential equations are integrated `backwards-in-time' from the observer image plane to the black hole, using a fifth-order Runge-Kutta-Fehlberg algorithm with adaptive step-size \citep[see][]{Press1992}. The centre of the observer's image plane is defined at some location $r_{\rm obs}, \theta_{\rm obs}$,  where $r_{\rm obs}$ is the distance from the black hole center and $\theta_{\rm obs}$ the angle from the positive black hole $z$-axis. Since the Kerr metric is axisymmetric we can set $\phi_{\rm obs}=0$. $r_{\rm obs}$ is chosen so as to be sufficiency large such that the observer's grid can be considered as a Euclidean grid with zero spacetime curvature, and all rays are perpendicularly incident on this grid. The coordinates $\alpha, \beta$ denote the location of a point on this grid location, where for $\theta_{\rm obs} = \pi/2$, $\beta$ is parallel to the positive black hole $z$-axis. The black hole coordinate system is right-handed with axes $\mathbf{x} = (x,y,z)^{\rm T}$. The observer's coordinate system is left-handed with axes $\mathbf{x'} = (x',y',z')^{\rm T}$ and the $z'$ axis oriented towards the centre of the black hole. Since we are interested in the plane coordinates $\alpha, \beta$ we can say $x' = \alpha, y' = \beta, z=0$. We determine the initial conditions of rays starting on the observer's grid. To do this we transform from $\mathbf{x'} \rightarrow \mathbf{x}$ via the methods outlined in \citet{YounsiThesis, Pu2016}:

\begin{enumerate}
	\item Rotate clockwise by $(\pi - \theta_{\rm obs})$ about the $x'$-axis ($R_{x'}$)
	\item Rotate clockwise by $(2\pi - \phi_{\rm obs})$ about the $z'$-axis ($R_{z'}$).
	\item Reflect in the plane $y'=x'$ ($A_{y'=x'}$).
	\item Translate $\mathbf{\bar{x'}}$ so that the origins of both coordinate systems coincide ($T_{x'\rightarrow x}$)
\end{enumerate}
The net transformation is then
\begin{eqnarray}
\mathbf{x} &= A_{y'=x'}R_{z'}R_{x'} \mathbf{x'} +T_{x'\rightarrow x} \\ 
&= \left( \begin{array}{c}
\mathcal{D}(y',z') \cos \phi_{\rm obs} -x' \sin\phi_{\rm obs}  \\
\mathcal{D}(y',z') \sin \phi_{\rm obs} +x' \cos\phi_{\rm obs}  \\
(r_{obs}-z') \cos \theta_{\rm obs} + y' \sin \theta_{\rm obs}  \end{array} \right)  \ ,
\end{eqnarray}
where $\mathcal{D} = (\sqrt{r_{\rm obs}^2 +a^2} -z') \sin \theta_{\rm obs} - y' \, \cos \theta_{\rm obs}$. We then transform from Cartesian to Boyer-Lindquist coordinates, 
\begin{eqnarray}
r = \frac{\sqrt{w+\sqrt{w^2+4a^2z^2}}}{2 } \ ;  \\ 
\theta = \arccos \left( \frac{z}{r}\right) \ ;   \\ 
\phi = \arctan2(y,x)  \ , 
\end{eqnarray}
where $w = x^2 +y^2 +z^2 -a^2$. 
This defines the initial $(r,\theta,\phi)$ for a photon on the observer grid.  

\noindent The initial velocities of the ray can then be determined. Since each ray arrives perpendicular to the image plane, 
$(\dot{x'}, \dot{y'}, \dot{z'}) = (0,0,1)$. 
Consequently, the velocity components in the black hole frame are given by 
\begin{eqnarray}
\dot{x} &= \left( \begin{array}{c}
-\sin \theta_{\rm obs} \cos \phi_{\rm obs}  \\
-\sin \theta_{\rm obs} \sin \phi_{\rm obs}  \\
-\cos \theta_{\rm obs}  \end{array} \right)   \ . 
\end{eqnarray}
Converting to Boyer-Lindquist coordinates gives expressions for $(\dot{r}, \dot{\theta}, \dot{\phi})$ 
in the black hole frame:
\begin{eqnarray}
\dot{r} = - \frac{-r\mathcal{R} \sin \theta \sin \theta_{\rm obs} \cos{\Phi} 
	+ \mathcal{R}^2\cos \theta \cos \theta_{\rm obs}}{\Sigma} \ , \\ 
\dot{\theta} = \frac{r \sin \theta \cos \theta_{\rm obs} 
	- \mathcal{R}\cos \theta \sin \theta_{\rm obs} \cos \Phi}{\Sigma} \ ,  \\ 
\dot{\phi} = \frac{\sin \theta_{\rm obs} \sin \Phi}{\mathcal{R} \sin \theta} \ ,
\end{eqnarray}   
where $\mathcal{R} = \sqrt{r^2 +a^2}$ and $\Phi = \phi - \phi_{\rm obs}$. 
This completely defines our initial conditions.
\section{MPD Spin-Interaction}

\subsection{Momentum-Velocity relation in the Tulczyjew-Dixon condition}
\noindent With the TD condition the pulsar mass is given by,
\begin{eqnarray}
m = \sqrt{- p^{\mu} p_{\mu}} \ ,
\end{eqnarray} 
The spin vector is given by,
\begin{eqnarray}
s_{\mu} = -\frac{1}{2m} \epsilon_{\mu \nu \alpha \beta} p^{\nu} s^{\alpha \beta} \ .
\end{eqnarray}
We can express the spin tensor in terms of the spin-vector,
\begin{eqnarray}
s^{\mu \nu} = \frac{1}{m} \epsilon^{\mu \nu \alpha \beta} p_{\alpha} s_{\beta} \ ,
\end{eqnarray}
and so it follows that,
\begin{eqnarray}
s^2 = s^{\mu} s_{\mu} = \frac{1}{2} s^{\mu \nu} s_{\mu \nu} \ ,
\end{eqnarray}
which is also a constant of the motion.  Contracting Eq. \ref{Eq:mpd2} with $u_{\nu}$ gives an expression for the momentum as,
\begin{eqnarray}
p^{\mu} = m u^{\mu} + u_{\nu}\frac{D S^{\mu \nu}}{d \tau} \ .
\end{eqnarray}

\subsection{Initialisation of MPD orbits}
We can initialize the $p^{\alpha}, s^{\alpha}, x^{\alpha}$ 4-vectors of our particle, which are related to the conserved quantities of the Kerr spacetime,  the mass $m$ of the pulsar, its energy $E$, angular momentum $L_z$ and Carter constant $Q$. More usefully, the conserved parameters $E, L_z, Q$ can be mapped to geometrical orbital parameters $p,e, z_-$ with the transformation described in \citet{Barausse2007} where $p$ is the semi-latus rectum, $e$ the eccentricity, $z_- = \cos^2 \theta_{\rm min}$ and $\theta_{\rm min}$ is the minimum angle reached by the pulsar. This framework is fundamentally an approximation since it does not include spin effects and so both the eccentricity and semi-latus rectum are not constant, as for a Keplerian orbit, but evolve with times \citep[see e.g.][]{Singh2014}. Nevertheless these variations are typically small they provide a decent first order approximation to the sorts of orbits that we want to model. 

\bsp	
\label{lastpage}
\end{document}